\documentclass[11pt,twoside]{report}
\usepackage[latin1]{standard}
\usepackage{pstricks,pst-coil}

\freestyle{\hspace*{-0.8ex}{\sl \chapterhead}\rule[-0.5ex]{0cm}{2ex}\hfill 
          \rm\thepage\vskip 3pt\hrule}{\rm \thepage\hfill 
	  \rule[-0.5ex]{0cm}{2ex}\sl \chapterhead\vskip 3pt\hrule}{\nop}{\nop}

\renewcommand{\chapappname}{}

\partstyle{\clearemptypage\thispagestyle{empty}}
       {\vspace*{7cm}\noindent\nop\hrulefill\nop\\}
       {\bf\sl\LV\partname\quad\thepart\quad}{\bf\sl\LV}
       {\nop\\[-0.4cm] \nop\hrulefill\nop\newpage\nop
       \thispagestyle{empty}\newpage}
\chapterstyle{\clearemptypage\thispagestyle{empty}}
       {\vspace*{10pt}\noindent\nop\hrulefill\nop\\}{\bf\sl\LV\thechapter\quad}
       {\bf\sl\LV}{\vspace*{-16pt}\nop\hrulefill\nop\\ \vspace*{30pt}}
\sectionstyle{-3.5ex plus-1pt minus-1pt}{\LII\bf\sl}{2.3ex plus 1pt}
\subsectionstyle{-3.25ex plus-1pt minus-1pt}{\LI\bf\sl}{1.5ex plus 1pt}
\subsubsectionstyle{-3.25ex plus -1ex minus-.2ex}{\bf\sl}{1.5ex plus.2ex}

\psset{linewidth=0.06cm}

\newcommand{\xloops}{{\md\up\ch xloops}\xsp}
\newcommand{\Cxx}{{\sf C\mbox{\raisebox{0.15ex}{\SII++}}}\xsp}
\newcommand{\maple}{{\sf Maple}\xsp}
\newcommand{\maplev}{{\sf Maple~V}\xsp}
\newcommand{\vegas}{{\sf VEGAS}\xsp}
\newcommand{\tcl}{{\sf Tcl/Tk}\xsp}
\newcommand{\bmm}[2]{\bm{#1\rule{0pt}{8pt}}^{\hspace*{-1.05ex}#2}}
\newcommand{\eps}{\varepsilon}
\newcommand{\Li}{\mbox{Li}_{2}}
\newcommand{\Tri}{\mbox{Li}_{3}}
\newcommand{\Spe}{\mbox{S}_{12}}

\newcommand{\eval}{\ovalbox{\rule[-0.25ex]{0cm}{1ex}{\SIII\sf Evaluate}}\xsp}
\newcommand{\evalf}{\ovalbox{\rule[-0.25ex]{0cm}{1ex}{\SIII\sf Eval. Full}}\xsp}
\newcommand{\ef}{\ovalbox{\rule[-0.25ex]{0cm}{1ex}{\SIII\sf E. Full}}\xsp}
\newcommand{\evalm}{\ovalbox{\rule[-0.25ex]{0cm}{1ex}{\SIII\sf Eval. More}}\xsp}
\newcommand{\evaln}{\ovalbox{\rule[-0.25ex]{0cm}{1ex}{\SIII\sf Eval. Numeric}}%
                    \xsp}

\newcommand{\Rfc}{${\mcl R}$ function}

\newcommand{\btt}[1]{\ovalbox{\rule[-0.25ex]{0cm}{1ex}{\SIII\sf #1}}\xsp}
\newcommand{\CC}[1]{\ifnum#1<10$\bigcirc$\hspace{-1.6ex}{\rm\SIII #1}\,\,\fi
                     \ifnum#1>9$\bigcirc$\hspace{-1.9ex}%
		     {\rm\SIII\scalebox{0.7 1}{#1}}\,\fi}
\newcommand{\Cp}{$\bigcirc$\hspace{-1.9ex}\raisebox{0.1ex}{\rm\SIII +}\,\,}
\newcommand{\Cm}{$\bigcirc$\hspace{-1.6ex}\raisebox{0.1ex}{\rm\SIII --}\,\,}

\indexstyle{3}{\addcontentsline{toc}{chapter}{\indexname}\typeout{\indexname}}
           {\SI\columnseprule0.4pt}
\makeindex

\begin{document}

\title{\bfr \NM MZ-TH 97-35 \\[1cm] \efr 
{\LVII\xloops} \\[2cm]
\bf A Program Package calculating \\[1cm]
One- and Two-Loop Feynman Diagrams \\[2cm]}
\author{L. Brücher, J. Franzkowski, D. Kreimer \\[1.5cm]} 
\date{October 1997 \\[1cm]} 

\maketitle

\thispagestyle{empty}
\noindent
L. Brücher\footnote{e-mail: \tt bruecher@dipmza.physik.uni-mainz.de},
J. Franzkowski\footnote{e-mail: \tt franzkowski@dipmza.physik.uni-mainz.de},
D. Kreimer\footnote{e-mail: \tt kreimer@dipmza.physik.uni-mainz.de} \\[0.4cm]
Institut für Physik \\
Johannes Gutenberg-Universität Mainz \\
Staudingerweg 7 \\
D-55099 Mainz \\
Germany

\tableofcontents

\chapter{Introduction}

\section{Intention}

The aim of \xloops is to calculate one-particle irreducible Feynman diagrams
with one or two closed loops for arbitrary processes in the Standard model of
particles and related theories. Up to now this aim is realized for all one-loop
diagrams with at most three external lines and for two-loop diagrams with two
external lines.

The results can be returned both algebraically and numerically. All necessary
tensor integrals are treated for arbitrary masses and momenta. Those two-loop
two-point functions for which no analytic result is known%
\index{two-loop functions!analytic result} are integrated numerically%
\index{numerical integration@numerical integration\\ \nop}%
\index{integration!numerical} by \xloops.

The calculations are performed in \maplev\index{Maple@\maple} -- a language
for symbolic computations \cite{Ma1,Ma2}. Numerical integrations are done using
\vegas\index{VEGAS@\vegas} -- a procedure for adaptive Monte Carlo integration%
\index{Monte Carlo integration}\index{integration!Monte Carlo} \cite{Ve1,Ve2}.
For convenient input \xloops has an {\sf Xwindows} interface%
\index{Xwindows interface@{\sf Xwindows} interface} based on \tcl%
\index{Tcl/Tk@\tcl} \cite{Tcl}.

The package consists of the following parts:
\bi
\item Input via {\sf Xwindows} interface: \\
  In a window the user selects the topology which shall be calculated. A
  Feynman diagram pops up in which the particle names have to be inserted.
\item Processing with \maple: \\
  The selected diagram is evaluated. The necessary steps for reducing the
  numerator ($SU(N)$ algebra, Dirac matrices) are performed. The result is
  expressed in terms of one- and two-loop integrals.
\item Evaluation of one-loop integrals: \\
  One-loop one-, two- and three-point integrals are calculated analytically or
  numerically to any tensor rank using \maple. This part was already released
  separately \cite{Fr4,Fr10}.
\item Evaluation of two-loop integrals: \\
  All two-loop two-point topologies including tensor integrals are supported
  by the \maple routines. For those topologies where no analytic result is
  known, \xloops creates either an analytic two-fold integral representation or
  integrates numerically with the help of \vegas using \Cxx\index{C@\Cxx}
  \cite{Cxx}. A parallelized implementation of the \vegas algorithm%
  \index{VEGAS@\vegas!parallelized implementation@parallelized\\ implementation}
  invented by R. Kreckel \cite{Kre1,Kre2} is part of this distribution.

  The other topologies can be calculated analytically or numerically like
  one-loop integrals.
\ei

In the second chapter we briefly denote how to install and start the program.
Chapter three gives several examples for quick and easy usage. A more detailed
description of the user-interface can be found in chapter four of this manual.

\section{Copyright}

The copyright\index{copyright} of this program package is owned by the Johannes
Gutenberg-Universität Mainz, Germany.

\xloops is ``{\it cite-ware}''. That means that you can freely use and
redistribute\index{redistribution} it, but you have to cite \xloops if you
publish any results that you achieved with the help of it.

Any further program development which intends to use the name \xloops or the
incorporation of \xloops or a part of the \xloops code into any other program
needs the permission of the authors.

With the receipt of \xloops the customer accepts that he uses the program
entirely on his own risk. Neither the authors nor the distributors of \xloops
are liable for any loss or damage as a result of any use which is made of this
program.

\section{Comments and bug reports}

\xloops is still under development. So if you have questions, comments%
\index{comments} or if you have found any bug -- please don't hesitate to
contact \\[0.4cm]
{\tt xloops@thep.physik.uni-mainz.de} \\[0.4cm]
or one of the authors.

\chapter{Preliminaries}

\section{Requirements}\index{requirements}

\xloops can be used on any computer where \maplev\index{Maple@\maple} is
working. If \maple is not yet installed on your computer system you have to do
that first. For numerical integration of two-loop integrals a compiler for
\Cxx\index{C@\Cxx} is necessary. The {\sf Xwindows} interface%
\index{Xwindows interface@{\sf Xwindows} interface} needs
\tcl\index{Tcl/Tk@\tcl} and is at present only available for {\sf
Unix}\index{Unix@{\sf Unix}}. On other platforms or if \tcl is not installed the
{\sf Xwindows} interface will not work. In this case you can use the \maple part
directly (cf. sect.~\ref{mapinst} and \ref{start}). The \maple part covers most
features of the \xloops package but the input may be found not as user-friendly
as if the interface is used.

\section{Installation}\index{installation} \label{inst}

The distribution\index{distribution} of the \xloops package is accessible%
\index{access} via WWW at \\[0.4cm]
\verb+http://wwwthep.physik.uni-mainz.de/~xloops/+ \\[0.4cm]
It consists of one packed (tared, gzipped) file \\[0.4cm]
{\tt xloops-1.0.tgz} \\[0.4cm]
This file contains all files needed for \xloops. To install the package unpack
{\tt xloops-1.0.tgz}: \\[0.4cm]
{\tt tar -xzvf xloops-1.0.tgz} \\[0.4cm]
\xloops then automatically creates a subdirectory of the current directory
which is called {\tt xloops}. The package is then deposited there. In addition
there will occur several subdirectories of {\tt xloops}. A complete list of all
files in the distribution is given in appendix~\ref{dist}. Moreover, there
exists an additional (tared, gzipped) file \\[0.4cm]
{\tt xloops-lib.tgz} \\[0.4cm]
in the same directory as {\tt xloops-1.0.tgz} which contains a library of
one-loop diagrams\index{one-loop diagrams!library} (cf. sect. \ref{lib}). For
convenience one should unpack this file in the same directory where {\tt
xloops-1.0.tgz} was unpacked before. \xloops will read from this library
whenever a one- or two-loop integral has to be evaluated. If this library is
missing \xloops will create the necessary files automatically, but this will
usually take some time. If you have the {\sf bash}\index{bash@{\sf bash}}
shell type \\[0.4cm]
{\tt cd xloops/} \\
{\tt configure}\index{configure@{\tt configure}} \\[0.4cm]
This will settle the configuration of the {\sf Xwindows} interface. {\tt
configure} will request
\bi
\item the release of \maplev\index{Maple@\maple!release} (1, 2, 3 or 4) you are
  using. There were subtle changes from release 2 to 3 in handling local and
  global variables, therefore it is important to select the correct option.
  Since we tested only \maplev release 1 and 3, for release 2 and 4 {\tt
  configure} internally does the following replacement: \\[0.4cm]
  release 2: use the same files as for release 1 \\
  release 4: use the same files as for release 3 (with some patches) \\[0.4cm]
  which is expected to be consistent.
\item the WWW browser\index{WWW browser} which should be used for the on-line
  html version\index{html version} of this manual. The name of the executable
  (cf. {\tt netscape}, {\tt mosaic}, \ldots) must be inserted.
\item the character encoding\index{character encoding} of your terminal
  (Isolatin 1\index{Isolatin 1} or standard ascii\index{ascii}).
\item whether multiprocessing\index{multiprocessing} is enabled. This question
  is necessary to choose between normal \vegas%
  \index{VEGAS@\vegas!normal implementation} or the parallelized version of
  \vegas%
  \index{VEGAS@\vegas!parallelized implementation@parallelized\\ implementation}
  running on multiprocessor computers only. If multiprocessing is enabled the
  number of processors is requested.
\item the platform. Several frequently used systems ({\sf Linux}, {\sf Digital
  Unix}, {\sf SUN Solaris}, {\sf IBM AIX})\index{Linux@{\sf Linux}}%
  \index{Digital Unix@{\sf Digital Unix}}\index{SUN Solaris@{\sf SUN Solaris}}%
  \index{IBM AIX@{\sf IBM AIX}} are distinguished from other
  {\sf Unix}\index{Unix@{\sf Unix}} to install the right {\tt Makefile} for
  the numerical integration%
  \index{numerical integration@numerical integration\\ \nop}%
  \index{integration!numerical}.
\item whether a library of one-loop integrals%
  \index{one-loop diagrams!library} should be build. It can take a lot of time
  (up to an hour) on slower computers to build the whole library. Therefore it
  is recommended not to do that, but to load a pre-built library {\tt
  xloops-lib.tgz} from our WWW server instead.
\ei
After that several checks are performed. First an embarassing bug occuring in
\maple releases up to release 3 is tested: If a file is read in a procedure, no
output of the procedure will appear. Therefore we have written a little
workaround for this bug, which can be installed by the user.

{\tt configure} continues with a test run of the {\tt Makefile}%
\index{Makefile@{\tt Makefile}} which is needed for numerical integrations.
This {\tt Makefile} depends on the platform. The run will take some time.

Finally you are asked to agree that automatically a mail is sent to the
developer team of the \xloops package. This mail is understood as
registration of your \xloops installation.

A typical run of {\tt configure}\index{configure@{\tt configure}} is given in
appendix \ref{instapp}. {\tt configure}\index{configure@{\tt configure}}
automatically creates the \xloops executable file\index{executable file}. 

If you have root access you can make it executable for all users. You then can
set a symbolic link\index{symbolic link} to {\tt /usr/bin}, i.e. \\[0.4cm]
\verb+cd /usr/bin+ \\
\verb+ln -s +$\langle${\it directory}$\rangle$\verb+/xloops/xloops .+ \\[0.4cm]
where $\langle${\it directory}$\rangle$ is the directory where \xloops was
installed.

If you don't have {\sf bash}\index{bash@{\sf bash}} you can find the necessary
instructions in the file {\tt README}\index{README@{\tt README}} which is part
of the \xloops distribution.

\section{Getting started}\index{program start}\index{start}

You can now type \\[0.4cm]
{\tt ./xloops} \\[0.4cm]
from the directory where \xloops was installed before. This starts your \xloops
session. The {\it main window}\index{main window@{\it main window}} of \xloops
(cf. fig. \ref{prog1}) will appear. If \xloops was made executable for all users
(cf. previous section) it can be started with \\[0.4cm]
{\tt xloops} \\[0.4cm]
from any directory.

There is on principle no restriction in the number of \xloops sessions which
can be used simultaneously\index{simultaneous sessions} -- by the same user or
by different users. On the other hand the numerical integration%
\index{numerical integration@numerical integration\\ \nop} with \vegas%
\index{VEGAS@\vegas} should be used only once by each user (cf. sect.
\ref{num}).

\section{Installing the {\md\maple} part only}
\index{maple part@\maple part!installation} \label{mapinst}

If you are only interested in the \maple routines\index{maple part@\maple part}
you have to start the installation as in sect. \ref{inst}, but you then skip
{\tt configure} and just copy all files with extension {\tt .ma}
\bi
\item loops.ma
\item fmrules.ma
\item fmuser.ma
\item evalproc.ma
\item oneloop.ma 
\item cfcn.ma 
\item pv.ma 
\item r.ma 
\item simple.ma 
\item twoloop.ma
\item numint.ma
\item values.ma
\item unvalue.ma
\item mess.ma
\ei
in one directory instead and delete the rest of the distribution. The correct
{\tt  .ma} files you can find in the subdirectory {\tt xloops/MapleVR1} -- if
you use release\index{Maple@\maple!release} 1 or 2 -- or in the subdirectory
{\tt xloops/MapleVR3} -- if you have release 3 or higher.

\section{Usage within a {\md\maple} session} \label{start}

You can use \xloops within an ordinary \maple session%
\index{Maple@\maple!session} as well: After having started \maple just type
the following lines: \\[0.4cm]
\verb+LoopPath:=+{$\langle$\it path$\rangle$}\verb+;+%
\index{LoopPath@{\tt LoopPath}} \\
\verb+read`+{$\langle$\it path$\rangle$}\verb+loops.ma`;+ \\[0.4cm]
{$\langle$\it path$\rangle$} describes the directory where the {\tt .ma} files
are located. If the \xloops package is installed properly this should be the
path of the \xloops directory to which you have to add {\tt MapleVR1} -- if you
are using release 1 or 2 of \maplev -- or {\tt MapleVR3} -- if you are using
release 3 or higher. If the {\tt .ma} files were copied to the current directory
before you can skip the assignment of \verb+LoopPath+ and only type \\[0.4cm]
\verb+read`loops.ma`;+ \\[0.4cm]
Now you can use the functions described in chapter \ref{detail}.

\chapter{Quick reference}

\section{{\it Main window}}

After having typed the command {\tt xloops} the {\it main window}%
\index{main window@{\it main window}} (cf. fig. \ref{prog1}) appears on the
screen. It consists of three parts:
\ben
\item The {\it menu bar}\index{menu bar@{\it menu bar}} which contains several
  menues for \xloops commands (for details cf. sect.~\ref{menu}).
\item The {\it topology bar}\index{topology bar@{\it topology bar}} which
  displays all possible topologies\index{topology} for a definite $m$-loop
  $n$-point function. The selection of $m$ closed loops\index{closed loops}
  and $n$ external lines\index{external lines} is done with the help of the
  menues {\sf Loops}\index{Loops@{\sf Loops}} and {\sf Ext. Lines}%
  \index{Ext. Lines@{\sf Ext. Lines}}.
\item The {\it text window}\index{text window@{\it text window}} with scroll
  bar which contains the results of calculations.
\een
\bfg[h]\bc
\epsfig{file=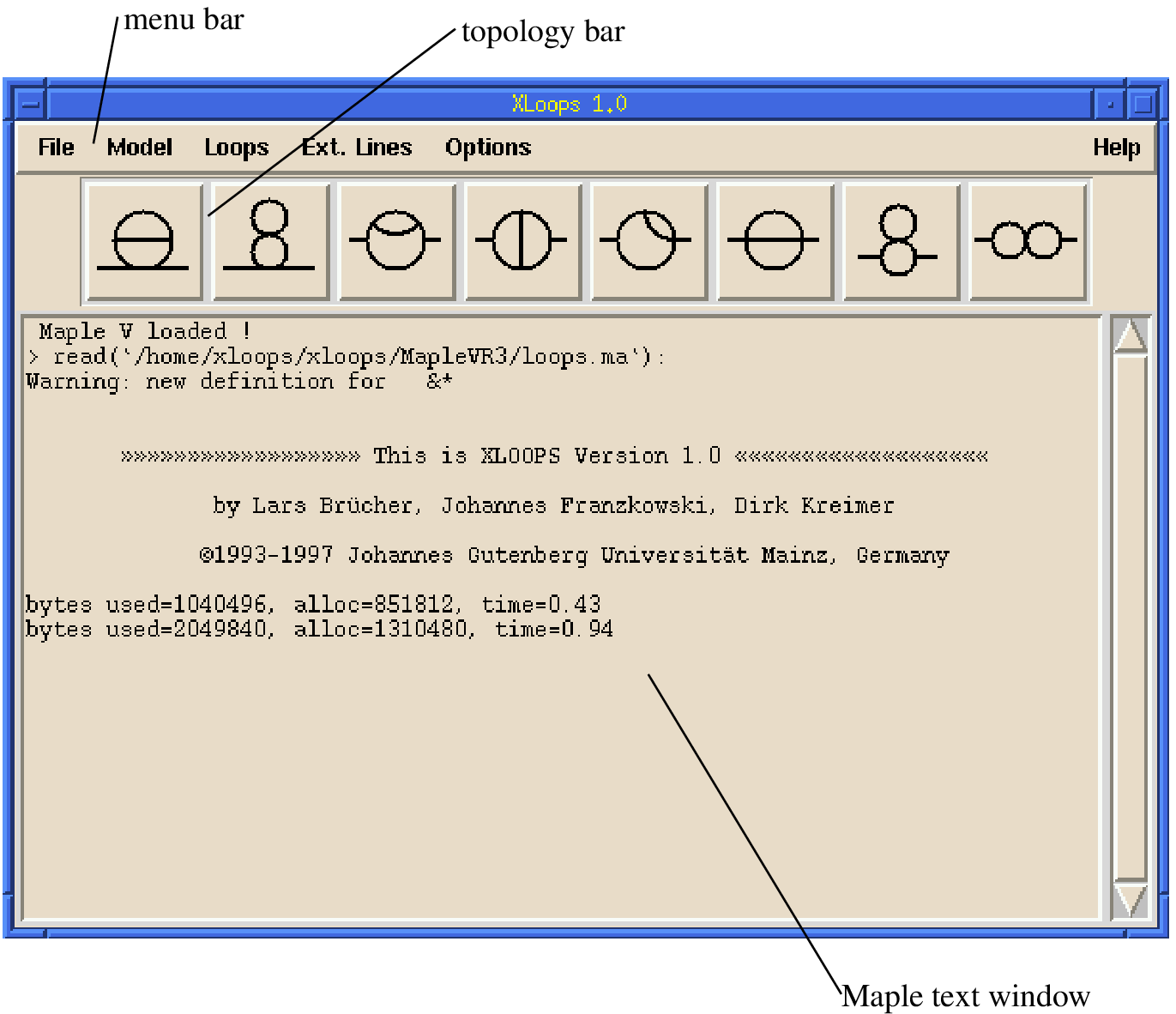,width=13.2cm}
\caption{{\it Main window}} \label{prog1}
\ec\efg

Before we will explain the functions and their meanings in detail in the next
chapter, we demonstrate here as an example how to use \xloops for the
calculation of a two-loop self-energy diagram and a one-loop vertex correction.

\section{Two-loop self-energy}

We now perform the calculation of a diagram which contributes to the Higgs
self-energy~$\Sigma$\index{Higgs self-energy} at the two-loop level:
\bc
{\setlength{\shadowsize}{0.1cm}\setlength{\fboxsep}{0.1cm}
\shadowbox{\rule{0cm}{2.8cm}\epsfig{file=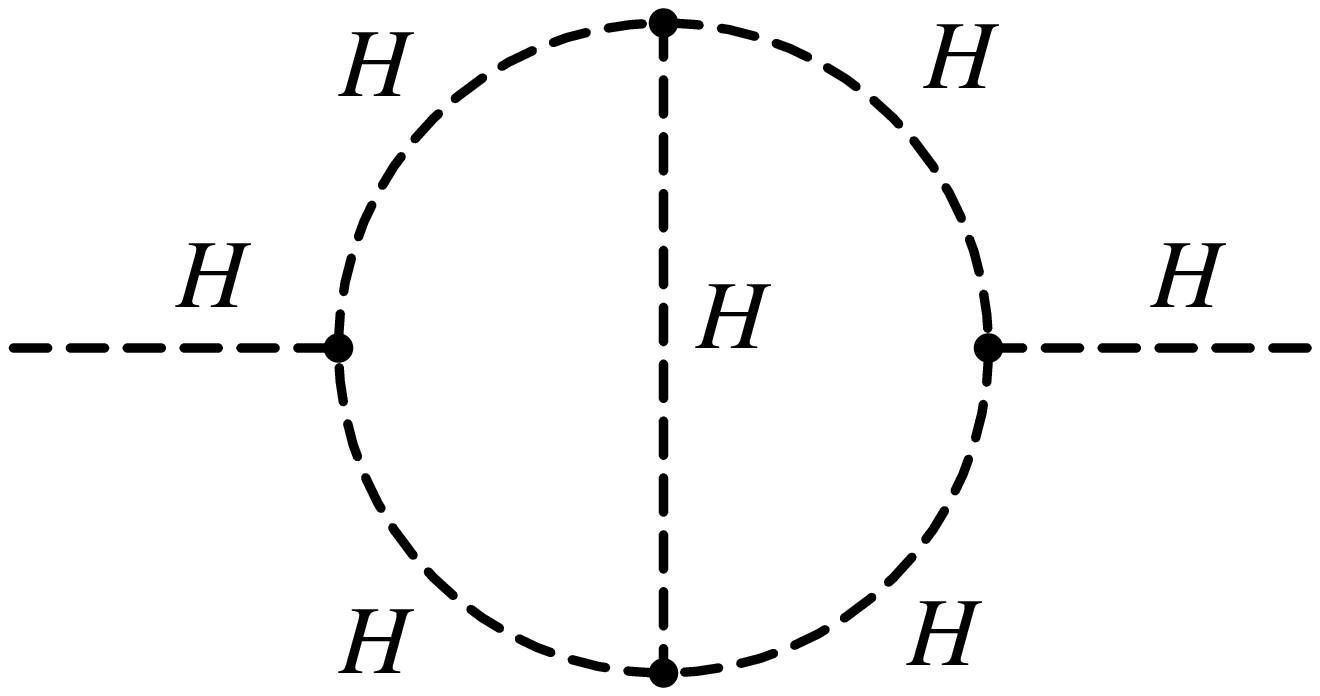,width=4.8cm}}}
\ec
The underlying topology is the master topology\index{master topology}, which
is located in the fourth place in the {\it topology bar}%
\index{topology bar@{\it topology bar}} (cf. fig. \ref{prog1}). In this case
the number of external lines and the number of closed loops is already fixed
correctly, so one now has just to click on the topology\index{topology}.

Then the master topology appears in a separate window, the {\it diagram window}%
\index{diagram window@{\it diagram window}\\ \nop} (cf. fig.~\ref{prog2}). In
this window for each internal and external particle the corresponding particle
name\index{particle names} has to be inserted. Only particles can occur here, no
antiparticles. The allowed terms for particle names are listed in sect.
\ref{masses}. They can be obtained alternatively with the menu {\sf Help}%
\index{Help@{\sf Help}}. The flow of charges\index{charge} and fermion%
\index{fermion number} or ghost numbers\index{ghost number} is determined by the
arrows\index{arrow} of the lines. The direction of an arrow%
\index{arrow!direction} is reverted with a click on it.
\bfg[h]\bc
\epsfig{file=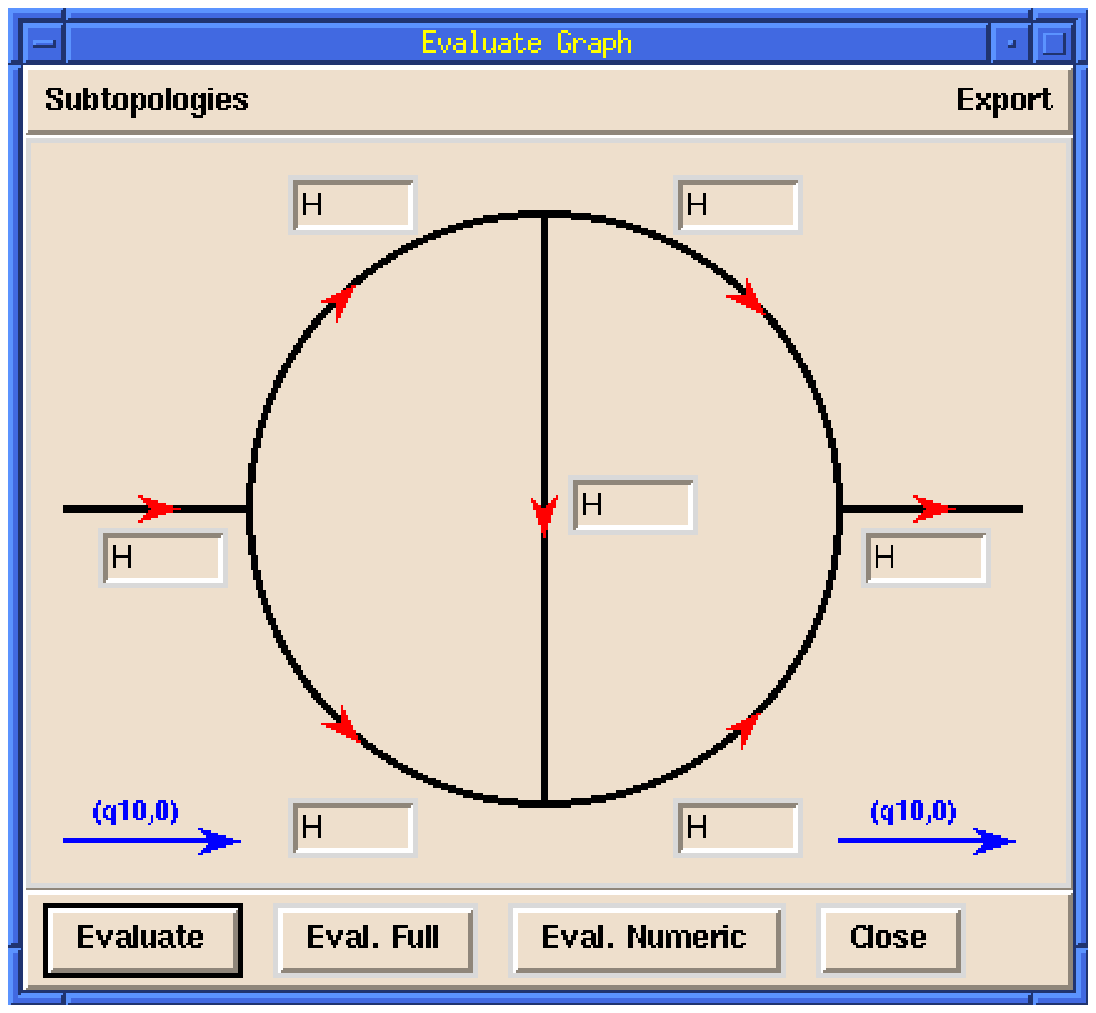,width=10cm}
\caption{The diagram for the two-loop Higgs self-energy} \label{prog2}
\ec\efg

Since in our example we have to deal only with chargeless scalars, the direction
of arrows is meaningless and we have to insert the name {\sf H} or {\sf higgs}
for each particle only.

For the sake of clearness, the flow of external momenta\index{external momenta}
in our conventions is displayed in blue. In this example there is only one
momentum
\bdm
\bm{q_1} = (q_{10},0,0,0) = (q_{10},0)\,.
\edm
Vanishing components are put together to a common zero-vector. In our case there
exists only one parallel space\index{parallel space} component $q_{10}$. The
orthogonal space\index{orthogonal space} is represented through a common $0$.

When all particles are inserted, \xloops is ready for calculation. The
{\it diagram window}\index{diagram window@{\it diagram window}\\ \nop} gives
you the possibility to select different modes of evaluation%
\index{evaluation!modes}.
\bi
\item Click on the \eval button\index{Evaluate@\eval}: The output --
  for simplicity -- is returned in terms of \verb+Oneloop+%
  \index{Oneloop@{\tt Oneloop}} and \verb+Twoloop+\index{Twoloop@{\tt Twoloop}}
  functions which correspond directly to one-loop and two-loop functions
  explained in sect. \ref{Oneloop} and \ref{Twoloop}. In our example the
  result consists only of the scalar master two-loop two-point function which
  \xloops calls \verb+Twoloop2Pt2+\index{Twoloop2Pt2@{\tt Twoloop2Pt2}}:

  {\SII
  \begin{verbatim}
> G1 := EvalGraph2 (7, [higgs,higgs,higgs,higgs,higgs,higgs,higgs,higgs,higgs,
> higgs,higgs,higgs] );

G1 := [

          81          8  4                   2   2         2        2   2 2
    C1 = ---- I Mhiggs  e  (1 + 2 eps Ln(4 Pi  MU ) + 2 eps  Ln(4 Pi  MU ) )
         8192

        Twoloop2Pt2(0, 0, 0, 0, 0, q10, Mhiggs, Mhiggs, Mhiggs, Mhiggs, Mhiggs)

           /         4   4   8
          /  (sin(tw)  Mw  Pi ),
         /

    C1]
\end{verbatim}}

  Generally, the output is decomposed in one or several form factors%
  \index{form factors} {\tt C1}, {\tt C2}, \ldots\ to reflect the Dirac
  $\gamma$\index{Dirac gamma@Dirac $\gamma$} or Lorentz structure%
  \index{Lorentz structure} of the result. Therefore the result\index{result}
  is a list of form factors. The last entry in this list is the defining
  equation for the form factors. In our example of a scalar self-energy there
  appears only one form factor {\tt C1} and the defining equation is just the
  scalar {\tt C1}. A more complicated situation can be found in the next
  section (for details cf. sect~\ref{output}).

  Moreover \xloops assigned the whole result to a variable {\tt G1}%
  \index{result!assignment to a variable}. Within an \xloops session each
  calculation will get such a name in an unique way. This gives the user the
  opportunity to perform further calculations with the results of diagrams,
  for instance to sum up all evaluated graphs which contribute to the same
  process.    
\item Click on the \evalf button\index{Eval. Full@\evalf}: \xloops
  evaluates directly the {\tt Oneloop}\index{Oneloop@{\tt Oneloop}} and {\tt
  Twoloop}\index{Twoloop@{\tt Twoloop}} integrals. Now each form factor%
  \index{form factors} {\tt C1}, \ldots\ is a Laurent expansion in terms of the
  ultraviolet regulator $\eps$\index{regulator!ultraviolet}, where the
  significant coefficients of this expansion -- ${\mcl O}(\eps^{-1})$, ${\mcl O}
  (\eps^0)$ in the one-loop case, ${\mcl O}(\eps^{-2})$, ${\mcl O}(\eps^{-1})$,
  ${\mcl O}(\eps^0)$ in the two-loop case -- are denoted in form of a list (cf.
  sect~\ref{output}).

  For those two-loop topologies\index{topology!two-loop} where no analytic
  result\index{result!non-analytic} is known -- like in the case of our
  example, the finite part contributes two entries to the output list. The
  first of them is a list itself which denotes an integral representation%
  \index{integral representation}, the second contains the part which is
  analytically calculable:

  {\SII
  \begin{verbatim}
GF1 := [

C1 = [0, 0,

               8  4
  81     Mhiggs  e
[---- ----------------,
 8192        4   4   4
      sin(tw)  Mw  Pi

                       2                                           2
             2   Mhiggs                        2             Mhiggs
  I Ln(Sqrt(x  - ------- + I rho) + %1 + Sqrt(y  - 2 y + 1 - ------- + I rho))
                      2                                           2
                   q10                                         q10
4 ----------------------------------------------------------------------------
                             2
                          q10  (- 2 x - 1) (- 2 y + 1)

                                       2
                   2             Mhiggs
    - 4 I Ln(Sqrt(x  + 2 x + 1 - ------- + I rho) + %1
                                      2
                                   q10

                                2
            2             Mhiggs              /     2
    + Sqrt(y  - 2 y + 1 - ------- + I rho))  /  (q10  (- 2 x - 1) (- 2 y + 1))
                               2            /
                            q10

                             2                                 2
                   2   Mhiggs                        2   Mhiggs
        I Ln(Sqrt(x  - ------- + I rho) + %1 + Sqrt(y  - ------- + I rho))
                            2                                 2
                         q10                               q10
    - 4 ------------------------------------------------------------------ + 4
                              2
                           q10  (- 2 x - 1) (- 2 y + 1)

                                  2                                 2
              2             Mhiggs                        2   Mhiggs
   I Ln(Sqrt(x  + 2 x + 1 - ------- + I rho) + %1 + Sqrt(y  - ------- + I rho))
                                 2                                 2
                              q10                               q10
   ----------------------------------------------------------------------------
                              2
                           q10  (- 2 x - 1) (- 2 y + 1)

   ,

x = - infinity .. infinity, y = - infinity .. infinity],

0],

C1]

                                                2
                         2            2   Mhiggs
%1 :=              Sqrt(x  + 2 x y + y  - ------- + I rho)
                                               2
                                            q10
\end{verbatim}}

  In our example the result is finite, therefore the first two entries in the
  list are zero. The only nonvanishing contribution comes from the integral
  representation. A more detailed description of the integral representation%
  \index{integral representation} is given in sect. \ref{twoout}.

  The result is a function of the Higgs mass\index{Higgs mass} \verb+Mhiggs+%
  \index{Mhiggs@{\tt Mhiggs}} and the external momentum%
  \index{external momenta} \verb+q10+. The coupling of the vertices involve
  the $W$ mass\index{W mass@$W$ mass} \verb+Mw+\index{Mw@{\tt Mw}}, the
  Weinberg angle\index{Weinberg angle} \verb+tw+\index{tw@{\tt tw}} and the
  electromagnetic coupling constant\index{couplings!electromagnetic} \verb+e+%
  \index{e@{\tt e}}. \verb+rho+ denotes the infinitesimal imaginary part%
  \index{masses!infinitesimal imaginary part@infinitesimal\\ imaginary part} of
  the masses in the denominator (cf. sect.~\ref{output}), \verb+x+ and
  \verb+y+ are the variables which have to be integrated out numerically.

\item Click on the \evaln button\index{Eval. Numeric@\evaln}:
  The integral representation\index{integral representation} described above
  is evaluated directly with the help of \vegas.\index{VEGAS@\vegas} This
  option exists only for two-loop integrals, because all one-loop integrals
  are analytically calculable.

  Please be aware that \xloops can only perform the numerical integration%
  \index{numerical integration@numerical integration\\ \nop} if all variables --
  \verb+Mhiggs+ and \verb+q10+ in our example -- are assigned with numerical
  values. For that purpose the menu {\sf File}\index{File@{\sf File}} provides
  the {\sf Insert Maple Command}%
  \index{Insert Maple Command@{\sf Insert Maple\\ Command}} entry. If this
  option is selected, a window (cf. fig. \ref{fig:insmap}) appears, where any
  command can be passed directly to \maple.\index{Maple@\maple} In our example
  we insert \\[0.4cm]
  {\tt q10:=91.17;} \\
  {\tt Mhiggs:=300;} \\[0.4cm]
  to evaluate the integral on the $Z$ resonance for an assumed value of 300
  GeV for the Higgs mass.
  \bfg[h]\bc
  \epsfig{file=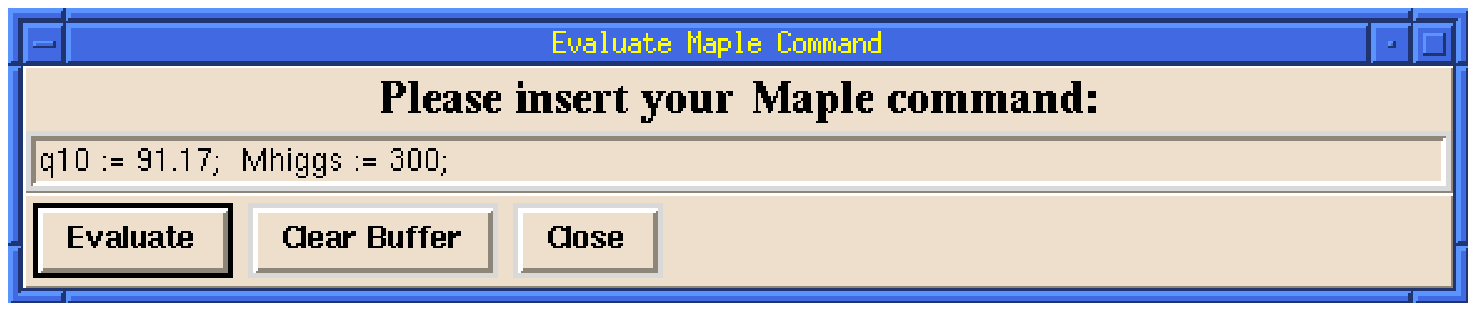,width=12cm}   
  \caption{Direct \maple input of masses and momenta} \label{fig:insmap}
  \ec\efg
  After clicking on \evaln \xloops starts the numerical integration. The result
  including the numerical uncertainty appears in the {\it main window}%
  \index{main window@{\it main window}} and reads:

  {\SII
  \begin{verbatim}
GM1 := [

C1 = [0, 0,

                                      4
                             16      e
    [.66598680119771701375*10   ------------,
                                       4   4
                                sin(tw)  Mw

                      -8             -5             -7             -7
        [ - .354349*10   + .880157*10   I, .13042*10   + .148817*10   I]], 0],

C1]
\end{verbatim}}

  In the result\index{result} -- compared to the \evalf\index{Eval. Full@\evalf}
  option -- the integral representation is now replaced by the numerical result%
  \index{result!numerical} of the integral. This result consists of a list: a
  pre-factor and another list. This list contains two entries: first the
  result of the \vegas\index{VEGAS@\vegas} calculation, then the uncertainty%
  \index{result!uncertainty}\index{numerical uncertainty} \vegas detected for
  this result (for details cf. sect.~\ref{output}).

  In our example the real part of the diagram is less than the numerical
  uncertainty, whereas the imaginary part is significantly larger. This means
  that the real part in fact is zero. Since we calculated the entire diagram
  which is not precisely $\Sigma$ but $-i\Sigma$, this is what one expects
  below threshold\index{threshold}.

  Alternatively to the direct input of \maple\index{Maple@\maple} commands
  described above it is possible to assign all known masses%
  \index{masses!assignment} of particles, couplings\index{couplings!assignment}
  and elements of the CKM matrix\index{CKM matrix!assignment} with the option
  {\sf Insert Particle Properties}%
  \index{Insert Particle Properties@{\sf Insert Particle\\ Properties}} of the
  {\sf Options}\index{Options@{\sf Options}} menu.

  The input information for \vegas\index{VEGAS@\vegas} can also be determined
  with the menu {\sf Options}: Click on {\sf Numeric}%
  \index{Numeric@{\sf Numeric}} and an additional window will appear where
  the input parameters for \vegas\index{VEGAS@\vegas!input parameters} can be
  fixed (for details cf. sect. \ref{winopt}).

  Finally the result can be saved\index{result!saving} in text mode to a file.
  Use in the menu {\sf File}\index{File@{\sf File}} the entry {\sf Save Maple
  Output}\index{Save Maple Output@{\sf Save Maple Output}}.
\item Click on the \btt{Close} button:\index{Close@\btt{Close}} The {\it
  diagram window}\index{diagram window@{\it diagram window}\\ \nop} disappears.
\ei
The four buttons described above are completely independent. You can for example
select \evaln\index{Eval. Numeric@\evaln} without having clicked on \eval%
\index{Evaluate@\eval} or \evalf\index{Eval. Full@\evalf} before.

\section{One-loop vertex correction}

What follows is the calculation of a one-loop contribution of the correction to
the $eeZ$ vertex\index{eeZ vertex@$eeZ$ vertex}:
\bc
{\setlength{\shadowsize}{0.1cm}\setlength{\fboxsep}{0.1cm}
\shadowbox{\rule{0cm}{2.8cm}\epsfig{file=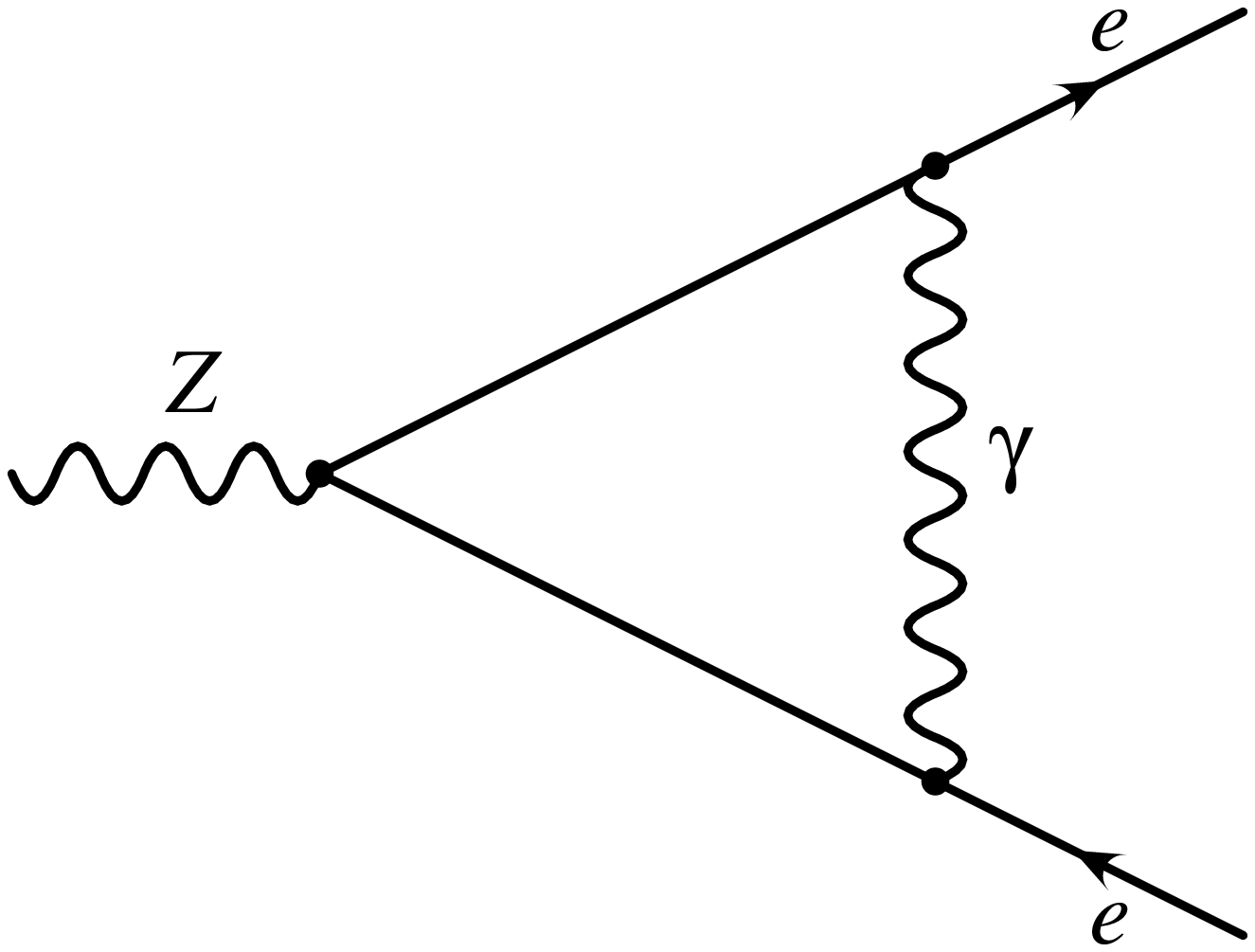,width=4.8cm}}}
\ec
First one has to select the correct values in the menues {\sf Loops}%
\index{Loops@{\sf Loops}} and {\sf Ext. Lines}%
\index{Ext. Lines@{\sf Ext. Lines}} of the {\it main window}%
\index{main window@{\it main window}} to get access to the one-loop
three-point topologies. A click on the correct topology\index{topology} opens
the {\it diagram window}\index{diagram window@{\it diagram window}\\ \nop} (cf.
fig.~\ref{prog3}) where the particle names\index{particle names} have to be
inserted. Again the correct particle names can be found in sect.
\ref{masses} or in the menu {\sf Help}\index{Help@{\sf Help}}. The
direction of the fermion line\index{fermion line!direction} must be adjusted
by clicking on the arrows\index{arrow}\index{arrow!direction}.
\bfg[t]\bc
\epsfig{file=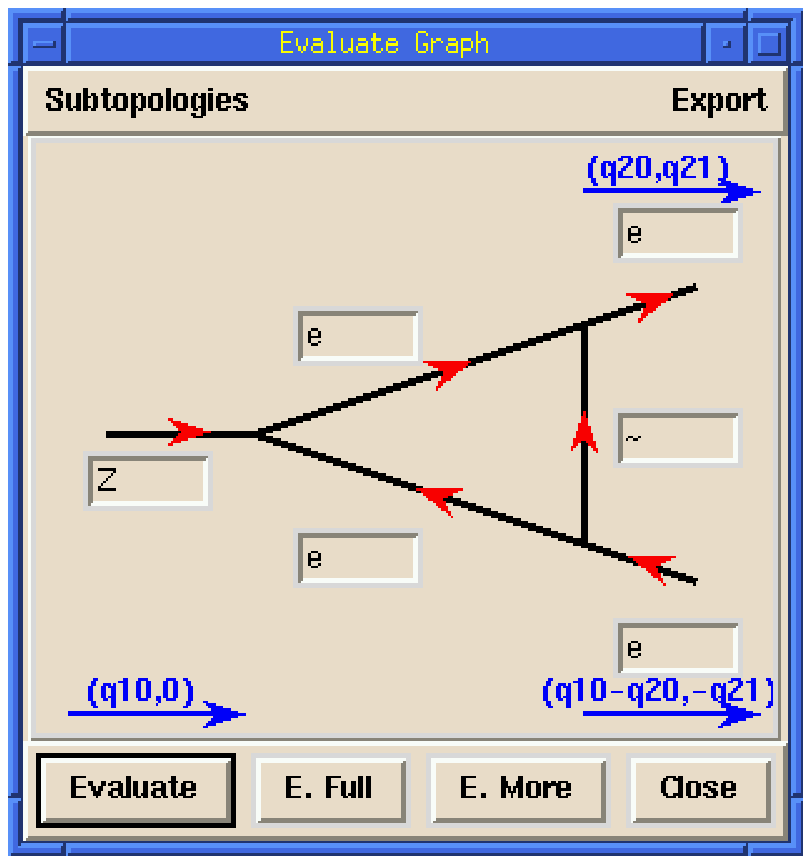,width=7cm}
\caption{The diagram for the $eeZ$ one-loop vertex correction} \label{prog3}
\ec\efg
The flow of external momenta\index{external momenta} is displayed in blue
again. The three momenta are
\beas
\bm{q_1}     & = & (q_{10},0,0,0) = (q_{10},0) \\
\bm{q_2}     & = & (q_{20},q_{21},0,0) = (q_{20},q_{21}) \\
\bm{q_1-q_2} & = & (q_{10}-q_{20},-q_{21},0,0) = (q_{10}-q_{20},-q_{21})
\eeas
where the vanishing orthogonal space\index{orthogonal space} components are
not displayed in the window. $q_{10},q_{20},q_{21}$ are the kinematical
variables\index{kinematical variables} \xloops uses for three-point functions.
To get a numerical result, we assign\index{kinematical variables!assignment}
these variables with \\[0.4cm]
\verb+q10:=Mz0; q20:=Mz0/2; q21:=evalf(sqrt(Mz0^2/4-Melec^2));+ \\[0.4cm]
using the option {\sf Insert Maple Command}%
\index{Insert Maple Command@{\sf Insert Maple\\ Command}} of the menu {\sf
File}\index{File@{\sf File}} so that all external particles\index{external
particles} are on-shell\index{external particles!on-shell}. Moreover the masses%
\index{masses!assignment} and couplings\index{couplings!assignment} are
assigned by the option {\sf Insert Particle Properties}%
\index{Insert Particle Properties@{\sf Insert Particle\\ Properties}} of the
{\sf Options}\index{Options@{\sf Options}} menu. The renormalization scale%
\index{renormalization scale} \verb+MU+\index{MU@{\tt MU}} and the unit
\verb+GeV+\index{GeV@{\tt GeV}} can be omitted by setting \\[0.4cm]
\verb+MU:=1; GeV:=1;+ \\[0.4cm]
Clicking on the \ef button\index{E. Full@\ef} gives the following result:

  {\SII
  \begin{verbatim}
GF2 := [

  C1 = [.000010058632263504308717 I,

       - .00086276496443282788638 - .0072236143340212361270 I],

     C2 = [.00010510130634321490041 I,

          - .0090149159887518402249 - .075478582279918574719 I],

     C3 = [0, .00083116483915048522171 + .0071524795241752712917 I],

     C4 = [0, .0086847304974275009526 + .074735304153769864092 I],

                                      -9                           -8
     C5 = [0, .40863107990668029069*10   + .35164209243516742716*10   I],

     C6 = [0, .000018229897665296319548 + .00015687498271956514041 I],

     C7 = [0, .00019048176819004810260 + .0016391657617525244669 I],

                                         -28                           -27
     C8 = [0,  - .25407817700937862122*10    - .21701516787925566779*10    I],

     C9 = [0, .000017536874322335126422 + .00015153965406412662322 I],

     C10 = [0, .00018324045975332954356 + .0015834176245538375947 I],

                                          -9                           -8
     C11 = [0,  - .20431553995334014636*10   - .17582104621758371369*10   I],

                                       -8                           -7
     C12 = [0, .22971899930888218199*10   + .19620932339352982103*10   I],

                                       -25
     C13 = [ - .24074590585723032270*10    I,

                                 -6                           -5
         .34651167097341202104*10   + .28882798115295165844*10   I],

                                       -5
     C14 = [0, .36206542130597753444*10   + .000030179250351750043557 I],

                                       -28                           -27
     C15 = [0, .12703908850468931061*10    + .10850758393962783388*10    I],

                                       -6                           -5
     C16 = [0, .34651167090811890878*10   + .26676643227928772485*10   I],

                                       -5
     C17 = [0, .36206542123775363980*10   + .000027874068547868333215 I],

     C1 (1 &* Dg(nu1)) + C2 &*(1, Dg5, Dg(nu1)) + C3 &*(1, Dg0, Dg1, Dg(nu1))

          + C4 &*(1, Dg5, Dg0, Dg1, Dg(nu1)) + C5 (1 &* ONE) q2(nu1)

          + C6 (1 &* Dg0) q2(nu1) + C7 &*(1, Dg5, Dg0) q2(nu1)

          + C8 &*(1, Dg0, Dg1) q2(nu1) + C9 (1 &* Dg1) q2(nu1)

          + C10 &*(1, Dg5, Dg1) q2(nu1) + C11 (1 &* ONE) q1(nu1)

          + C12 (1 &* Dg5) q1(nu1) + C13 (1 &* Dg0) q1(nu1)

          + C14 &*(1, Dg5, Dg0) q1(nu1) + C15 &*(1, Dg0, Dg1) q1(nu1)

          + C16 (1 &* Dg1) q1(nu1) + C17 &*(1, Dg5, Dg1) q1(nu1)             ]
\end{verbatim}}

The output\index{output} is decomposed in several form factors%
\index{form factors} {\tt C1}, {\tt C2}, \ldots\ to reflect the Dirac $\gamma$%
\index{Dirac gamma@Dirac $\gamma$} and Lorentz structure%
\index{Lorentz structure} of the result. The last entry in this list is the
defining equation for the form factors (cf. sect.~\ref{output}).

\chapter{Detailed description} \label{detail}

There are two intentions\index{intentions} covered by \xloops:
\bi
\item Either the calculation of complete Feynman diagrams. Here you need the
  {\sf Evaluate Graph} window (see sect. \ref{win}) where you can insert
  particles into a Feynman diagram. Alternatively you can use the {\tt
  EvalGraph} procedures in \maple described in sect. \ref{evalgraph}.
\item Or the solution of single integrals occurring in the calculation of one-
  and two-loop diagrams. Then you can make use of the {\tt OneLoop} and {\tt
  TwoLoop} procedures explained in sect. \ref{Oneloop} and~\ref{Twoloop}.
\ei
The creation of a library for the integrals follows in sect. \ref{lib}. The
numerical integration of two-loop integrals is explained in sect. \ref{num}.

\section{The {\md\sf Xwindows} interface}%
\index{Xwindows interface@{\sf Xwindows} interface} \label{win}

\subsection{The functions of the {\it menu bar}}\index{menu bar@{\it menu bar}}
\label{menu}

The {\it menu bar} provides a variety of different functions for the user. These
functions are combined to groups which appear as menues if the user clicks on
it. The list below gives a short overview. Additional descriptions can be
found in the following sections.
\bfg[t]\bc
\epsfig{file=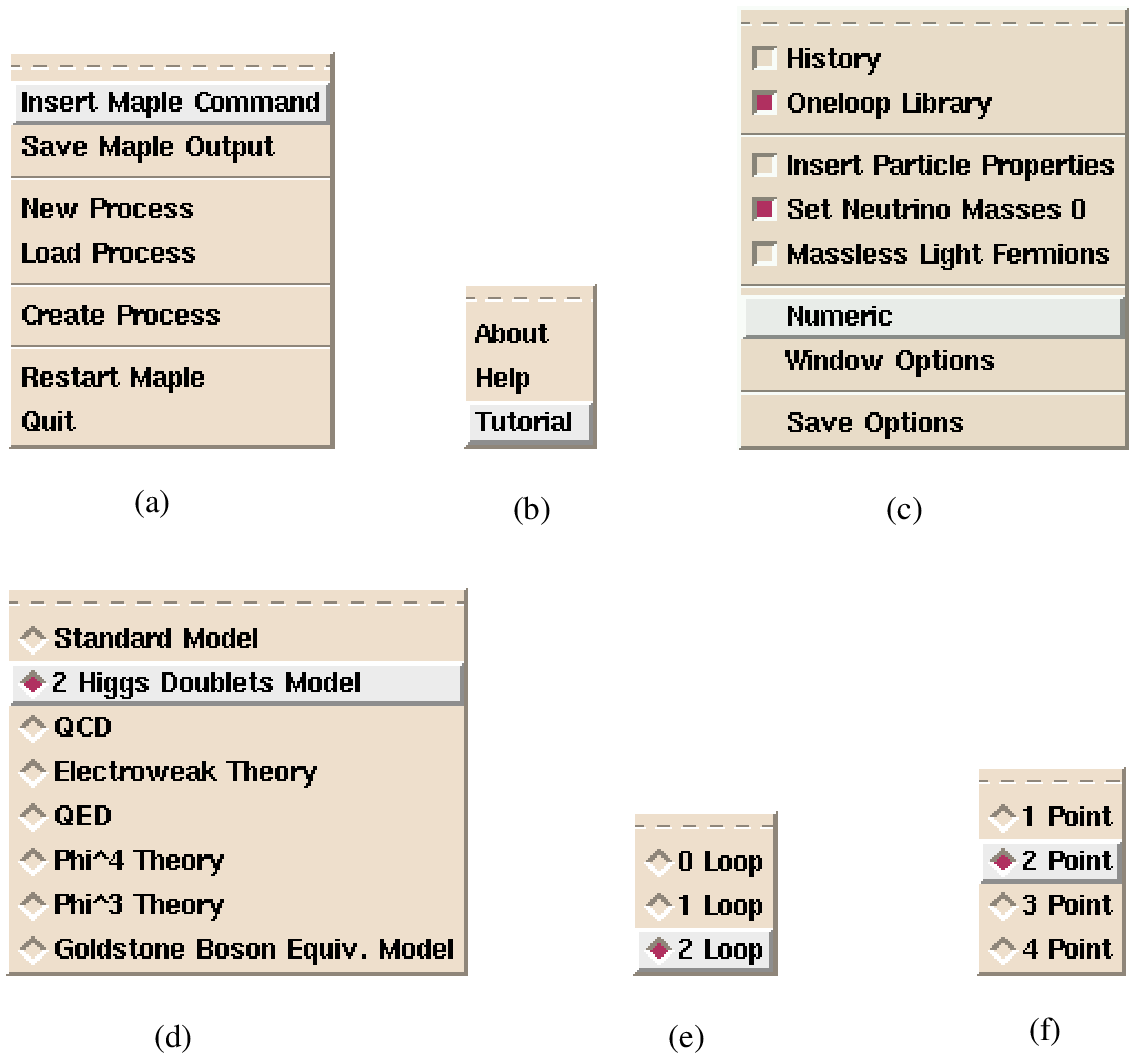}    
\caption{The menues of the {\it menu bar}} \label{fig:menues}
\ec\efg
\bi
\item The menu {\sf File}\index{File@{\sf File}} (cf. fig. \ref{fig:menues}a)
  has the following entries:
  \bi
  \item {\sf Insert Maple Command}%
    \index{Insert Maple Command@{\sf Insert Maple\\ Command}} allows the
    insertion of an arbitrary command which is passed directly to \maple%
    \index{Maple@\maple} (cf. fig. \ref{fig:insmap}). Especially the numerical
    input\index{numerical input} of masses\index{masses!numerical input},
    momenta\index{momenta!numerical input} and renormalization conditions%
    \index{renormalization conditions!numerical input} can be done easily this
    way. 
  \item {\sf Save Maple Output}\index{Save Maple Output@{\sf Save Maple Output}}
    saves the output as it is denoted in the \maple\ {\it text window}%
    \index{text window@{\it text window}} to an extra file. The name of the
    file has to be determined in a separate window, which appears if the user
    clicks on this entry.
  \item {\sf New Process}\index{New Process@{\sf New Process}} allows the user
    to collect a set of graphs and save them with a common process name.
    Detailed explanation can be found in section \ref{procsave}.
  \item With {\sf Load Process}\index{Load Process@{\sf Load Process}}
    processes which were declared with {\sf New Process} before can be reloaded
    into the current \xloops session (cf. sect. \ref{procsave}). 
  \item {\sf Create Process}\index{Create Process@{\sf Create Process}} will
    call the graph generator to produce all graphs to a given number of loops
    and external lines in a pre-defined model (in future versions of \xloops
    only).
  \item {\sf Restart Maple}\index{Restart Maple@{\sf Restart Maple}} terminates
    the current \maple session\index{Maple@\maple!session} and removes all
    assigned variables\index{variables!removal}. \maple and \xloops are
    started again.
  \item {\sf Quit}\index{Quit@{\sf Quit}} terminates \xloops.
  \ei
\item The menu {\sf Model}\index{Model@{\sf Model}} (cf. fig. \ref{fig:menues}d)
  offers several physical models\index{physical models} for calculation (for
  details cf. sect. \ref{models}).
\item The menu {\sf Loops}\index{Loops@{\sf Loops}} (cf. fig. \ref{fig:menues}e)
  selects the number of closed loops\index{closed loops} for the calculation. 
  The user can choose between 0, 1 and 2. A change of this value results in
  different topologies\index{topology} displayed in the {\it topology bar}%
  \index{topology bar@{\it topology bar}}.
\item The menu {\sf Ext. Lines}\index{Ext. Lines@{\sf Ext. Lines}} (cf. fig.
  \ref{fig:menues}f) allows the user to choose the number of external lines%
  \index{external lines}. At the moment the upper limit is three. The change of
  this number results in a different {\it topology bar} as well.
\item The menu {\sf Options}\index{Options@{\sf Options}} (cf. fig.
  \ref{fig:menues}c) contains the following entries:
  \bi
  \item If {\sf History}\index{History@{\sf History}} is selected \xloops saves
    all calculations\index{result!saving} under a common process name%
    \index{process name}. Those diagrams which were already saved before are
    read from disk\index{result!reading}. Otherwise, if {\sf History}
    is not selected they are recalculated (cf. sect. \ref{procsave}).
  \item With the selection of {\sf Oneloop Library}%
    \index{Oneloop Library@{\sf Oneloop Library}} \xloops takes all one-loop
    functions from a pre-built library\index{one-loop diagrams!library}.
    Otherwise the one-loop functions are calculated every time they are needed
    (cf. sect. \ref{lib} and appendix \ref{alltable}).
  \item If {\sf Insert Particle Properties}%
    \index{Insert Particle Properties@{\sf Insert Particle\\ Properties}} is
    selected the values of masses\index{masses!values} and couplings%
    \index{couplings!values} reported by the Particle Data Group%
    \index{Particle Data Group} \cite{PDG} are inserted (cf. sect.
    \ref{masses}).
  \item {\sf Set Neutrino Masses 0}%
    \index{Set Neutrino Masses 0@{\sf Set Neutrino Masses 0}} assigns all
    neutrino masses\index{neutrinos}\index{masses!of neutrinos} to zero
    (default) or, if not selected, leaves the neutrino masses unassigned. In
    this case an undefined neutrino mixing matrix\index{mixing matrix} exists
    as well (cf. sect. \ref{masses}). 
  \item {\sf Massless Light Fermions}%
    \index{Massless Light Fermions@{\sf Massless Light Fermions}}
    assigns the masses of all fermions\index{fermions}%
    \index{masses!of light fermions} except the $b$ and $t$ quark to zero (cf.
    sect. \ref{masses}).
  \item {\sf Numeric}\index{Numeric@{\sf Numeric}} defines the number of
    points and the number of iterations\index{iterations} for the numerical
    integration\index{numerical integration@numerical integration\\ \nop} with
    \vegas\index{VEGAS@\vegas} (cf. sect. \ref{num}).
  \item {\sf Window Options}\index{Window Options@{\sf Window Options}} allows
    the user to adjust the appearance of the {\sf Xwindows} interface%
    \index{Xwindows interface@{\sf Xwindows} interface}, especially the
    placement of the red arrows\index{arrow!placement} of the {\it diagram
    window}\index{diagram window@{\it diagram window}\\ \nop} (cf. sect.
    \ref{winopt}).
  \item {\sf Save Options}\index{Save Options@{\sf Save Options}} saves all
    options in the file {\tt .xloops}\index{xloops@{\tt .xloops}} in the home
    directory of the user.
  \ei
\item The menu {\sf Help}\index{Help@{\sf Help}} (cf. fig. \ref{fig:menues}b)
  has three entries:
  \bi
  \item {\sf About}\index{About@{\sf About}} shows a brief information on
    \xloops.
  \item {\sf Help} gives a short help where especially the notation of all
    particle names in \xloops is written.
  \item {\sf Tutorial}\index{Tutorial@{\sf Tutorial}} calls a WWW browser%
    \index{WWW browser} which displays the on-line html version%
    \index{html version} of this manual.
  \ei
\ei
 
\subsection{{\it Topology bar} and {\it diagram window}}
\index{topology bar@{\it topology bar}}%
\index{diagram window@{\it diagram window}\\ \nop} \label{topol}

According to the pre-defined values for the numbers of closed loops and external
lines the {\it topology bar} displays all possible topologies. If the user
clicks on one of them this topology\index{topology} appears in a separate
window, the {\it diagram window}%
\index{diagram window@{\it diagram window}\\ \nop}.
This window consists of three parts: a {\it menu bar}%
\index{menu bar@{\it menu bar}}, an area for inserting particles ({\it diagram
area})\index{diagram area@{\it diagram area}} and a {\it command bar}%
\index{command bar@{\it command bar}} with different buttons for evaluation
(cf. fig. \ref{fig:graphwin}).

The {\it menu bar} provides two options. The right one, {\sf Export}%
\index{Export@{\sf Export}}, gives the user the possibility to export the
graph to a Postscript\index{Postscript} file. Only the {\it diagram area} of
the {\it diagram window} is exported, {\it menu bar} and {\it command bar} are
not included.

The left option, {\sf Subtopologies}, will only be available in future versions.
It will provide the possibility to switch between several subtopologies, which
are obtained when external lines are interchanged. The upper topology in fig.
\ref{fig:graphwin} may serve as an example. This topology can be rotated by
$120^{\circ}$ or $240^{\circ}$ which corresponds to a cyclic permutation of the
external lines. This changes the topology to the same type of topology but with
different momentum flow. In contrast, the lower topology in fig.
\ref{fig:graphwin} does not change if it is rotated by $120^{\circ}$ or
$240^{\circ}$. So this topology does not exhibit different subtopologies.

In the {\it diagram area}\index{diagram area@{\it diagram area}} the topology%
\index{topology} is displayed graphically. For every propagator an entry for
the insertion of particles is located. The correct terms for particle names are
written in section \ref{masses}. They can also be obtained from the {\sf
Help}\index{Help@{\sf Help}} function of the {\it main window}%
\index{main window@{\it main window}}. The flow of charge and fermion and ghost
numbers are determined by the red arrows\index{arrow} on the propagator lines.
Clicking on an arrow changes its direction. The correct flow of charge%
\index{charge} and fermion\index{fermion number} and ghost numbers%
\index{ghost number} can be settled in this way. Only particles, no
antiparticles\index{antiparticles} can be inserted. Charge conjugation%
\index{charge!conjugation} may be achieved by inversion of arrows%
\index{arrow!inversion}. Please keep in mind that the arrows don't describe the
direction of momenta. The convention of momentum flow\index{momentum flow} is
shown graphically in the {\it diagram window} in blue. It is defined for all
diagrams in the same way, to facilitate a final summation of all calculated
diagrams.

If the user has inserted all particles and checked the charge flow, the {\it
command area} gives him several possibilities for the evaluation of the
diagram. They correspond to the different modes of evaluation as explained in
section \ref{output}.
\bfg[p]\bc
\epsfig{file=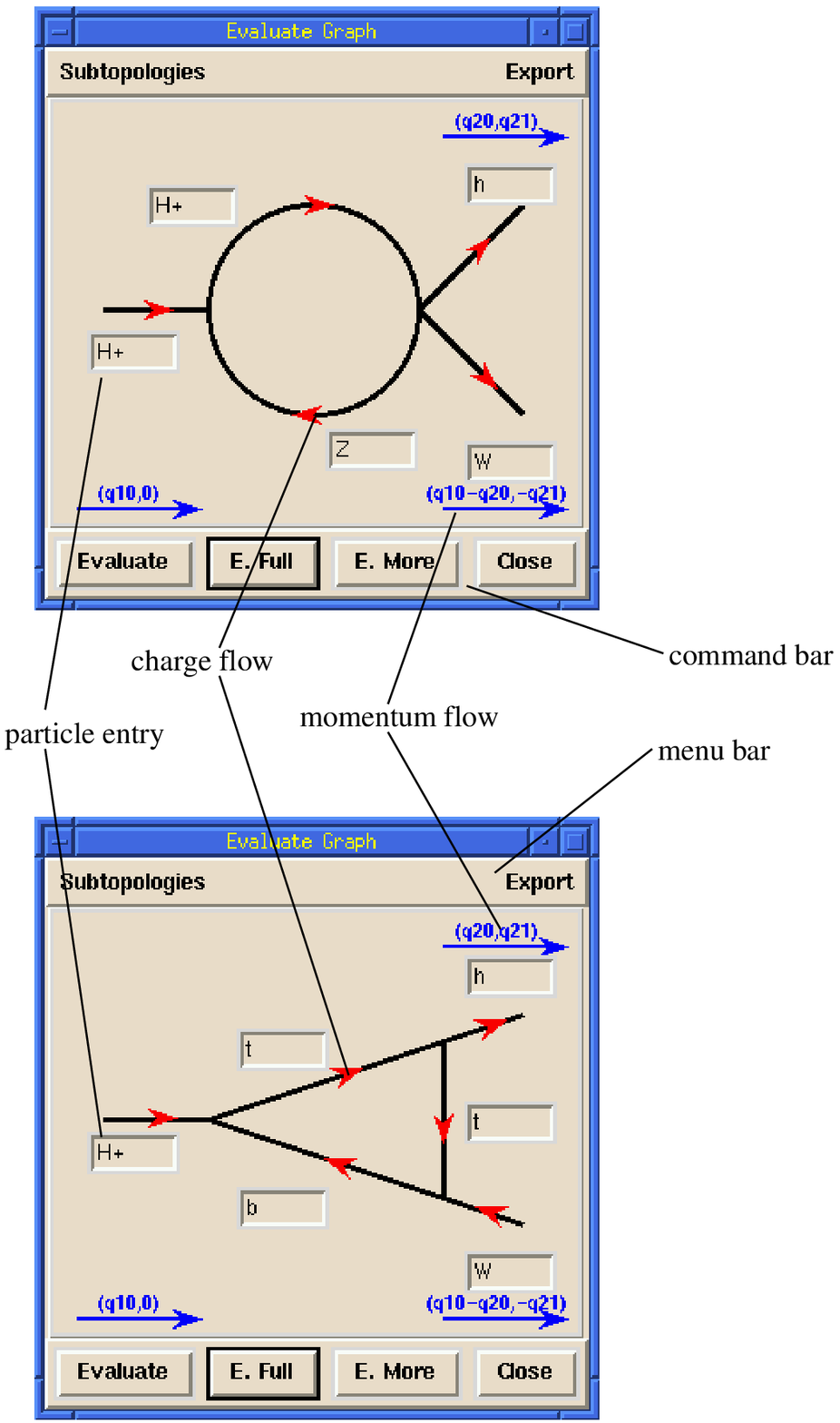,height=20cm}
\caption{Different {\it diagram windows}} \label{fig:graphwin}
\ec\efg

\subsection{Loading and saving of processes} \label{procsave}

For the evaluation of complete processes\index{process!saving} it is possible to
save the contributing graphs with a common process name\index{process!name}.
Therefore \xloops provides the {\sf File}\index{File@{\sf File}} menu entries
{\sf Load Process}\index{Load Process@{\sf Load Process}} and {\sf New Process}%
\index{New Process@{\sf New Process}}.

If {\sf Load Process} is selected the {\it file window}%
\index{file window@{\it file window}} (cf. fig. \ref{fig:filebox}) appears on
the screen.
\bfg[h]\bc
\epsfig{file=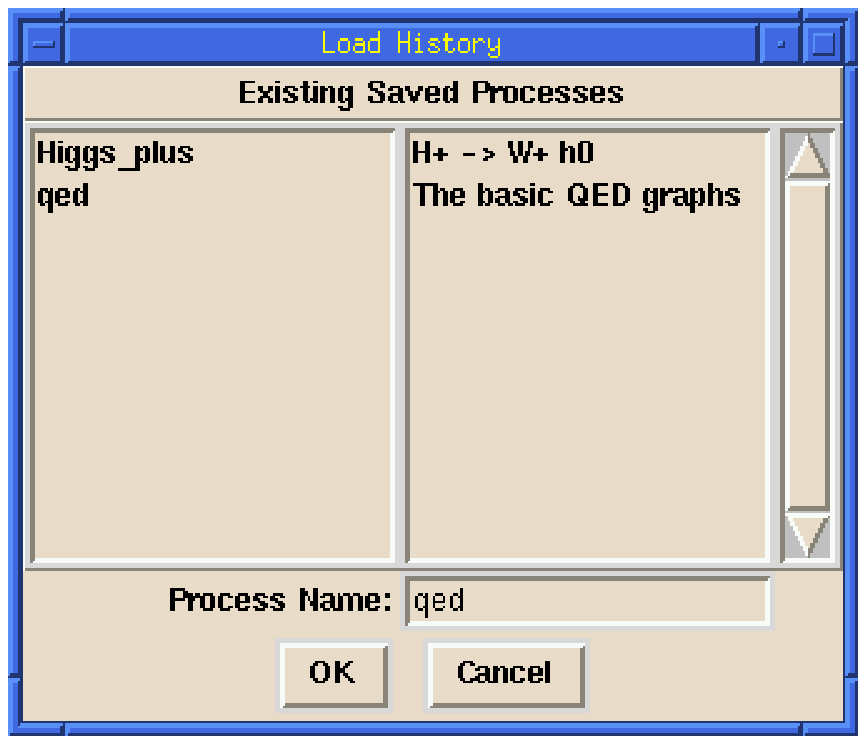,width=8cm}\hfill
\caption{The {\sf Load Process} {\it file window}} \label{fig:filebox}
\ec\efg
In this window the processes which were already saved are listed on the left
hand side. In addition for every process a user specified description%
\index{process!description} is given on the right hand side. This may help to
reidentify a process later. To select a specific process one clicks on a process
name on the left. After clicking on the \btt{OK}\index{OK@\btt{OK}} button the
{\it history table}\index{history table@{\it history table}} (cf. fig
\ref{fig:proc}) of the selected process appears whereas \btt{Cancel}%
\index{Cancel@\btt{Cancel}} closes the {\it file window}.
\bfg[p]\bc
\epsfig{file=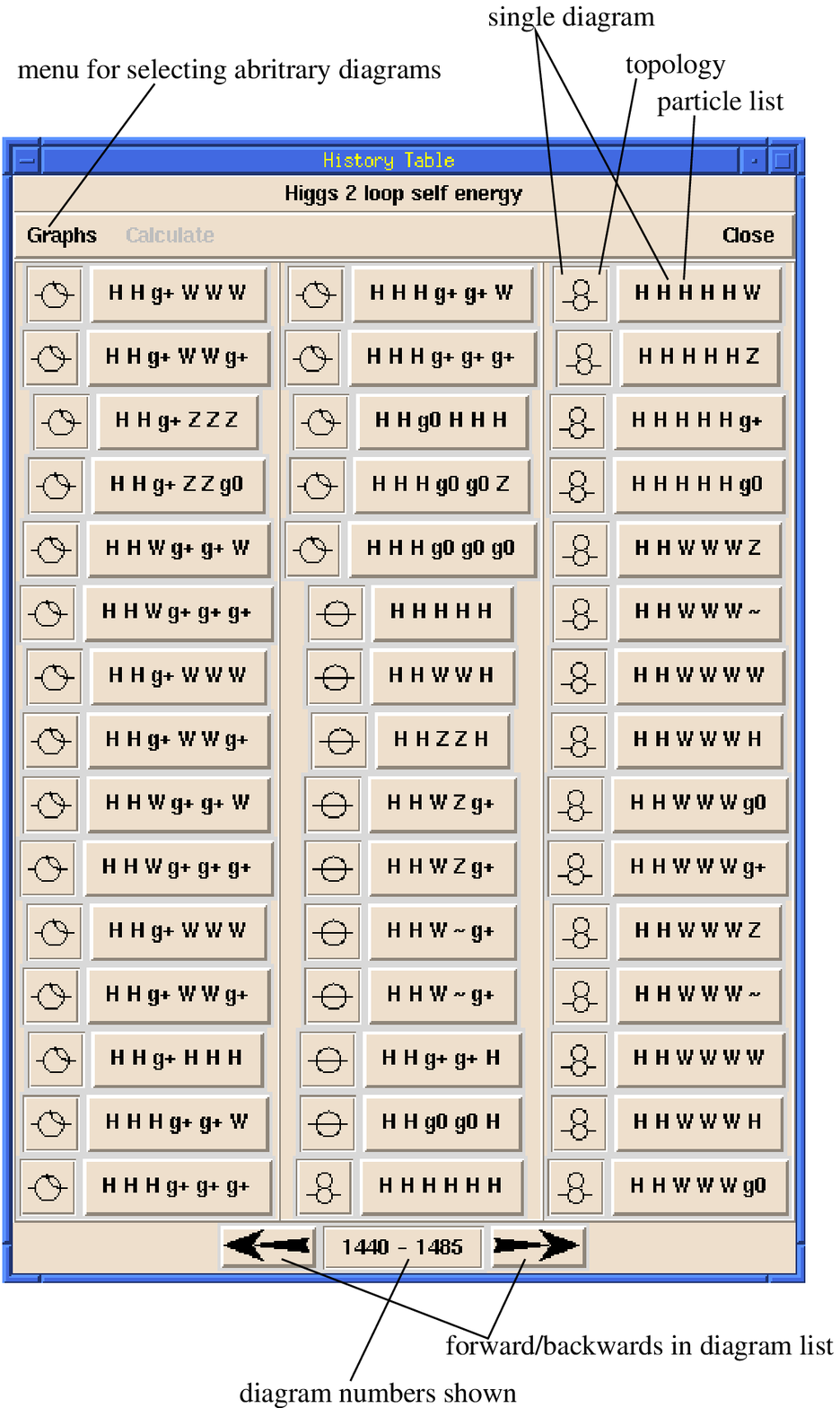}    
\caption{The {\it history table} containing diagrams of the process $H\to H$}
\label{fig:proc}
\ec\efg
In the {\it history table} all diagrams which were included in the corresponding
process occur with a button. A click on such a button opens a {\it diagram
window}\index{diagram window@{\it diagram window}\\ \nop} (cf. fig
\ref{fig:graphwin}) with which the Feynman diagram can be evaluated. The {\it
history table} contains two menues:
\bi
\item In {\sf Graphs}\index{Graphs@{\sf Graphs}} one has the opportunity -- if
  the number of diagrams exceeds the capacity of the window -- to switch to
  another sheet of diagrams. The same effect have the arrows at the bottom of
  the {\it history table}.
\item {\sf Close}\index{Close@{\sf Close}} lets the {\it history table}
  disappear.
\ei
If the entry {\sf History}\index{History@{\sf History}} of the {\sf Options}%
\index{Options@{\sf Options}} menu is selected all results are saved%
\index{result!saving}. With {\sf History} all calculations which were already
saved in this or another \xloops session are read in\index{result!reading}
automatically and not calculated again. For each process \xloops automatically
creates a subdirectory of the directory {\tt xloops\_user}%
\index{xloops_user@{\tt xloops\_user}} where the results are saved to different
files.

If {\sf History} is selected without having specified a process before \xloops
assumes that a new process shall be added to the list and opens a similar {\it
file window}\index{file window@{\it file window}} (cf. fig \ref{fig:filebox2})
-- the same window appears if {\sf New Process}%
\index{New Process@{\sf New Process}} is selected.
Two entries are provided for inserting a new process name\index{process!name}
and description\index{process!description} at the bottom of the window. After
the user inserted the name and description for his process and clicked on the
\btt{OK}\index{OK@\btt{OK}} button a new -- and therefore empty -- {\it history
table}\index{history table@{\it history table}}
appears on the screen. All diagrams which are calculated from now are included
in the {\it history table} as additional entries. The topology\index{topology}
of the graph and the list of interacting particles are displayed.
\bfg[t]\bc
\epsfig{file=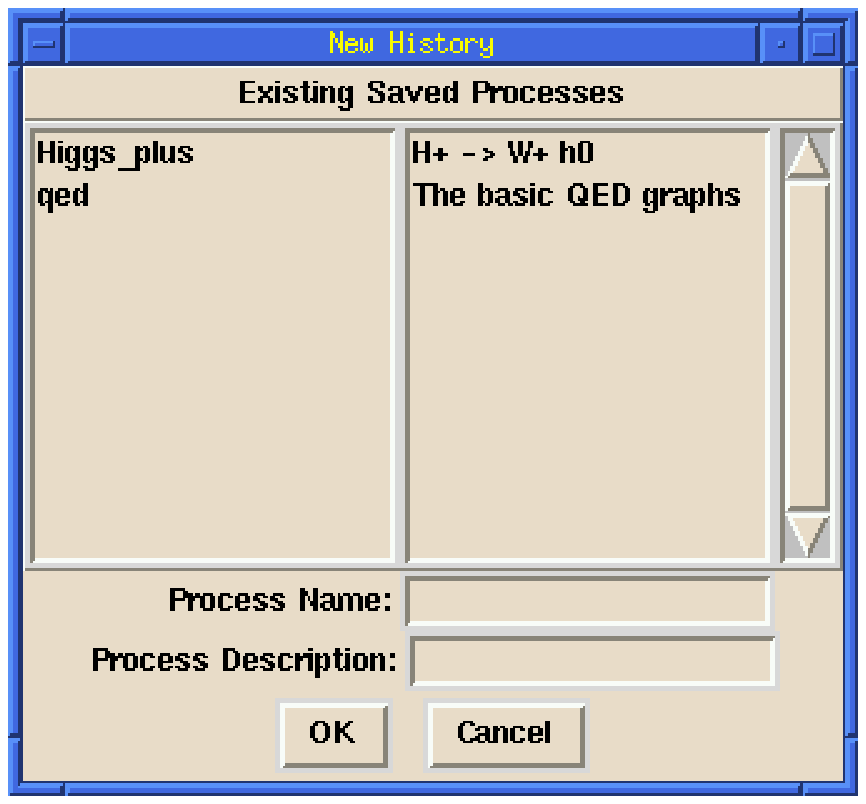,width=8cm}
\caption{The {\sf New Process} {\it file window}} \label{fig:filebox2}
\ec\efg

\subsection{Window options} \label{winopt}

With the entry {\sf Window Options} of the {\sf Options} menu the appearance of
some windows of \xloops can be adjusted to fit the terminal characteristics (cf.
fig. \ref{fig:opts}).
\bfg[h]\bc
\epsfig{file=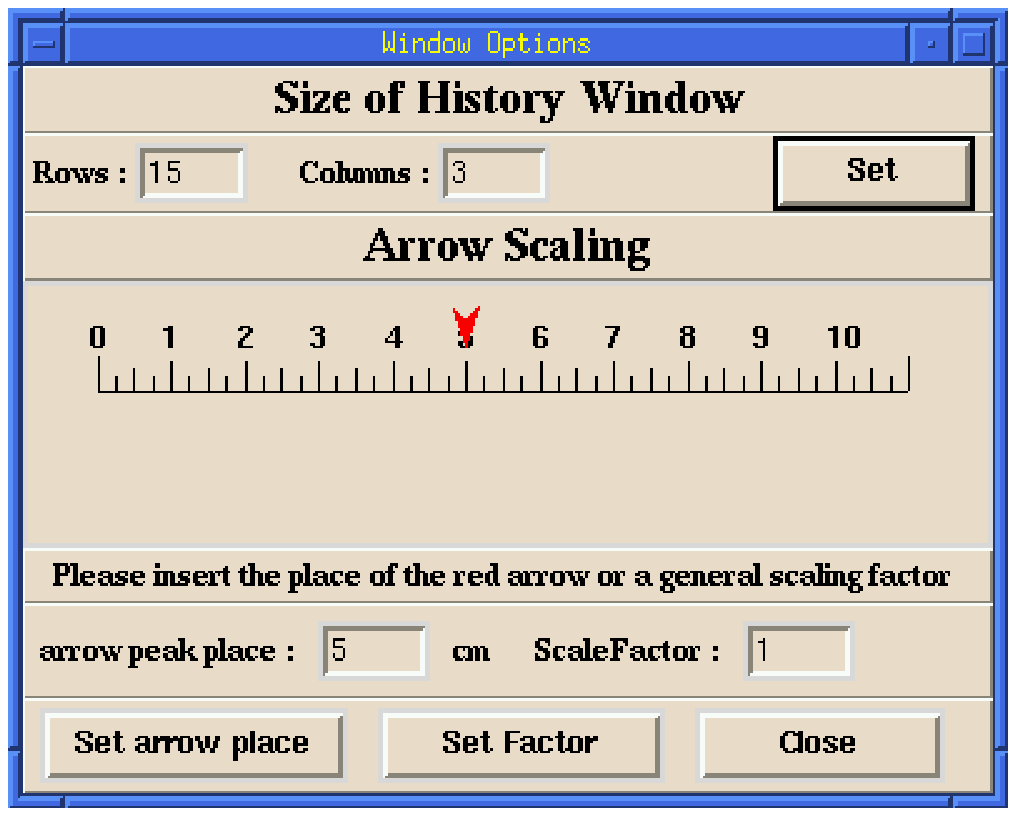,width=8cm}    
\caption{Window options of \xloops}
\label{fig:opts}
\ec\efg
First, the size of the {\it history table}%
\index{history table@{\it history table}!size} (cf. fig. \ref{fig:proc}) can
be changed by decreasing or increasing the numbers of the {\sf Rows}%
\index{Rows@{\sf Rows}} and {\sf Columns}\index{Columns@{\sf Columns}} entry in
fig. \ref{fig:opts}. Then, the position of the arrows\index{arrow!position} of
the propagators in the {\it diagram window}%
\index{diagram window@{\it diagram window}\\ \nop} can be corrected. This
might be necessary on some terminal types, where the arrows%
\index{arrow!misplaced} of the propagators are misplaced. This has to do with
the fact that the relation between screen dots (as used by {\sf Tk}%
\index{Tk@{\sf Tk}}) and centimeters is different on different {\sf X}
terminals\index{X terminals@{\sf X} terminals}. The embarrassing consequence
can be that the arrows of the propagators are misplaced. The problem can be
avoided by inserting a global {\sf Scale Factor}%
\index{Scale Factor@{\sf Scale Factor}}.
Alternatively, if this factor is not obvious, \xloops will calculate it.
Therefore the user must insert the position of the arrow%
\index{arrow!position on the ruler} on the ruler\index{ruler} in the {\sf Arrow
Scaling}\index{Arrow Scaling@{\sf Arrow Scaling}} field of fig.
\ref{fig:opts}.\footnote{If the scaling factor is already correct, the arrow
should point on the 5.}

\section{Evaluation of Feynman diagrams} \label{evalgraph}

\subsection{Input}

To evaluate a particular Feynman diagram \xloops needs to know four
ingredients:
\ben
\item The number of closed loops.
\item The number of external lines.
\item The topology, that means the information how the lines of the diagram are
  connected.
\item The particles of the diagram.
\een
The first two numbers have to be selected with the help of the {\it menu bar}
entries {\sf Loops} and {\sf Ext. Lines} (cf. sect. \ref{menu}). Then the
topology can be selected on the {\it topology bar} so that the {\it diagram
window} pops up where the particles can be inserted (cf. sect. \ref{topol}).

Internally the input from the {\sf Xwindows} interface is converted to a
procedure which performs the calculations in \maple. It is also possible to
start directly from this routine for \maple -- either with the {\sf Insert Maple
Command} of the {\sf File} menu or within an ordinary \maple session where the
\maple part of \xloops was read in before (cf. sect. \ref{start}).

For that purpose the internal \maple commands will be mentioned in the
following. All commands which follow without reference to the {\sf Xwindows}
interface have to be passed to \maple.

The input is then of the following form
\bi
\item for tree diagrams:
  \verb+EvalGraph0(+$n$\verb+,+$\langle${\it list}$\rangle$\verb+);+
\item for one-loop diagrams:
  \verb+EvalGraph1(+$n$\verb+,+$\langle${\it list}$\rangle$\verb+);+
\item for two-loop diagrams:
  \verb+EvalGraph2(+$n$\verb+,+$\langle${\it list}$\rangle$\verb+);+
\ei
The different topologies with the same number of loops are numbered. The number
of the desired topology has to be inserted in the function call for $n$. The
correct number for each topology will be given below in sect. \ref{not}.
$\langle${\it list}$\rangle$ describes the list of particles involved in the
Feynman diagram. The conventions for the ordering of the particles are also
explained for every topology in \ref{not}.

\subsection{Internal Notation} \label{not}

What follows is a list of all topologies which at present can be solved by
\xloops. This list is only of interest for direct \maple input. To fix the flow
of charges and of fermion or ghost numbers in an unique way, each internal
propagator is assigned with two particle names. For example \verb+[up,upbar]+
and \verb+[upbar,up]+ describe the up-quark propagator but with different
direction of the quark line arrow. External lines are described by only one
particle name.

As a convention all particles at each vertex are incoming. The arrows of the
diagrams displayed below only describe the flow of momenta. Of course these
arrows need not to coincide with the arrows of fermions or other charged
particles.

For each topology we give the correct {\tt EvalGraph} command as well as the
command for the corresponding {\tt OneLoop} or {\tt TwoLoop} integral.

\subsection*{Tree level}

\subsubsection*{Three-point diagrams}

\noindent
{\setlength{\fboxsep}{0cm}%
\shadowbox{\parbox{14.6cm}{\bc
{\tt EvalGraph0$($1}$,[$\CC{1},\CC{2},\CC{3}$])$ \\
\parbox{10cm}{\bp(10,5)
\psline{->}(2,2.5)(3.7,2.5)
\psline(3.5,2.5)(5,2.5)
\pscircle[linewidth=0.1](5.03,2.5){0.1}
\psline(5,2.5)(6.7,1.65)
\psline{->}(8,1)(6.6,1.7)
\psline(8,4)(6.3,3.15)
\psline{->}(5,2.5)(7,3.5)
\put(4,2.07){\CC{1}}
\put(5.25,3){\CC{3}}
\put(5.25,1.8){\CC{2}}
\put(3.3,2.77){$\bm{q_{1}}$}
\put(6.6,3){$\bm{q_{2}}$}
\put(6.6,1.9){$\bm{q_{2}-q_{1}}$}
\ep} 
\\
\nop
\ec}}}

\subsubsection*{Four-point diagrams}

\noindent
{\setlength{\fboxsep}{0cm}%
\shadowbox{\parbox{14.6cm}{\bc
{\tt EvalGraph0$($2}$,[$\CC{1},\CC{2},\CC{3},\CC{4}$])$ \\
\parbox{10cm}{\bp(10,5)
\pscircle[linewidth=0.1](5,2.5){0.1}
\psline{->}(3.5,1)(4.3,1.8)
\psline{->}(3.95,1.45)(6,3.5)
\psline(5.6,3.1)(6.5,4)
\psline{->}(3.5,4)(4.3,3.2)
\psline(3.9,3.6)(6.1,1.4)
\psline{->}(6.5,1)(5.7,1.8)
\put(4.45,3.34){\CC{1}}
\put(4.45,1.46){\CC{2}}
\put(5.25,1.46){\CC{3}}
\put(5.25,3.34){\CC{4}}
\put(6.1,3.2){$\bm{q_{3}}$}
\put(6.1,1.7){$\bm{q_{3}-q_{2}}$}
\put(2.64,1.66){$\bm{q_{2}-q_{1}}$}
\put(3.54,3.14){$\bm{q_{1}}$}
\ep}
\\
\nop
\ec}}}

\newpage

\subsection*{One-loop level}

\subsubsection*{One-point diagrams}

\noindent
{\setlength{\fboxsep}{0cm}%
\shadowbox{\parbox{14.6cm}{\bc
{\tt EvalGraph1$($1}$,[$\CC{1},\CC{2},\CC{3}$])$ \\
\parbox{10cm}{\bp(10,5)
\pscircle(5,2.5){1.5}
\psline{->}(6.47,2.6)(6.47,2.4)
\pscircle[linewidth=0.1](3.53,2.5){0.1}
\psline(1,2.5)(3.5,2.5)
\put(2.5,2.07){\CC{1}}
\put(3.96,3.14){\CC{2}}
\put(3.96,1.66){\CC{3}}
\put(6.7,2.5){$\bm{l}$}
\put(5.9,2.5){$m$}
\ep} 
\\
{\tt OneLoop1Pt}$(p,m)$
\ec}}}

\subsubsection*{Two-point diagrams}

\noindent
{\setlength{\fboxsep}{0cm}%
\shadowbox{\parbox{14.6cm}{\bc
{\tt EvalGraph1$($2}$,[$\CC{1},\CC{2},\CC{3},\CC{4}$])$ \\
\parbox{10cm}{\bp(10,5)
\pscircle(5,2.5){1.5}
\psline{->}(5.1,3.97)(4.9,3.97)
\pscircle[linewidth=0.1](5,1.03){0.1}
\psline{->}(1,1.03)(3.2,1.03)
\psline{->}(3,1.03)(7.2,1.03)
\psline(7,1.03)(9,1.03)
\put(3.5,0.6){\CC{1}}
\put(6.1,0.6){\CC{2}}
\put(5.64,1.66){\CC{3}}
\put(3.96,1.66){\CC{4}}
\put(5,4.2){$\bm{l}$}
\put(4.9,3.6){$m$}
\put(2.8,1.3){$\bm{q_{1}}$}
\put(6.8,1.3){$\bm{q_{1}}$}
\ep}
\\
{\tt OneLoop1Pt}$(p,m)$
\ec}}
\newpage
\noindent
\shadowbox{\parbox{14.6cm}{\bc
{\tt EvalGraph1$($3}$,[$\CC{1},\CC{2},\CC{3},\CC{4},\CC{5},\CC{6}$])$ \\
\parbox{10cm}{\bp(10,5)
\pscircle(5,2.5){1.5}
\psline{->}(5.1,3.97)(4.9,3.97)
\psline{->}(4.9,1.03)(5.1,1.03)
\pscircle[linewidth=0.1](3.53,2.5){0.1}
\pscircle[linewidth=0.1](6.47,2.5){0.1}
\psline{->}(1,2.5)(2.45,2.5)
\psline(2.25,2.5)(3.5,2.5)
\psline{->}(6.5,2.5)(7.95,2.5)
\psline(7.75,2.5)(9,2.5)
\put(2.5,2.07){\CC{1}}
\put(7.1,2.07){\CC{2}}
\put(3.96,3.14){\CC{3}}
\put(3.96,1.66){\CC{4}}
\put(5.64,1.66){\CC{5}}
\put(5.64,3.14){\CC{6}}
\put(4.85,1.3){$m_{1}$}
\put(4.85,3.6){$m_{2}$}
\put(4.6,0.53){$\bm{l+q_{1}}$}
\put(5,4.2){$\bm{l}$}
\put(2.05,2.77){$\bm{q_{1}}$}
\put(7.55,2.77){$\bm{q_{1}}$}
\ep}
\\
{\tt OneLoop2Pt}$(p_{0},p_{\bot},q_{10},m_{1},m_{2})$
\ec}}}

\subsubsection*{Three-point diagrams}

\noindent
{\setlength{\fboxsep}{0cm}%
\shadowbox{\parbox{14.6cm}{\bc
{\tt EvalGraph1$($4}$,[$\CC{1},\CC{2},\CC{3},\CC{4},\CC{5},\CC{6},\CC{7}$])$ \\
\parbox{10cm}{\bp(10,5)
\psarc(5,2.5){1.485}{45}{315}
\psline{->}(5.1,3.985)(4.9,3.985)
\psline{->}(4.9,1.015)(5.1,1.015)
\pscircle[linewidth=0.1](3.515,2.5){0.1}
\pscircle[linewidth=0.1](7.1,2.5){0.1}
\psline{->}(1,2.5)(2.45,2.5)
\psline(2.25,2.5)(3.5,2.5)
\psline{->}(6.05,1.45)(8.1,3.5)
\psline(7.7,3.1)(8.6,4)
\psline(6.05,3.55)(8.2,1.4)
\psline{->}(8.6,1)(7.8,1.8)
\put(2.5,2.07){\CC{1}}
\put(7.15,1.66){\CC{2}}
\put(7.15,3.14){\CC{3}}
\put(3.96,3.14){\CC{4}}
\put(3.96,1.66){\CC{5}}
\put(5.64,1.66){\CC{6}}
\put(5.64,3.14){\CC{7}}
\put(4.6,0.53){$\bm{l+q_{1}}$}
\put(5,4.2){$\bm{l}$}
\put(4.85,1.3){$m_{1}$}
\put(4.85,3.6){$m_{2}$}
\put(2.05,2.77){$\bm{q_{1}}$}
\put(8.2,3.2){$\bm{q_{2}}$}
\put(8.2,1.7){$\bm{q_{2}-q_{1}}$}
\ep}
\\
{\tt OneLoop2Pt}$(p_{0},p_{\bot},q_{10},m_{1},m_{2})$
\ec}}
\\[0.5cm]
\shadowbox{\parbox{14.6cm}{\bc
{\tt EvalGraph1$($5}$,[$\CC{1},\CC{2},\CC{3},\CC{4},\CC{5},\CC{6},\CC{7},\CC{8}%
,\CC{9}$]
)$ \\
\parbox{10cm}{\bp(10,5)
\pscircle[linewidth=0.1](3.53,2.5){0.1}
\pscircle[linewidth=0.1](6.5,1){0.1}
\pscircle[linewidth=0.1](6.5,4){0.1}
\psline{->}(1,2.5)(2.45,2.5)
\psline(2.25,2.5)(3.5,2.5)
\psline{->}(6.5,4)(7.95,4)
\psline(9,4)(7.55,4)
\psline(7.75,1)(6.5,1)
\psline{->}(9,1)(7.55,1)
\psline{->}(3.5,2.5)(5.2,1.65)
\psline(5.1,1.7)(6.5,1)
\psline{->}(6.5,4)(4.8,3.15)
\psline(5.1,3.3)(3.5,2.5)
\psline{->}(6.5,1)(6.5,2.7)
\psline(6.5,2.4)(6.5,4)
\put(2.5,2.07){\CC{1}}
\put(7.1,0.57){\CC{2}}
\put(7.1,4.2){\CC{3}}
\put(3.75,3){\CC{4}}
\put(3.75,1.8){\CC{5}}
\put(5.6,0.9){\CC{6}}
\put(6,1.8){\CC{7}}
\put(6,3){\CC{8}}
\put(5.6,3.9){\CC{9}}
\put(4.1,1.35){$\bm{l+q_{1}}$}
\put(6.715,2.35){$\bm{l+q_{2}}$}
\put(4.8,3.35){$\bm{l}$}
\put(4.85,2){$m_{1}$}
\put(5.8,2.35){$m_{2}$}
\put(4.85,2.9){$m_{3}$}
\put(2.15,2.77){$\bm{q_{1}}$}
\put(7.65,3.6){$\bm{q_{2}}$}
\put(7.1,1.2){$\bm{q_{2}-q_{1}}$}
\ep}
\\
{\tt OneLoop3Pt}$(p_{0},p_{1},p_{\bot},q_{10},q_{20},q_{21},m_{1},
m_{2},m_{3})$
\ec}}}
\newpage

\subsection*{Two-loop level}

\subsubsection*{Two-point diagrams}

\noindent
{\setlength{\fboxsep}{0cm}%
\shadowbox{\parbox{14.6cm}{\bc
{\tt EvalGraph2$($6}$,[$\CC{1},\CC{2},\CC{3},\CC{4},\CC{5},\CC{6},\CC{7},\CC{8}%
,\CC{9},\CC{10},\CC{11},\CC{12}$])$ \\[0.5cm]
\parbox{11cm}{\bp(11,6)
\pscircle(5.5,3){2.5}
\psarc(5.5,6.493){2.47}{225}{315}
\psline{->}(5.4,5.47)(5.6,5.47)
\psline{->}(5.4,0.53)(5.6,0.53)
\psline{->}(5.6,4.03)(5.4,4.03)
\psline{->}(7.821,3.857)(7.729,4.057)
\psline{->}(3.271,4.057)(3.169,3.857)
\pscircle[linewidth=0.1](3.03,3){0.1}
\pscircle[linewidth=0.1](7.97,3){0.1}
\pscircle[linewidth=0.1](7.243,4.75){0.1}
\pscircle[linewidth=0.1](3.757,4.75){0.1}
\psline{->}(0.5,3)(1.95,3)
\psline(1.75,3)(3,3)
\psline{->}(8,3)(9.45,3)
\psline(9.25,3)(10.5,3)
\put(2,2.57){\CC{1}}
\put(8.6,2.57){\CC{2}}
\put(3.2,3.24){\CC{3}}
\put(3.3,2.26){\CC{4}}
\put(7.3,2.26){\CC{5}}
\put(7.4,3.24){\CC{6}}
\put(7,4.14){\CC{7}}
\put(6.2,4.9){\CC{8}}
\put(6.2,4.4){\CC{9}}
\put(4.4,4.4){\CC{10}}
\put(4.4,4.9){\CC{11}}
\put(3.6,4.14){\CC{12}}
\put(1.55,3.27){$\bm{q_{1}}$}
\put(9.05,3.27){$\bm{q_{1}}$}
\put(5.5,5.7){$\bm{k}$}
\put(5.1,0.03){$\bm{l+q_{1}}$}
\put(8.1,3.9){$\bm{l}$}
\put(2.8,3.9){$\bm{l}$}
\put(5.1,3.54){$\bm{l+k}$}
\put(5.35,0.8){$m_{1}$}
\put(7.2,3.8){$m_{2}$}
\put(3.4,3.8){$m_{3}$}
\put(5.35,4.19){$m_{4}$}
\put(5.35,5.1){$m_{5}$}
\ep} 
\\[0.5cm]
{\tt TwoLoop2Pt1}$(p_{0},p_{\bot},r_{0},r_{\bot},s,q_{10},m_{1},m_{2},m_{3},
m_{4},m_{5})$
\ec}}
\\[0.5cm]
\shadowbox{\parbox{14.6cm}{\bc
{\tt EvalGraph2$($7}$,[$\CC{1},\CC{2},\CC{3},\CC{4},\CC{5},\CC{6},\CC{7},\CC{8}%
,\CC{9},\CC{10},\CC{11},\CC{12}$])$ \\[0.5cm]
\parbox{11cm}{\bp(11,6)
\pscircle(5.5,3){2.5}
\psline{->}(5.5,0.5)(5.5,3.2)
\psline(5.5,5.5)(5.5,3)
\psline{->}(3.823,4.817)(3.683,4.677)
\psline{->}(3.683,1.323)(3.823,1.183)
\psline{->}(7.177,4.817)(7.317,4.677)
\psline{->}(7.317,1.323)(7.177,1.183)
\pscircle[linewidth=0.1](5.5,5.47){0.1}
\pscircle[linewidth=0.1](5.5,0.53){0.1}
\pscircle[linewidth=0.1](3.03,3){0.1}
\pscircle[linewidth=0.1](7.97,3){0.1}
\psline{->}(0.5,3)(1.95,3)
\psline(1.75,3)(3,3)
\psline{->}(8,3)(9.45,3)
\psline(9.25,3)(10.5,3)
\put(2,2.57){\CC{1}}
\put(8.6,2.57){\CC{2}}
\put(3.3,3.54){\CC{3}}
\put(3.3,2.26){\CC{4}}
\put(4.6,0.9){\CC{5}}
\put(6,0.9){\CC{6}}
\put(5.65,1.5){\CC{7}}
\put(5.65,4.3){\CC{8}}
\put(6,4.9){\CC{9}}
\put(4.6,4.9){\CC{10}}
\put(7.3,2.26){\CC{11}}
\put(7.3,3.54){\CC{12}}
\put(1.55,3.27){$\bm{q_{1}}$}
\put(9.05,3.27){$\bm{q_{1}}$}
\put(7.4,0.95){$\bm{k-q_{1}}$}
\put(7.4,4.75){$\bm{k}$}
\put(3.5,4.75){$\bm{l}$}
\put(2.7,0.95){$\bm{l+q_{1}}$}
\put(4.4,2.85){$\bm{l+k}$}
\put(3.8,1.45){$m_{1}$}
\put(3.8,4.4){$m_{2}$}
\put(5.715,2.85){$m_{3}$}
\put(6.7,1.45){$m_{4}$}
\put(6.7,4.4){$m_{5}$}
\ep} 
\\[0.5cm]
{\tt TwoLoop2Pt2}$(p_{0},p_{\bot},r_{0},r_{\bot},s,q_{10},m_{1},m_{2},m_{3},
m_{4},m_{5})$
\ec}}
\newpage
\noindent
\shadowbox{\parbox{14.6cm}{\bc
{\tt EvalGraph2$($8}$,[$\CC{1},\CC{2},\CC{3},\CC{4},\CC{5},\CC{6},\CC{7},\CC{8}%
,\CC{9},\CC{10}$])$ \\[0.5cm]
\parbox{11cm}{\bp(11,6)
\pscircle(5.5,3){2.5}
\psarc(3,5.5){2.5}{270}{360}
\psline{->}(5.4,0.53)(5.6,0.53)
\psline{->}(3.823,4.817)(3.683,4.677)
\psline{->}(7.317,4.677)(7.177,4.817)
\psline{->}(4.697,3.663)(4.837,3.803)
\pscircle[linewidth=0.1](3.03,3){0.1}
\pscircle[linewidth=0.1](7.97,3){0.1}
\pscircle[linewidth=0.1](5.5,5.47){0.1}
\psline{->}(0.5,3)(1.95,3)
\psline(1.75,3)(3,3)
\psline{->}(8,3)(9.45,3)
\psline(9.25,3)(10.5,3)
\put(2,2.57){\CC{1}}
\put(8.6,2.57){\CC{2}}
\put(3.3,3.54){\CC{3}}
\put(3.85,2.8){\CC{4}}
\put(3.3,2.26){\CC{5}}
\put(7.3,2.26){\CC{6}}
\put(7.3,3.54){\CC{7}}
\put(6,4.9){\CC{8}}
\put(5.4,4.3){\CC{9}}
\put(4.6,4.9){\CC{10}}
\put(1.55,3.27){$\bm{q_{1}}$}
\put(9.05,3.27){$\bm{q_{1}}$}
\put(7.4,4.75){$\bm{l}$}
\put(5.1,0.03){$\bm{l+q_{1}}$}
\put(4.8,3.4){$\bm{k}$}
\put(2.7,4.75){$\bm{l+k}$}
\put(5.35,0.8){$m_{1}$}
\put(6.7,4.4){$m_{2}$}
\put(3.75,4.4){$m_{3}$}
\put(4.1,3.823){$m_{4}$}
\ep} 
\\[0.5cm]
{\tt TwoLoop2Pt3}$(p_{0},p_{\bot},r_{0},r_{\bot},s,q_{10},m_{1},m_{2},m_{3},
m_{4})$
\ec}}
\\[0.5cm]
\shadowbox{\parbox{14.6cm}{\bc
{\tt EvalGraph2$($9}$,[$\CC{1},\CC{2},\CC{3},\CC{4},\CC{5},\CC{6},\CC{7}%
,\CC{8}$])$
\\[0.5cm]
\parbox{11cm}{\bp(11,6)
\pscircle(5.5,3){2.5}
\psline{->}(5.6,5.47)(5.4,5.47)
\psline{->}(5.4,0.53)(5.6,0.53)
\pscircle[linewidth=0.1](3.03,3){0.1}
\pscircle[linewidth=0.1](7.97,3){0.1}
\psline{->}(1.75,3)(5.6,3)
\psline{->}(5.3,3)(9.45,3)
\psline{->}(0.5,3)(1.95,3)
\psline(9.25,3)(10.5,3)
\put(2,2.57){\CC{1}}
\put(8.6,2.57){\CC{2}}
\put(3.3,3.54){\CC{3}}
\put(3.85,3.2){\CC{4}}
\put(3.3,2.26){\CC{5}}
\put(7.3,2.26){\CC{6}}
\put(6.75,3.2){\CC{7}}
\put(7.3,3.54){\CC{8}}
\put(1.55,3.27){$\bm{q_{1}}$}
\put(9.05,3.27){$\bm{q_{1}}$}
\put(5.1,0.03){$\bm{l+q_{1}}$}
\put(5.1,5.7){$\bm{l+k}$}
\put(5.4,2.5){$\bm{k}$}
\put(5.35,0.8){$m_{1}$}
\put(5.35,5.1){$m_{2}$}
\put(5.35,3.2){$m_{3}$}
\ep} 
\\[0.5cm]
{\tt TwoLoop2Pt4}$(p_{0},p_{\bot},r_{0},r_{\bot},s,q_{10},m_{1},m_{2},m_{3})$
\ec}}
\newpage
\noindent
\shadowbox{\parbox{14.6cm}{\bc
{\tt EvalGraph2$($10}$,[$\CC{1},\CC{2},\CC{3},\CC{4},\CC{5},\CC{6},\CC{7}%
,\CC{8},\CC{9},\CC{10}$])$ \\[0.5cm]
\parbox{11cm}{\bp(11,6)
\pscircle(5.5,4.47){1.5}
\pscircle(5.5,1.53){1.5}
\psline{->}(5.4,5.94)(5.6,5.94)
\psline{->}(5.4,0.06)(5.6,0.06)
\psline{->}(0.5,1.5)(2.7,1.5)
\psline{->}(6.639,2.469)(6.439,2.669)
\psline{->}(4.561,2.669)(4.361,2.469)
\psline(2.5,1.5)(4.03,1.5)
\psline{->}(6.97,1.5)(8.7,1.5)
\psline(8.5,1.5)(10.5,1.5)
\pscircle[linewidth=0.1](5.5,3){0.1}
\pscircle[linewidth=0.1](4.03,1.5){0.1}
\pscircle[linewidth=0.1](6.97,1.5){0.1}
\put(3,1.1){\CC{1}}
\put(7.6,1.1){\CC{2}}
\put(4.2,1.7){\CC{3}}
\put(4.49,0.69){\CC{4}}
\put(6.17,0.69){\CC{5}}
\put(6.4,1.7){\CC{6}}
\put(5.8,2.5){\CC{7}}
\put(6.17,3.63){\CC{8}}
\put(4.49,3.63){\CC{9}}
\put(4.8,2.5){\CC{10}}
\put(2.3,1.8){$\bm{q_{1}}$}
\put(8.3,1.8){$\bm{q_{1}}$}
\put(5.5,6.17){$\bm{k}$}
\put(5.1,-0.47){$\bm{l+q_{1}}$}
\put(4.2,2.6){$\bm{l}$}
\put(6.8,2.6){$\bm{l}$}
\put(5.35,0.3){$m_{1}$}
\put(6.1,2.2){$m_{2}$}
\put(4.5,2.2){$m_{3}$}
\put(5.4,5.57){$m_{4}$}
\ep} 
\\[0.5cm]
{\tt TwoLoop2Pt5}$(p_{0},p_{\bot},r_{0},r_{\bot},s,q_{10},m_{1},m_{2},m_{3},
m_{4})$
\ec}}
\\[0.5cm]
\shadowbox{\parbox{14.6cm}{\bc
{\tt EvalGraph2$($11}$,[$\CC{1},\CC{2},\CC{3},\CC{4},\CC{5},\CC{6},\CC{7}%
,\CC{8},\CC{9},\CC{10}$])$ \\[0.5cm]
\parbox{11cm}{\bp(11,6)
\pscircle(4.03,3){1.5}
\pscircle(6.97,3){1.5}
\pscircle[linewidth=0.1](2.56,3){0.1}
\pscircle[linewidth=0.1](8.44,3){0.1}
\pscircle[linewidth=0.1](5.5,3){0.1}
\psline{->}(4.13,4.47)(3.93,4.47)
\psline{->}(3.93,1.53)(4.13,1.53)
\psline{->}(6.87,4.47)(7.07,4.47)
\psline{->}(7.07,1.53)(6.87,1.53)
\psline{->}(0.5,3)(1.6,3)
\psline(1.4,3)(2.47,3)
\psline{->}(8.53,3)(9.8,3)
\psline(9.6,3)(10.5,3)
\put(1.5,2.57){\CC{1}}
\put(9.1,2.57){\CC{2}}
\put(2.99,3.64){\CC{3}}
\put(2.99,2.16){\CC{4}}
\put(4.67,2.16){\CC{5}}
\put(5.93,2.16){\CC{6}}
\put(5.93,3.64){\CC{7}}
\put(4.67,3.64){\CC{8}}
\put(7.61,2.16){\CC{9}}
\put(7.61,3.64){\CC{10}}
\put(1.25,3.27){$\bm{q_{1}}$}
\put(9.35,3.27){$\bm{q_{1}}$}
\put(6.87,4.7){$\bm{k}$}
\put(6.47,1.03){$\bm{k-q_{1}}$}
\put(3.93,4.7){$\bm{l}$}
\put(3.53,1.03){$\bm{l+q_{1}}$}
\put(3.83,1.8){$m_{1}$}
\put(3.83,4.1){$m_{2}$}
\put(6.77,1.8){$m_{3}$}
\put(6.77,4.1){$m_{4}$}
\ep} 
\\[0.5cm]
{\tt TwoLoop2Pt6}$(p_{0},p_{\bot},r_{0},r_{\bot},s,q_{10},m_{1},m_{2},m_{3},
m_{4})$
\ec}}}
\\[1cm]
For each encircled number one particle has to be inserted. The correct symbols
for all allowed particles are listed in the next section. The conventions for
the {\tt OneLoop} and {\tt TwoLoop} functions are denoted in sect.
\ref{Oneloop} and \ref{Twoloop}.

\subsection{Models} \label{models}

\xloops distinguishes several models. The menu {\sf Model} of the {\it
menu bar} serves for the selection of the desired model. Internally the global
variable \verb+Model+ selects the model which shall be used. The following
models are declared in \xloops:
\bi
\item {\sf Standard Model} -- the (minimal) standard model (electroweak and
  QCD). This corresponds to \verb+Model:=SM;+ and is the default value
\item {\sf 2 Higgs Doublets Model} -- the standard model with two Higgs
  doublets which corresponds to five physical Higgs particles;
  \verb+Model:=THDM;+
\item {\sf QCD} -- QCD with three generations of Quarks; \verb+Model:=QCD;+
\item {\sf Electroweak Theory} -- the electroweak sector of the standard model;
  \verb+Model:=EW;+
\item {\sf QED} -- QED with three generations of Leptons and Quarks;
  \verb+Model:=QED;+
\item {\sf Phi\^{ }4 Theory} -- $\phi^4$-theory in four dimensions;
  \verb+Model:=phifour;+
\item {\sf Phi\^{ }3 Theory} -- $\phi^3$-theory in four dimensions;
  \verb+Model:=phithree;+
\item {\sf Goldstone Boson Equiv. Model} -- the equivalence model for
  Goldstone bosons -- standard model vector bosons are replaced by scalar
  Goldstone modes\cite{Lee,Vel,Rie}; \verb+Model:=GBEM;+
\item a user-defined model where arbitrary Feynman rules can be declared. To
  work with this option the user has to set \verb+Model:=user;+ and to read an
  additional file \\[0.4cm]
  \verb+read`+{$\langle$\it path$\rangle$}\verb+fmuser.ma`;+ \\[0.4cm]
  where {$\langle$\it path$\rangle$} is the same path used for the assignment
  of \verb+LoopPath+ at the beginning of the \maple session (cf. sect.
  \ref{start}). In the file \verb+fmuser.ma+ the user-defined rules have to be
  inserted. Initially all QCD Feynman rules are denoted there as an example. The
  conventions of declaration can be taken from this example.
\ei

\subsection{Particles and Masses} \label{masses}

What follows is a list which contains all particles \xloops knows. The names
valid in the {\sf Xwindows} interface are denoted (second column from the left)
as well as the names for internal use within \maple (third column from the
left). In addition for each particle the models in which it is declared (right
column) and the name which \xloops uses for the corresponding particle mass
(second column from the right) are listed.

\newpage

{\SI\setlength{\fboxsep}{0cm}\psset{linewidth=0.05cm}
\setlength{\tabcolsep}{0.15cm}\setlength{\unitlength}{0.9cm}\psset{unit=0.9cm}
\noindent
\rotateleft{
\shadowbox{\parbox{18.4cm}{\bc\tt \btb{p{4.5cm}p{3.0cm}p{3.6cm}p{3cm}p{2.8cm}}
\parbox{4.5cm}{\bp(5,1)\put(0.4,0.43){\epsfig{file=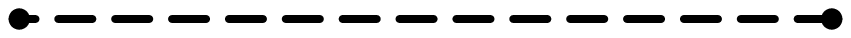,width=3.8cm}}
\put(1.1,0.07){\CC{1}}\put(3.5,0.07){\CC{2}}\put(2.3,0.8){$m_{\phi}$}\ep}
        & {\sf f, phi}
	& \parbox{3.6cm}{\CC{1} $=$ phi   \\ \CC{2} $=$ phi}
             & $m_{\phi}=$ Mphi & \parbox{2.8cm}{phithree, \\ phifour} \\[1.8cm]
\parbox{4.5cm}{\bp(5,1)\put(0.4,0.43){\epsfig{file=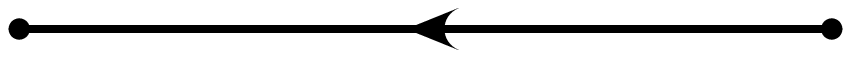,width=3.8cm}}
\put(1.1,0.07){\CC{1}}\put(3.5,0.07){\CC{2}}\put(2.3,0.8){$m_{\nu_{e}}$}\ep}
        & {\sf ve, nuelec}
	& \parbox{3.6cm}{\CC{1} $=$ nuelec \\ \CC{2} $=$ nuelecbar} 
           & $m_{\nu_{e}}=$ Mnuelec & \parbox{2.8cm}{SM, EW, QED, \\ THDM, GBEM}
	      \\[1cm]
\parbox{4.5cm}{\bp(5,1)\put(0.4,0.43){\epsfig{file=prop4.eps,width=3.8cm}}
\put(1.1,0.07){\CC{1}}\put(3.5,0.07){\CC{2}}\put(2.3,0.8){$m_{e}$}\ep}
        & {\sf e, elec}
	& \parbox{3.6cm}{\CC{1} $=$ elec    \\ \CC{2} $=$ elecbar}
	      & $m_{e}=$ Melec & \parbox{2.8cm}{SM, EW, QED, \\ THDM, GBEM}
	      \\[1cm]
\parbox{4.5cm}{\bp(5,1)\put(0.4,0.43){\epsfig{file=prop4.eps,width=3.8cm}}
\put(1.1,0.07){\CC{1}}\put(3.5,0.07){\CC{2}}\put(2.3,0.8){$m_{\nu_{\mu}}$}\ep}
        & {\sf v$\mu$, numu}
	& \parbox{3.6cm}{\CC{1} $=$ numu    \\ \CC{2} $=$ numubar} 
           & $m_{\nu_{\mu}}=$ Mnumu & \parbox{2.8cm}{SM, EW, QED, \\ THDM, GBEM}
	      \\[1cm]
\parbox{4.5cm}{\bp(5,1)\put(0.4,0.43){\epsfig{file=prop4.eps,width=3.8cm}}
\put(1.1,0.07){\CC{1}}\put(3.5,0.07){\CC{2}}\put(2.3,0.8){$m_{\mu}$}\ep}
        & {\sf $\mu$, mu}
	& \parbox{3.6cm}{\CC{1} $=$ mu      \\ \CC{2} $=$ mubar}
	      & $m_{\mu}=$ Mmu & \parbox{2.8cm}{SM, EW, QED, \\ THDM, GBEM}
	      \\[1cm]
\parbox{4.5cm}{\bp(5,1)\put(0.4,0.43){\epsfig{file=prop4.eps,width=3.8cm}}
\put(1.1,0.07){\CC{1}}\put(3.5,0.07){\CC{2}}\put(2.3,0.8){$m_{\nu_{\tau}}$}\ep}
        & {\sf vy, nutau}
	& \parbox{3.6cm}{\CC{1} $=$ nutau   \\ \CC{2} $=$ nutaubar}
	 & $m_{\nu_{\tau}}=$ Mnutau & \parbox{2.8cm}{SM, EW, QED, \\ THDM, GBEM}
	      \\[1cm]
\parbox{4.5cm}{\bp(5,1)\put(0.4,0.43){\epsfig{file=prop4.eps,width=3.8cm}}
\put(1.1,0.07){\CC{1}}\put(3.5,0.07){\CC{2}}\put(2.3,0.8){$m_{\tau}$}\ep}
        & {\sf y, tau}
	& \parbox{3.6cm}{\CC{1} $=$ tau     \\ \CC{2} $=$ taubar}
	      & $m_{\tau}=$ Mtau & \parbox{2.8cm}{SM, EW, QED, \\ THDM, GBEM}
	      \\[1cm]
\parbox{4.5cm}{\bp(5,1)\put(0.4,0.43){\epsfig{file=prop4.eps,width=3.8cm}}
\put(1.1,0.07){\CC{1}}\put(3.5,0.07){\CC{2}}\put(2.3,0.8){$m_{u}$}\ep}
        & {\sf u, up}
	& \parbox{3.6cm}{\CC{1} $=$ up      \\ \CC{2} $=$ upbar}
	      & $m_{u}=$ Mup & \parbox{2.8cm}{SM, EW, QED, \\ QCD, THDM, GBEM}
\etb\ec}}}
\newpage
\noindent
\rotateleft{
\shadowbox{\parbox{18.4cm}{\bc\tt \btb{p{4.5cm}p{3.0cm}p{3.6cm}p{3cm}p{2.8cm}}
\parbox{4.5cm}{\bp(5,1)\put(0.4,0.43){\epsfig{file=prop4.eps,width=3.8cm}}
\put(1.1,0.07){\CC{1}}\put(3.5,0.07){\CC{2}}\put(2.3,0.8){$m_{d}$}\ep}
        & {\sf d, down}
	& \parbox{3.6cm}{\CC{1} $=$ down    \\ \CC{2} $=$ downbar}
	      & $m_{d}=$ Mdown & \parbox{2.8cm}{SM, EW, QED, \\ QCD, THDM, GBEM}
	      \\[1cm]
\parbox{4.5cm}{\bp(5,1)\put(0.4,0.43){\epsfig{file=prop4.eps,width=3.8cm}}
\put(1.1,0.07){\CC{1}}\put(3.5,0.07){\CC{2}}\put(2.3,0.8){$m_{c}$}\ep}
        & {\sf c, charm}
	& \parbox{3.6cm}{\CC{1} $=$ charm   \\ \CC{2} $=$ charmbar}
	    & $m_{c}=$ Mcharm & \parbox{2.8cm}{SM, EW, QED, \\ QCD, THDM, GBEM}
	      \\[1cm]
\parbox{4.5cm}{\bp(5,1)\put(0.4,0.43){\epsfig{file=prop4.eps,width=3.8cm}}
\put(1.1,0.07){\CC{1}}\put(3.5,0.07){\CC{2}}\put(2.3,0.8){$m_{s}$}\ep}
        & {\sf s, strange}
	& \parbox{3.6cm}{\CC{1} $=$ strange \\ \CC{2} $=$ strangebar} 
	   & $m_{s}=$ Mstrange & \parbox{2.8cm}{SM, EW, QED, \\ QCD, THDM, GBEM}
	      \\[1cm]
\parbox{4.5cm}{\bp(5,1)\put(0.4,0.43){\epsfig{file=prop4.eps,width=3.8cm}}
\put(1.1,0.07){\CC{1}}\put(3.5,0.07){\CC{2}}\put(2.3,0.8){$m_{t}$}\ep}
        & {\sf t, top}
	& \parbox{3.6cm}{\CC{1} $=$ top	   \\ \CC{2} $=$ topbar}
	      & $m_{t}=$ Mtop & \parbox{2.8cm}{SM, EW, QED, \\ QCD, THDM, GBEM}
	      \\[1cm]
\parbox{4.5cm}{\bp(5,1)\put(0.4,0.43){\epsfig{file=prop4.eps,width=3.8cm}}
\put(1.1,0.07){\CC{1}}\put(3.5,0.07){\CC{2}}\put(2.3,0.8){$m_{b}$}\ep}
        & {\sf b, bottom}
	& \parbox{3.6cm}{\CC{1} $=$ bottom  \\ \CC{2} $=$ bottombar}
	    & $m_{b}=$ Mbottom & \parbox{2.8cm}{SM, EW, QED, \\ QCD, THDM, GBEM}
	      \\[1.8cm]
\parbox{4.5cm}{\bp(5,1)\put(0.23,0.43){\epsfig{file=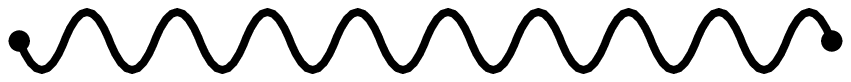,width=4.07cm}}
\put(1.1,-0.08){\CC{1}}\put(3.5,-0.08){\CC{2}}\put(2.3,0.95){$m_{\gamma}$}\ep}
        & {\sf \~{}, gamma}
	& \parbox{3.6cm}{\CC{1} $=$ gamma   \\ \CC{2} $=$ gamma}
	& $m_{\gamma}=$ Mgamma $=0$ & \parbox{2.8cm}{SM, EW, QED, \\ THDM, GBEM}
               \\[1cm]
\parbox{4.5cm}{\bp(5,1)\put(0.23,0.43){\epsfig{file=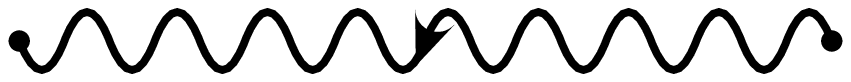,width=4.07cm}}
\put(1.1,-0.08){\CC{1}}\put(3.5,-0.08){\CC{2}}\put(2.3,0.95){$m_{W}$}\ep}
        & {\sf W, wp}
	& \parbox{3.6cm}{\CC{1} $=$ wp      \\ \CC{2} $=$ wm}
	      & $m_{W}=$ Mw & \parbox{2.8cm}{SM, EW, \\ THDM, GBEM} \\[1cm]
\parbox{4.5cm}{\bp(5,1)\put(0.23,0.43){\epsfig{file=prop1.eps,width=4.07cm}}
\put(1.1,-0.08){\CC{1}}\put(3.5,-0.08){\CC{2}}\put(2.3,0.95){$m_{Z}$}\ep}
        & {\sf Z, z0}
	& \parbox{3.6cm}{\CC{1} $=$ z0      \\ \CC{2} $=$ z0}
	      & $m_{Z}=$ Mz0 & \parbox{2.8cm}{SM, EW, \\ THDM, GBEM}
\etb\ec}}}
\newpage
\noindent
\rotateleft{
\shadowbox{\parbox{18.4cm}{\bc\tt \btb{p{4.5cm}p{3.0cm}p{3.6cm}p{3cm}p{2.8cm}}
\parbox{4.5cm}{\bp(5,1)\put(0.23,0.43){\epsfig{file=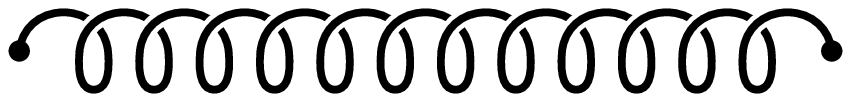,width=4.07cm}}
\put(1.1,-0.08){\CC{1}}\put(3.5,-0.08){\CC{2}}\put(2.3,1.05){$m_{g}$}\ep}
        & {\sf g, gluon}
	& \parbox{3.6cm}{\CC{1} $=$ gluon   \\ \CC{2} $=$ gluon}
	     & $m_{g}=$ Mgluon $=0$ & SM, QCD, THDM \\[1.8cm]
\parbox{4.5cm}{\bp(5,1)\put(0.4,0.43){\epsfig{file=prop6.eps,width=3.8cm}}
\put(1.1,0.07){\CC{1}}\put(3.5,0.07){\CC{2}}\put(2.3,0.8){$m_{H}$}\ep}
        & {\sf H, higgs}
	& \parbox{3.6cm}{\CC{1} $=$ higgs   \\ \CC{2} $=$ higgs}
	      & $m_{H}=$ Mhiggs & SM, EW, GBEM \\[1.8cm]
\parbox{4.5cm}{\bp(5,1)\put(0.4,0.43){\epsfig{file=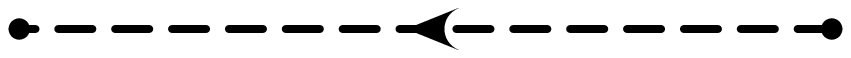,width=3.8cm}}
\put(1.1,0.07){\CC{1}}\put(3.5,0.07){\CC{2}}\ep}
        & {\sf g+, goldp}
	& \parbox{3.6cm}{\CC{1} $=$ goldp   \\ \CC{2} $=$ goldm}
	      & {\rm ---} & SM, EW, THDM \\[1cm]
\parbox{4.5cm}{\bp(5,1)\put(0.4,0.43){\epsfig{file=prop6.eps,width=3.8cm}}
\put(1.1,0.07){\CC{1}}\put(3.5,0.07){\CC{2}}\ep}
        & {\sf g0, gold0}
	& \parbox{3.6cm}{\CC{1} $=$ gold0   \\ \CC{2} $=$ gold0}
	      & {\rm ---} & SM, EW, THDM \\[1.8cm]
\parbox{4.5cm}{\bp(5,1)\put(0.4,0.43){\epsfig{file=prop6.eps,width=3.8cm}}
\put(1.1,0.07){\CC{1}}\put(3.5,0.07){\CC{2}}\put(2.3,0.8){$m_{H_0}$}\ep}
        & {\sf H, Higgs0}
	& \parbox{3.6cm}{\CC{1} $=$ Higgs0 \\ \CC{2} $=$ Higgs0}
	      & $m_{H_0}=$ MHiggs0 & THDM \\[1cm]
\parbox{4.5cm}{\bp(5,1)\put(0.4,0.43){\epsfig{file=prop6.eps,width=3.8cm}}
\put(1.1,0.07){\CC{1}}\put(3.5,0.07){\CC{2}}\put(2.3,0.8){$m_{h_0}$}\ep}
        & {\sf h, higgs0}
	& \parbox{3.6cm}{\CC{1} $=$ higgs0 \\ \CC{2} $=$ higgs0}
	      & $m_{h_0}=$ Mhiggs0 & THDM
\etb\ec}}}
\newpage
\noindent
\rotateleft{
\shadowbox{\parbox{18.4cm}{\bc\tt \btb{p{4.5cm}p{3.0cm}p{3.6cm}p{3cm}p{2.8cm}}
\parbox{4.5cm}{\bp(5,1)\put(0.4,0.43){\epsfig{file=prop5.eps,width=3.8cm}}
\put(1.1,0.07){\CC{1}}\put(3.5,0.07){\CC{2}}\put(2.3,0.8){$m_{H}$}\ep}
        & {\sf H+, Higgsp}
	& \parbox{3.6cm}{\CC{1} $=$ Higgsp  \\ \CC{2} $=$ Higgsm}
	      & $m_{H}=$ MHiggs & THDM \\[1cm]
\parbox{4.5cm}{\bp(5,1)\put(0.4,0.43){\epsfig{file=prop6.eps,width=3.8cm}}
\put(1.1,0.07){\CC{1}}\put(3.5,0.07){\CC{2}}\put(2.3,0.8){$m_{A_0}$}\ep}
        & {\sf A0, a0}
	& \parbox{3.6cm}{\CC{1} $=$ a0   \\ \CC{2} $=$ a0}
	      & $m_{A_0}=$ Ma0 & THDM \\[1.8cm]
\parbox{4.5cm}{\bp(5,1)\put(0.4,0.43){\epsfig{file=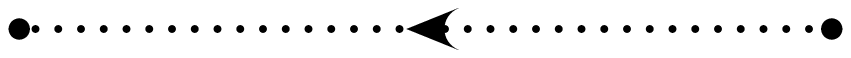,width=3.8cm}}
\put(1.1,0.07){\CC{1}}\put(3.5,0.07){\CC{2}}\ep}
        & {\sf G\~{}, ghostgamma}
	& \parbox{3.6cm}{\CC{1} $=$ ghostgamma \\ \CC{2} $=$ ghostgammabar} 
	      & {\rm ---} & SM, EW, THDM \\[1cm]
\parbox{4.5cm}{\bp(5,1)\put(0.4,0.43){\epsfig{file=prop7.eps,width=3.8cm}}
\put(1.1,0.07){\CC{1}}\put(3.5,0.07){\CC{2}}\ep}
        & {\sf GZ, ghostz0}
	& \parbox{3.6cm}{\CC{1} $=$ ghostz0 \\ \CC{2} $=$ ghostz0bar} 
	      & {\rm ---} & SM, EW, THDM \\[1cm]
\parbox{4.5cm}{\bp(5,1)\put(0.4,0.43){\epsfig{file=prop7.eps,width=3.8cm}}
\put(1.1,0.07){\CC{1}}\put(3.5,0.07){\CC{2}}\put(2.4,0.75){\Cp}\ep}
        & {\sf GW+, ghostwp}
	& \parbox{3.6cm}{\CC{1} $=$ ghostwp \\ \CC{2} $=$ ghostwmbar} 
	      & {\rm ---} & SM, EW, THDM \\[1cm]
\parbox{4.5cm}{\bp(5,1)\put(0.4,0.43){\epsfig{file=prop7.eps,width=3.8cm}}
\put(1.1,0.07){\CC{1}}\put(3.5,0.07){\CC{2}}\put(2.4,0.75){\Cm}\ep}
        & {\sf GW--, ghostwm}
	& \parbox{3.6cm}{\CC{1} $=$ ghostwm \\ \CC{2} $=$ ghostwpbar} 
	      & {\rm ---} & SM, EW, THDM \\[1cm]
\parbox{4.5cm}{\bp(5,1)\put(0.4,0.43){\epsfig{file=prop7.eps,width=3.8cm}}
\put(1.1,0.07){\CC{1}}\put(3.5,0.07){\CC{2}}\ep}
        & {\sf Gg, ghostgluon}
	& \parbox{3.6cm}{\CC{1} $=$ ghostgluon \\ \CC{2} $=$ ghostgluonbar} 
	      & {\rm ---} & SM, QCD, THDM
\etb\ec}}}}

\newpage

\noindent
Please note that the masses of the neutrinos are set to zero by default.
To keep them as variables, one can switch off the entry {\sf Set Neutrino Mass
0} of the {\sf Options} menu or use \\[0.4cm]
\verb+UnsetNeutrinoMass0();+ \\[0.4cm]
within \maple which removes any assignment of {\tt Mnuelec}, {\tt Mnumu} and
{\tt Mnutau}. As a consequence now a leptonic CKM matrix exists which behaves
analogous to the usual CKM matrix of quarks (cf. sect. \ref{mixing}). The
reverse command \\[0.4cm]
\verb+SetNeutrinoMass0();+ \\[0.4cm]
or the reselection of {\sf Set Neutrino Mass 0} sets all neutrino masses back
to zero. The leptonic CKM matrix equals now the unit matrix.

The masses of photon and gluon are also set to zero by default. To keep them as
variables you have to type \\[0.4cm]
\verb+Mgamma:='Mgamma'; Mgluon:='Mgluon';+ \\[0.4cm]
The masses of leptons and light quarks -- all standard model fermions except $b$
and $t$ quark -- can collectively be set to zero with the entry {\sf Massless
Light Fermions} of the {\sf Options} menu or with \\[0.4cm]
\verb+SetLightMass0();+ \\[0.4cm]
and be unassigned with \\[0.4cm]
\verb+UnsetLightMass0();+ \\[0.4cm]
or another click on {\sf Massless Light Fermions}. Apart from this, masses,
momenta and couplings -- as they are explained in the following sections -- can
be assigned by the user.

Moreover, if you want to assign all particle masses\footnote{Unfortunately the
Higgs mass is not yet included} as well as the couplings,
the Cabibbo-Kobayashi-Maskawa matrix (CKM matrix) and the Weinberg angle with
their experimentally determined values, you can use the entry {\sf Insert
Particle Properties} of the {\sf Options} menu or type \\[0.4cm]
\verb+read`+{$\langle$\it path$\rangle$}\verb+values.ma`;+ \\[0.4cm]
where {$\langle$\it path$\rangle$} is the same path used for the assignment
of \verb+LoopPath+ at the beginning of the \maple session (cf. sect.
\ref{start}). The file \verb+values.ma+ contains the parameters according to
the currently published data of the Particle Data Group \cite{PDG}. To remove
all information of \verb+values.ma+ in a \maple session you can type \\[0.4cm]
\verb+read`+{$\langle$\it path$\rangle$}\verb+unvalue.ma`;+ \\[0.4cm]
or switch off the {\sf Insert Particle Properties} entry. Instead of the
fermion names which were already introduced one can also use the more general
terms
\bdm
\verb+up1, up2, up3, down1, down2, down3+\,.
\edm
The difference is, that for these terms no electric charge {\tt Q} and no weak
isospin {\tt T3l} is defined. While \xloops assumes
\bdm
\verb+Q(elec)+ = -1\,; \qquad \verb+T3l(elec)+ = -\Ts{\frac{1}{2}}\,; \qquad
\verb+Q(up)+ = \Ts{\frac{2}{3}}\,; \qquad \verb+T3l(up)+ = \Ts{\frac{1}{2}}\,;
\qquad \ldots
\edm
it has no values for \verb+Q(up1)+, \verb+T3l(up1)+, \ldots\ and the user has
the opportunity to make use of this freedom.

\subsection{Feynman rules}

Since the conventions for Feynman rules vary in the literature, it is important
to explain one's own rules: We use Feynman gauge, for electroweak processes
according to the rules given in \cite{Aok}, but with all three-vertices and all
propagators multiplied by $-i$ and all four-vertices and the integration measure
multiplied by $i$. This convention coincides with the rules of \cite{Nac} -- as
far as these are applicable in Feynman gauge. From \cite{Nac} we took the rules
for QCD and the notation of the CKM matrix, too. The rules for the Two Higgs
doublet model are from \cite{Gun}.

Anyway, you receive the rule for one particular vertex easily if you just use
the tree level topologies or call the \verb+EvalGraph0+ procedure.

In our convention the electromagnetic coupling {\tt e} and the strong coupling
{\tt gs} are positive (the electric charge of the electron is {\tt -e}). \xloops
automatically inserts a factor $(2\pi)^{2\eps-4}$ for every closed loop and an
additional $-1$ for each fermion and ghost loop. This means that the dimension
-- which \xloops calls {\tt D} -- equals {\tt 4 - 2 eps}. The correct symmetry
factor of the diagram is also included. The scale which keeps the coupling
constants dimensionless is called {\tt MU}.

\subsection{Mixing angles and CKM matrix} \label{mixing}

Three mixing angles are known by \xloops: The usual Weinberg angle {\tt tw} and
-- only relevant in the case of the Two Higgs doublet model -- the angles {\tt
alpha} and {\tt beta} (cf. \cite{Gun}). The elements of the CKM matrix are
denoted by \verb+CKM(up,down)+ \ldots, its complex conjugates are
\verb+CKMC(up,down)+ \ldots. If neutrino masses are not set equal to zero there
exists also a leptonic CKM matrix with elements like \verb+CKM(nuelec,elec)+.
If the neutrino masses are set to zero -- which is the default (cf. sect.
\ref{masses}) -- this matrix equals the unit matrix.

\subsection{Momenta and Metric}

The external momenta \verb+q1(nu1), q2(nu1),+ \ldots\ may occur in two different
contexts,
\ben
\item as Lorentz vectors \verb+q1(nu1)+, \ldots\ with Lorentz indices
  \verb+nu1+, \ldots,
\item contracted with another Lorentz vector.
\een
In the first case the momentum will appear in the output as it is. In the second
case the contraction will be written in terms of parallel space components:
\be\ba{rclcrclcrcl}
\verb+q10+ & = & \Ds{\sqrt{\bmm{q_{1}}{2}}}\,; & &
\verb+q20+ & = & \Ds{\frac{\bm{q_{1} \cdot q_{2}}}{\sqrt{\bmm{q_{1}}{2}}}}\,;
& &
\verb+q21+ & = & \Ds{\sqrt{\frac{(\bm{q_{1} \cdot q_{2}})^{2}}{\bmm{q_{1}}{2}}
  - \bmm{q_{2}}{2}}}\,; \\[0.6cm]
\verb+q30+ & = & \Ds{\frac{\bm{q_{1} \cdot q_{3}}}{\sqrt{\bmm{q_{1}}{2}}}}\,;
& &
\verb+q31+ & = & \Ds{- \,\frac{\bm{\hat{q}_{2} \cdot \bm{q}_{3}}}{\sqrt{-
  \bmm{\hat{q}_{2}}{2}}}}\,; & &
\verb+q32+ & = & \Ds{\sqrt{\frac{(\bm{q_{1} \cdot q_{3}})^{2}}{\bmm{q_{1}}{2}}
  + \frac{(\bm{\hat{q}_{2} \cdot q_{3}})^{2}}{\bmm{\hat{q}_{2}}{2}} - 
  \bmm{q_{3}}{2}}}\,,
\ea\ee
where the auxiliary property
\be
\label{aux}
\bm{\hat{q}_{2}} = \bm{q_{2}} - \frac{\bm{q_{1} \cdot q_{2}}}{\bmm{q_{1}}{2}} \,
\bm{q_{1}}
\ee
was used. In appendix \ref{test} we demonstrate how this works in practice.

The metric tensor of \xloops is \verb+G(nu1,nu2)+. It obeys the usual
conventions
\bdm
\verb+G(0,0)+ = 1\,; \qquad \verb+G(1,1)+ = -1 \qquad \ldots
\edm
The contraction of \verb+G(nu1,nu2)+ depends on whether the indices {\tt nu1}
and {\tt nu2} are orthogonal space indices -- then $\verb+G(nu1,nu1)+={\tt
D}-\mbox{\tt DimP}$ -- or belong to the entire Minkowski space -- this means
that $\verb+G(nu1,nu1)+={\tt D}$.

The totally antisymmetric tensor is \verb+Eps(nu1,nu2,nu3,nu4)+ with
$\verb+Eps(0,1,2,3)+=1$.

\subsection{Dirac matrices}

Dirac $\gamma$-matrices \xloops calls \verb+Dg(mu1)+, \dots. Like the external
momenta they can appear either as Lorentz vector as they are, or contracted by
another Lorentz vector. In the latter case they are denoted by 
\be
\verb+Dg0+ = \frac{\bm{\slash{q}_{1}}}{\sqrt{\bmm{q_{1}}{2}}}\,; \quad
\verb+Dg1+ = - \,\frac{\bm{\slash{\hat{q}}_{2}}}{\sqrt{- \bmm{\hat{q}_{2}}
      {2}}}\,; \quad
\verb+Dg2+ = - \,\frac{\bm{\slash{\hat{q}}_{3}}}{\sqrt{- \bmm{\hat{q}_{3}}{2}}}
\ee
with
\be
\bm{\hat{q}_{3}} = \bm{q_{3}} - \frac{\bm{q_{1} \cdot q_{3}}}{\bmm{q_{1}}{2}} \,
\bm{q_{1}} - \frac{\bm{\hat{q}_{2} \cdot q_{3}}}{\bmm{\hat{q}_{2}}{2}} \, 
\bm{\hat{q}_{2}}
\ee
and the abbreviation (\ref{aux}). The Dirac structure is written as a string of
the non-commutative product \verb+&*+. Since the program is also designed for
diagrams with more than one fermion line it labels each string of Dirac matrices
with a number. For example in the case of fermionic self-energies
\bdm
\verb+&*(1,ONE)+\,; \qquad \verb+&*(1,Dg0)+\,; \qquad  \verb+&*(1,Dg5)+\,;
\qquad \verb+&*(1,Dg5,Dg0)+
\edm
will occur in the output -- the latter two of course only in the case of $W, Z$
or Goldstone boson exchange. \verb+ONE+ represents the identity of the Clifford
algebra, \verb+Dg5+ is $\gamma_5$. The first entry in the non-commutative string
gives the number of the fermion line. The 1 means that the string belongs to the
first fermion line which in this case is the only fermion line. In appendix
\ref{test} this treatment can be seen in practice.

\subsection[$SU(N_c)$ algebra]{$\bm{SU(N_c)}$ algebra}

In QCD calculations \xloops keeps the colour number {\tt Nc} unassigned by
default, but with \\[0.4cm]
\verb+Nc:=3;+ \\[0.4cm]
the number is fixed to its standard model value. The $SU(N_c)$ algebra needs as
ingredients
\bi
\item the identity matrix of the $SU(N_c)$ algebra \verb+delta3(ta1,ta2)+ which
  should not be mixed with Kronecker \verb+delta8(ta1,ta2)+ which contracts the
  Gell-Mann matrices with the labels {\tt ta1} and {\tt ta2}. Of course
  \verb+delta3(ta1,ta2)+ and \verb+delta8(ta1,ta2)+ are both diagonal with
  entries 1 in the diagonal. This means that
  \bdm
  \sum_{\mbox{\SIII\tt ta1}} \verb+delta3(ta1,ta1)+ = N_c = 3 \,; \qquad 
  \sum_{\mbox{\SIII\tt ta1}} \verb+delta8(ta1,ta1)+ = N_c^2-1 = 8 \,.
  \edm
\item the Gell-Mann matrices \verb+Ts(ta1,ta2,ta3)+ themselves, where {\tt ta1}
  -- which is the number of the matrix -- can take values from 1 to $N_c^2-1$.
  {\tt ta2} and {\tt ta3} are the indices of a particular Gell-Mann matrix and
  therefore have values from 1 to $N_c$. The \verb+Ts(ta1,ta2,ta3)+ are
  normalized by the following trace condition:
  \bdm
  \sum_{\mbox{\SIII\tt ta3,ta4}} \verb+Ts(ta1,ta3,ta4) Ts(ta2,ta4,ta3)+ =
  2 \, \verb+delta8(ta1,ta2)+ \,.
  \edm
\item the usual structure functions \verb+fs(ta1,ta2,ta3)+ and
  \verb+ds(ta1,ta2,ta3)+ where all three indices vary from 1 to $N_c^2-1$. They
  are defined by the Lie algebra relation
  \beas
  \lefteqn{\sum_{\mbox{\SIII\tt ta5}} \mbox{\tt Ts(ta1,ta3,ta5)
  Ts(ta2,ta5,ta4)}} \\
  & = & \frac{2}{\mbox{\tt Nc}} \, \mbox{\tt delta8(ta1,ta2) delta3(ta3,ta4)} \\
  & & + \sum_{\mbox{\SIII\tt ta5}} \verb+ds(ta1,ta2,ta5) Ts(ta5,ta3,ta4)+ \\
  & & + i \sum_{\mbox{\SIII\tt ta5}} \verb+fs(ta1,ta2,ta5) Ts(ta5,ta3,ta4)+\,,
  \eeas
  where the \verb+fs(ta1,ta2,ta5)+ are completely antisymmetric and the
  \verb+ds(ta1,ta2,+ \verb+ta5)+ are completely symmetric.
\ei

\subsection{Output} \label{output}

The output of the diagrams is decomposed in one or several form factors {\tt
C1}, {\tt C2}, \ldots\ to reflect the Dirac $\gamma$ or Lorentz structure of
the result. Therefore a list of form factors is returned. The last entry in
this list is the defining equation for the form factors. The result is
presented in the \maple\ {\it text window} of the {\it main window} and has the
following structure:  
\bdm
\mbox{\tt G1} := [\underbrace{\mbox{\tt C1}=\ldots,\,
\ldots,\mbox{{\tt C}{\it n}}=\ldots}_{\mbox{the form factors
{\tt C}{\it i}}},\,\underbrace{\mbox{\tt C1 q1(nu1) q2(nu2)} + \cdots +
\mbox{\tt C{\rm\it n} G(nu1,nu2)}\ldots}_{\mbox{defining equation for form
factors}}]
\edm

Let us consider for instance the self-energy of a vector boson.
\xloops finds as Lorentz structure $\mbox{\tt C1 G(nu1,nu2)} + \mbox{\tt C2
q1(nu1) q1(nu2)}$ and the output roughly looks like
\bdm
[\mbox{\tt C1} = \mbox{\em term1}\,,\,\mbox{\tt C2} = \mbox{\em term2}\,,\,
\mbox{\tt C1 G(nu1,nu2)} + \mbox{\tt C2 q1(nu1) q1(nu2)}]\,,
\edm
where {\em term1} and {\em term2} denote the form factors. In the case of a
fermionic self-energy the result carries no Lorentz index. Instead, it is
decomposed with respect to \verb+&*(1,ONE)+, \verb+&*(1,Dg5)+, \verb+&*(1,Dg0)+
and \verb+&*(1,Dg5,Dg0)+ and in the general case the result is of the form
\beas
\lefteqn{[\mbox{\tt C1} = \mbox{\em term1}\,,\,\mbox{\tt C2} = \mbox{\em term2}
\,,\,\mbox{\tt C3} = \mbox{\em term3}\,,\,\mbox{\tt C4} = \mbox{\em term4}\,,}
\\
& & \mbox{\tt  C1 (1 \&* ONE)} + \mbox{\tt C2 (1 \&* Dg5)} + \mbox{\tt C3 (1 \&*
Dg0)} + \mbox{\tt C4 \&*(1, Dg5, Dg0)}]\,.
\eeas
Here all $\gamma$-matrices are contracted. In vertex functions with a fermion
line and a vector boson there is also a {\tt Dg(nu1)} appearing. In appendix
\ref{test} several examples are given in more detail. In the case of a scalar
self-energy there appears only one form factor and the result reads
\bdm
[\mbox{\tt C1} = \mbox{\em term1}\,,\,\mbox{\tt  C1}]\,.
\edm
The form factors can be written in several modes of evaluation which correspond
to the different buttons of the {\it diagram windows}:
\bi
\item With \eval or if no qualifier is added to the {\tt EvalGraph} procedures
  the output -- for simplicity -- is returned in terms of unevaluated
  \verb+Oneloop+ and \verb+Twoloop+ functions which correspond directly to the
  \verb+OneLoop+ and \verb+TwoLoop+ functions explained in sect. \ref{Oneloop}
  and \ref{Twoloop}.
\item To get complete answers, \evalf or in the case of \verb+EvalGraph1+ and
  \verb+EvalGraph2+ the qualifier {\tt full} forces \xloops to evaluate
  directly the {\tt Oneloop} and {\tt Twoloop} integrals. To be precise, you
  have to type \\[0.4cm]
  \verb+EvalGraph+{\it m}\verb+(+$n$\verb+,+$\langle${\it
  list}$\rangle$\verb+,full);+ \\[0.4cm]
  where $m$ and $n$ are numbers. Now each form factor {\tt C1}, \ldots\ is a
  Laurent expansion in terms of the ultraviolet regulator $\eps = (4 - D)/2$,
  where the significant coefficients of this expansion --
  ${\mcl O}(\eps^{-1})$, ${\mcl O} (\eps^0)$ in the one-loop case,
  ${\mcl O}(\eps^{-2})$, ${\mcl O}(\eps^{-1})$,
  ${\mcl O}(\eps^0)$ in the two-loop case -- are denoted in form of a list. The
  bosonic self-energy we introduced before now has the structure
  \beas
  \lefteqn{[\mbox{\tt C1} = [\eps^{-1}\mbox{-{\em term}},\eps^0\mbox{-{\em
  term}}]\,,} \\
  & & \mbox{\tt C2} = [\eps^{-1}\mbox{-{\em term}},\eps^0\mbox{-{\em term}}]\,,
  \,\mbox{\tt C1 G(nu1,nu2)} + \mbox{\tt C2 q1(nu1) q1(nu2)}]
  \eeas
  for one-loop contributions and
  \beas
  \lefteqn{[\mbox{\tt C1} = [\eps^{-2}\mbox{-{\em term}},\eps^{-1}\mbox{-{\em
  term}}, \eps^0\mbox{-{\em term}}]\,,} \\
  & & \mbox{\tt C2} = [\eps^{-2}\mbox{-{\em term}}, \eps^{-1}\mbox{-{\em term}},
  \eps^0\mbox{-{\em term}}]\,,\,\mbox{\tt C1 G(nu1,nu2)} + \mbox{\tt C2
  q1(nu1) q1(nu2)}]
  \eeas
  for two-loop diagrams. $\eps^{-2}$-{\em term} and $\eps^{-1}$-{\em term} are
  meant to represent the divergent part (the coefficients of $\frac{1}{\eps^2}$
  and $\frac{1}{\eps}$) whereas $\eps^{0}$-{\em term} is describing the finite
  part of the corresponding form factor. For those two-loop topologies where no
  analytic result is known, the finite part contributes two entries to the
  output list. The first of them is a list itself which denotes an integral
  representation, the second contains the part which is analytically calculable:
  \beas
  \lefteqn{[\mbox{\tt C1} = [\eps^{-2}\mbox{-{\em term}},\eps^{-1}\mbox{-{\em
  term}}, \mbox{{\em numeric}}, \eps^0\mbox{-{\em term}}]\,,} \\
  & & \mbox{\tt C2} = [\eps^{-2}\mbox{-{\em term}}, \eps^{-1}\mbox{-{\em term}},
  \mbox{{\em numeric}}, \eps^0\mbox{-{\em term}}]\,, \\
  & & \mbox{\tt C1 G(nu1,nu2)} + \mbox{\tt C2 q1(nu1) q1(nu2)}]\,.
  \eeas
  A more detailed description of the integral representation -- which was called
  here {\em numeric} -- is given in sect. \ref{twoout}.
\item The integral representation is evaluated directly with the help of \vegas
  if \evaln is used or if the qualifier {\tt full} replaced by two lists which
  contain the input information for \vegas: \\[0.4cm]
  \verb+EvalGraph+{\it m}\verb+(+$n$\verb+,+$\langle${\it list}$\rangle$
  \verb+,[+$m_1,m_2$\verb+],[+$n_1,n_2$\verb+]);+ \\[0.4cm]
  Within an ordinary \maple session one has in addition to assign the variable
  \verb+NumPath+ as described in sect.~\ref{num}. The variables $m_1,m_2,n_1,
  n_2$ represent the grids where the integrand is evaluated: first $m_1$
  iterations with $m_2$ points -- \verb+itmx+$=m_1$ and \verb+ncall+$=m_2$ are
  passed to \vegas -- and then $n_1$ iterations with $n_2$ points -- this
  corresponds to \verb+itmx+$=n_1$ and \verb+ncall+$=n_2$. In the {\sf Xwindow}
  interface these values can be changed with the entry {\sf Numeric} of the
  {\sf Options} menu (cf. sect. \ref{num}). Real and imaginary part are
  integrated one after the other.

  The output has then the following structure:
  \beas
  \lefteqn{[\mbox{\tt C1} = [\eps^{-2}\mbox{-{\em term}},\eps^{-1}\mbox{-{\em
  term}}, [\mbox{{\em factor}},[\mbox{\em result},\mbox{\em error}]], \eps^0
  \mbox{-{\em term}}]\,,} \\
  & & \mbox{\tt C2} = [\eps^{-2}\mbox{-{\em term}}, \eps^{-1}\mbox{-{\em term}},
  [\mbox{{\em factor}},[\mbox{\em result},\mbox{\em error}]], \eps^0\mbox{-{\em
  term}}]\,, \\
  & & \mbox{\tt C1 G(nu1,nu2)} + \mbox{\tt C2 q1(nu1) q1(nu2)}]
  \eeas
  where {\em factor} denotes an analytically evaluated pre-factor of the
  integrand. The result of the integral itself is called {\em result} here, {\em
  error} describes the uncertainty of the numerical integration which is
  returned by \vegas. To be specific: the entire ${\mcl O}(\eps^0)$ contribution
  is obtained by
  \bdm
  \mbox{\em factor}\times\mbox{\em result}+\eps^0\mbox{-{\em term}},
  \edm
  the complete uncertainty is
  \bdm
  \mbox{\em factor}\times\mbox{\em error},
  \edm
  where the uncertainty of $\eps^{0}$-{\em term} is neglected.

  If not all parameters of the integrand carry numerical values the numerical
  integration cannot be performed. Therefore the same output as for the
  qualifier {\tt full} is returned.

\item With \evalm or in the case of \verb+EvalGraph1+, with the qualifier {\tt
  more} instead of {\tt full}, \xloops evaluates even the ${\mcl O}(\eps)$ of
  one-loop integrals. This
  becomes relevant if the one-loop integral is part of a two-loop calculation,
  because then it may be multiplied with a divergent one-loop $Z$-factor. This
  changes the output of a bosonic self-energy:
  \beas
  \lefteqn{[\mbox{\tt C1} = [\eps^{-1}\mbox{-{\em term}},\eps^0\mbox{-{\em
  term}},\eps^1\mbox{-{\em term}}]\,,} \\
  & & \mbox{\tt C2} = [\eps^{-1}\mbox{-{\em term}},\eps^0\mbox{-{\em term}},
  \eps^1\mbox{-{\em term}}]\,,\,\mbox{\tt C1 G(nu1,nu2)} + \mbox{\tt C2
  q1(nu1) q1(nu2)}]\,.
  \eeas
\ei
As far as the masses and momenta which enter the {\tt EvalGraph} functions are
symbols the output is given algebraically, whereas if they are all assigned with
numbers the result will be numerical.

Generally, in the numerical case, the program sets the value of the imaginary
part of the propagators, five digits higher than the numerical accuracy of the
whole calculation, for example to $10^{-15}$ if one calculates with 20 digits.
When reading the program \maple sets the number of digits to 20 -- except when
the user specified a higher accuracy before, for instance: \\[0.4cm]
\verb+Digits:=40;+ \\[0.4cm]
In the algebraic case just {\tt rho} is returned. The output is facilitated by
the usage of several abbreviations (\ref{abbrev-1}, \ref{abbrev-2},
\ref{abbrev-3}, \ref{abbrev-4}) for certain one-loop contributions which are
introduced in detail in the context of the {\tt OneLoop} procedure in sect.
\ref{rex}. The abbreviations are substituted if the {\tt evalRex} command is
used (cf. sect. \ref{rex}).

\subsection{Special relations} \label{rel}

If external momenta are zero or parallel, \xloops omits those Lorentz vectors
which are no longer independent and therefore redundant. The number of form
factors -- and the dimension of the parallel space -- may decrease. If this is
the case the form factors will no longer be expressed in terms of proper {\tt
Oneloop{\rm\it n}Pt} and {\tt Two}\-{\tt loop{\rm\it n}Pt{\rm \it m}} functions,
but with the help of its variants {\tt Oneloop{\rm\it k}\_{\rm\it n}Pt} and {\tt
Twoloop{\rm\it k}\_{\rm\it n}Pt{\rm\it m}} which are reduced in dimension of
parallel space. The {\tt Oneloop{\rm\it k}\_{\rm\it n}Pt} functions correspond
to the {\tt OneLoop{\rm\it k}\_{\rm\it n}Pt} functions from sect. \ref{Oneloop}.

If for instance in the case of the two-point function the external momentum {\tt
q1} shall vanish, one has to make the following assignment before starting the
calculation: \\[0.4cm]
\verb+q10:=0;+ \\[0.4cm]
Now the bosonic self-energy reduces to
\bdm
[\mbox{\tt C1} = \mbox{\em term1}\,,\,\mbox{\tt C1 G(nu1,nu2)}]
\edm
which means that one form factor vanished. The only remaining form factor is
expressed by a \verb+Oneloop1_2Pt+ function -- this is a two-point function with
a parallel space of dimension zero. The notation ``{\tt 1\_2Pt}'' reflects the
fact that this function carries properties of the two-point function as well as
of the one-point function.

If in the case of a three-point function both external momenta {\tt q1} and {\tt
q2} have a common rest frame one has to make the assignment
\\[0.4cm]
\verb+q21:=0;+ \\[0.4cm]
Now the momentum {\tt q2} will disappear from the equation which defines the
form factors since it is no longer independent of {\tt q1} (``zero recoil
limit''). In addition the proper \verb+Oneloop3Pt+ will switch to
\verb+Oneloop2_3Pt+ functions -- a mixing of two- and three-point functions.

It is important here to make the assignment first and then to evaluate the
diagram, because otherwise number and shape of form factors may be wrong.
Moreover, if special relations exist which fix the variables of the {\tt
EvalGraph} functions or connect them with each other -- like for instance the
on-shell condition which connects masses and momenta -- it is important first to
declare these relations and then evaluate the functions.

Otherwise, if the functions are evaluated first analytically and then numbers or
relations between variables are substituted in the result, \maple has to
do more work than necessary. Moreover, it might happen
in some special cases that an error occurs -- something like ``{\tt division by
0}''. Usually this sort of divergence is artificial and not an error \xloops is
responsible for -- at the moment of calculation the parameters which cause the
error message were general analytical expressions. The problem can be overcome
in either expanding in the critical parameters before the insertion of values,
or -- and this is recommended here -- in calling the diagram window or the {\tt 
EvalGraph} routines with the final values for the arguments from the very
beginning, because the program was taught to detect a lot of this critical
points and to avoid such artificial infinities in the result.

\subsection{Fast evaluation}

If the entry {\sf Oneloop Library} of the {\sf Options} menu is selected or --
within an ordinary \maple session -- the variable {\tt LibPath} is assigned
with some path, the procedures assume
that a library of {\tt OneLoop} functions is stored in the assigned directory --
as described in sect. \ref{lib}. The necessary integrals are then read from this
directory -- which of course is much faster (see sect. \ref{run}). Otherwise,
if {\tt LibPath} is not assigned, the procedures are forced to recalculate the
necessary {\tt OneLoop} functions at any occurrence.

\section{The {\md\tt OneLoop} procedures} \label{Oneloop}

\subsection{Input}

There exist nine basic functions for the calculation of one-loop integrals. The
first three functions
\bi
\item {\tt OneLoop1Pt}$(p,m,t)$
\item {\tt OneLoop2Pt}$(p_0,p_\bot,q_{10},m_1,m_2,t_1,t_2)$
\item {\tt OneLoop3Pt}$(p_0,p_1,p_\bot,q_{10},q_{20},q_{21},
           m_1,m_2,m_3,t_1,t_2,t_3)$
\ei
correspond to the notation in parallel and orthogonal space variables. The next
three functions
\bi
\item {\tt OneLoop1\_2Pt}$(p,m_1,m_2,t_1,t_2)$
\item {\tt OneLoop1\_3Pt}$(p,m_1,m_2,m_3,t_1,t_2,t_3)$
\item {\tt OneLoop2\_3Pt}$(p_0,p_\bot,q_{10},q_{20},m_1,m_2,m_3,t_1,t_2,t_3)$
\ei
are used in the case of reduced parallel space. All abovementioned functions
require the tensor degree, the external momenta $\bm{q_n}$ and the masses $m_n$
as input. The powers $t_n$ of the propagator terms in the denominator are
optional. The remaining three functions
\bi
\item {\tt OneLoopTens1Pt}$(i,m,t)$
\item {\tt OneLoopTens2Pt}$(i,\bmm{q_1}{2},m_1,m_2,t_1,t_2)$
\item {\tt OneLoopTens3Pt}$(i,\bmm{q_1}{2},\bmm{q_2}{2},(\bm{q_2-q_1})^2,
           m_1,m_2,m_3,t_1,t_2,t_3)$
\ei
require the squared external momenta as input variables and return the complete
rank-{\it i} tensor. The powers $t_n$ of the propagator terms in the denominator
again are optional.

\subsection{Notation}

In detail the basic functions directly correspond to the following integrals:
{\psset{linewidth=0.048cm}\setlength{\unitlength}{0.8cm}\psset{unit=0.8cm}
\bi
\item One-point function: \\
      \nop\hfill
      \parbox{10.6cm}{\bc\SII
      \parbox{10cm}{\bp(10,4)
      \pscircle(5,2.5){1.5}
      \psline{->}(6.47,2.6)(6.47,2.4)
      \pscircle[linewidth=0.1](3.53,2.5){0.1}
      \psline(1,2.5)(3.5,2.5)
      \put(6.7,2.5){$\bm{l}$}
      \put(5.9,2.5){$m$}
      \ep} 
      \ec} \vspace{-1cm} \nop
      \hfill\nop
      \bea
      \lefteqn{\mbox{\tt OneLoop1Pt}(p,m,t) = A^{(p)(t)}(m) = \int \! d^{D}l
      \,\, \frac{(\bm{l}^2)^{\frac{p}{2}}}{[\bm{l}^{2} - m^{2} + i \varrho]^t}}
      & & \rule{12.5cm}{0cm}
      \\[0.4cm]
      \label{one-1}
      \lefteqn{\mbox{\tt OneLoop1Pt}(p,m) = \mbox{\tt OneLoop1Pt}(p,m,1)}
      \\[0.4cm]
      \lefteqn{\mbox{\tt OneLoopTens1Pt}(i,m,t) = A_{\mu_1 \cdots \mu_i}
      ^{(t)}(m) = \int \! d^{D}l \,\, \frac{l_{\mu_1} \cdots l_{\mu_i}}
      {[\bm{l}^{2} - m^{2} + i \varrho]^t}} \\[0.4cm]
      \label{one-2}
      \lefteqn{\mbox{\tt OneLoopTens1Pt}(i,m) = \mbox{\tt OneLoopTens1Pt}
      (i,m,1)\,.} \eea \\[-0.2cm] \nop
\item Two-point function: \\
      \nop\hfill
      \parbox{10.6cm}{\bc\SII
      \parbox{10cm}{\bp(10,4)
      \pscircle(5,2.5){1.5}
      \psline{->}(5.1,3.97)(4.9,3.97)
      \psline{->}(4.9,1.03)(5.1,1.03)
      \pscircle[linewidth=0.1](3.53,2.5){0.1}
      \pscircle[linewidth=0.1](6.47,2.5){0.1}
      \psline{->}(1,2.5)(2.45,2.5)
      \psline(2.25,2.5)(3.5,2.5)
      \psline{->}(6.5,2.5)(7.95,2.5)
      \psline(7.75,2.5)(9,2.5)
      \put(4.85,1.3){$m_{1}$}
      \put(4.85,3.6){$m_{2}$}
      \put(4.6,0.53){$\bm{l+q_{1}}$}
      \put(5,4.2){$\bm{l}$}
      \put(2.05,2.77){$\bm{q_{1}}$}
      \put(7.55,2.77){$\bm{q_{1}}$}
      \ep}
      \ec} \vspace{-0.5cm} \nop
      \hfill\nop
      \bea
      \lefteqn{\mbox{\tt OneLoop2Pt}(p_0,p_\bot,q_{10},m_1,m_2,t_1,t_2) = 
      B^{(p_0 p_\bot)(t_{1} t_{2})}(q_{10}, m_{1}, m_{2})} & &
      \rule{12.5cm}{0cm} \nonumber \\[0.2cm]
      & = & \int \! d^{D}l \,\, \frac{(l_0)^{p_{0}} \, (l_{\bot})^{p_\bot}}
      {[(\bm{l+q_1})^{2}-m_{1}^{2} + i \varrho]^{t_1} \, [\bm{l}^{2} -
      m_{2}^{2} + i \varrho]^{t_2}} \hspace*{2.9cm} \nop \\[0.4cm]
      \label{two-1}
      \lefteqn{\mbox{\tt OneLoop2Pt}(p_0,p_\bot,q_{10},m_1,m_2) =
      \mbox{\tt OneLoop2Pt}(p_0,p_\bot,q_{10},m_1,m_2,1,1)} \\[0.4cm]
      \lefteqn{\mbox{\tt OneLoop1\_2Pt}(p,m_1,m_2,t_1,t_2)} \nonumber \\[0.2cm]
      & = & \int \! d^{D}l \,\, \frac{(\bm{l}^2)^{\frac{p}{2}}}
      {[\bm{l}^{2}-m_{1}^{2} + i \varrho]^{t_1} \, [\bm{l}^{2} -
      m_{2}^{2} + i \varrho]^{t_2}} \hspace*{2.9cm} \nop \\[0.4cm]
      \lefteqn{\mbox{\tt OneLoop1\_2Pt}(p,m_1,m_2) =
      \mbox{\tt OneLoop1\_2Pt}(p,m_1,m_2,1,1)} \\[0.4cm]
      \lefteqn{\mbox{\tt OneLoopTens2Pt}(i,\bmm{q_1}{2},m_1,m_2,t_1,t_2) = 
      B_{\mu_1 \cdots \mu_i}^{(t_1,t_2)}(\bmm{q_1}{2}, m_{1}, m_{2})} \nonumber
      \\[0.2cm]
      & = & \int \! d^{D}l \,\, \frac{l_{\mu_1} \cdots l_{\mu_i}}
      {[(\bm{l+q_1})^{2} - m_{1}^{2} + i \varrho]^{t_1} \, [\bm{l}^{2} -
      m_{2}^{2} + i \varrho]^{t_2}} \hspace*{2.9cm} \nop \\[0.4cm]
      \label{two-2}
      \lefteqn{\mbox{\tt OneLoopTens2Pt}(i,\bmm{q_1}{2},m_1,m_2) =
      \mbox{\tt OneLoopTens2Pt}(i,\bmm{q_1}{2},m_1,m_2,1,1)\,.}
      \eea \\[-0.2cm]
      Abbreviations:
      \be
      \label{split1} 
      \Ds{q_{10}} = \Ds{\sqrt{\bmm{q_{1}}{2}}}\,; \quad \Ds{l_{0}}
      = \Ds{\frac{\bm{l \cdot q_{1}}}{\sqrt{\bmm{q_{1}}{2}}}}\,; \quad
      l_{\bot} = \sqrt{l_0^2 - \bm{l}^2}\,.
      \ee \\[-0.2cm] \nop
\item Three-point function: \\
      \nop\hfill
      \parbox{10.6cm}{\bc\SII
      \parbox{10cm}{\bp(10,4)
      \pscircle[linewidth=0.1](3.53,2.5){0.1}
      \pscircle[linewidth=0.1](6.5,1){0.1}
      \pscircle[linewidth=0.1](6.5,4){0.1}
      \psline{->}(1,2.5)(2.45,2.5)
      \psline(2.25,2.5)(3.5,2.5)
      \psline{->}(6.5,4)(7.95,4)
      \psline(9,4)(7.55,4)
      \psline(7.75,1)(6.5,1)
      \psline{->}(9,1)(7.55,1)
      \psline{->}(3.5,2.5)(5.2,1.65)
      \psline(5.1,1.7)(6.5,1)
      \psline{->}(6.5,4)(4.8,3.15)
      \psline(5.1,3.3)(3.5,2.5)
      \psline{->}(6.5,1)(6.5,2.7)
      \psline(6.5,2.4)(6.5,4)
      \put(4.1,1.35){$\bm{l+q_{1}}$}
      \put(6.715,2.35){$\bm{l+q_{2}}$}
      \put(4.8,3.35){$\bm{l}$}
      \put(4.85,2){$m_{1}$}
      \put(5.8,2.35){$m_{2}$}
      \put(4.85,2.9){$m_{3}$}
      \put(2.15,2.77){$\bm{q_{1}}$}
      \put(7.65,3.6){$\bm{q_{2}}$}
      \put(7.1,1.2){$\bm{q_{2}-q_{1}}$}
      \ep}
      \ec} \vspace{-0.9cm} \nop
      \hfill\nop
      \bea
      \lefteqn{\mbox{\tt OneLoop3Pt}(p_0,p_1,p_\bot,q_{10},q_{20},q_{21},m_1,
      m_2,m_3,t_1,t_2,t_3)} & & \rule{12.5cm}{0cm} \nonumber \\[0.2cm]
      & = & C^{(p_{0} p_{1} p_\bot)(t_1,t_2,t_3)}(q_{10},q_{20},q_{21},m_{1},
      m_{2},m_{3}) \\[0.2cm]
      & = & \int \! d^{D}l \,\, \frac{(l_{0})^{p_{0}} \, (l_{1})^{p_{1}} \, 
      (l_{\bot})^{p_\bot}}{[(\bm{l+q_1})^{2} - m_{1}^{2} + i \varrho]^{t_1} \,
      [(\bm{l+q_2})^{2} - m_{2}^{2} + i \varrho]^{t_2} \,
      [\bm{l}^{2} - m_{3}^{2} + i \varrho]^{t_3}} \nonumber \\[0.4cm]
      \lefteqn{\mbox{\tt OneLoop3Pt}(p_0,p_1,p_\bot,q_{10},q_{20},q_{21},m_1,
      m_2,m_3)} \nonumber \\[0.4cm]
      \label{three-1}
      & = & \mbox{\tt OneLoop3Pt}(p_0,p_1,p_\bot,q_{10},q_{20},q_{21},m_1,
      m_2,m_3,1,1,1) \\[0.4cm]
      \lefteqn{\mbox{\tt OneLoop2\_3Pt}(p_0,p_\bot,q_{10},q_{20},m_1,
      m_2,m_3,t_1,t_2,t_3)} \nonumber \\[0.2cm]
      & = & \int \! d^{D}l \,\, \frac{(l_{0})^{p_{0}} \,
      (l_{\bot})^{p_\bot}}{[(\bm{l+q_1})^{2} - m_{1}^{2} + i \varrho]^{t_1} \,
      [(\bm{l+q_2})^{2} - m_{2}^{2} + i \varrho]^{t_2} \,
      [\bm{l}^{2} - m_{3}^{2} + i \varrho]^{t_3}} \nonumber \\[0.4cm]
      \lefteqn{\mbox{\tt OneLoop2\_3Pt}(p_0,p_\bot,q_{10},q_{20},m_1,
      m_2,m_3)} \nonumber \\[0.4cm]
      & = & \mbox{\tt OneLoop2\_3Pt}(p_0,p_\bot,q_{10},q_{20},m_1,
      m_2,m_3,1,1,1) \\[0.4cm]
      \lefteqn{\mbox{\tt OneLoop1\_3Pt}(p,m_1,
      m_2,m_3,t_1,t_2,t_3)} \nonumber \\[0.2cm]
      & = & \int \! d^{D}l \,\, \frac{(\bm{l}^2)^{\frac{p}{2}}}
      {[\bm{l}^{2} - m_{1}^{2} + i \varrho]^{t_1} \,
      [\bm{l}^{2} - m_{2}^{2} + i \varrho]^{t_2} \,
      [\bm{l}^{2} - m_{3}^{2} + i \varrho]^{t_3}} \nonumber \\[0.4cm]
      \lefteqn{\mbox{\tt OneLoop1\_3Pt}(p,m_1,m_2,m_3) =
      \mbox{\tt OneLoop1\_3Pt}(p,m_1,m_2,m_3,1,1,1)} \\[0.4cm]
      \lefteqn{\mbox{\tt OneLoopTens3Pt}(i,\bmm{q_1}{2},\bmm{q_2}{2},(\bm{q_2-
      q_1})^2,m_1,m_2,m_3,t_1,t_2,t_3)} \nonumber \\[0.2cm]
      & = & C_{\mu_1 \cdots \mu_i}^{(t_1,t_2,t_3)}(\bmm{q_1}{2},\bmm{q_2}{2},
      (\bm{q_2-q_1})^2,m_{1}, m_{2}, m_{3}) \\[0.2cm]
      & = & \int \! d^{D}l \,\, \frac{l_{\mu_1} \cdots l_{\mu_i}}
      {[(\bm{l+q_1})^{2} - m_{1}^{2} + i \varrho]^{t_1} \, [(\bm{l
      +q_2})^{2} - m_{2}^{2} + i \varrho]^{t_2} \, [\bm{l}^{2} - m_{3}^{2} +
      i \varrho]^{t_3}} \nonumber \\[0.4cm]
      \lefteqn{\mbox{\tt OneLoopTens3Pt}(i,\bmm{q_1}{2},\bmm{q_2}{2},(\bm{q_2-
      q_1})^2,m_1,m_2,m_3)} \nonumber \\[0.4cm]
      \label{three-2}
      & = & \mbox{\tt OneLoopTens3Pt}(i,\bmm{q_1}{2},\bmm{q_2}{2},(\bm{q_2}-
      \bm{q_1})^2,m_1,m_2,m_3,1,1,1)\,.
      \eea \\[-0.2cm]
      Abbreviations -- using the auxiliary property (\ref{aux}):
      \be\ba{rclcrclcrcl}
      \label{split2} 
      \Ds{q_{10}} & = & \Ds{\sqrt{\bmm{q_{1}}{2}}}\,; & & \Ds{q_{20}} & =
      & \Ds{\frac{\bm{q_{1} \cdot q_{2}}}{\sqrt{\bmm{q_{1}}{2}}}}\,; & &
      \Ds{q_{21}} & = & \Ds{\sqrt{\frac{(\bm{q_{1} \cdot q_{2}})^{2}}
      {\bmm{q_{1}}{2}} - \bmm{q_{2}}{2}}}\,; \\[0.6cm]
      \Ds{l_{0}} & = & \Ds{\frac{\bm{l \cdot q_{1}}}{\sqrt{\bmm{q_{1}}{2}}}}\,;
      & & \Ds{l_{1}} & = & \Ds{- \,\frac{\bm{l \cdot \hat{q}_{2}}}{\sqrt{-
      \bmm{\hat{q}_{2}}{2}}}}\,; & & l_{\bot} & = & \sqrt{l_0^2 - l_1^2 -
      \bm{l}^2}\,.
      \ea\ee
\ei}
Our notation for the {\tt OneLoop{\rm\it n}Pt} functions distinguishes between
parallel and orthogonal space which is reflected by the definitions of
(\ref{split1}) and (\ref{split2}) for the momentum components. Please keep in
mind the fact that in our notation the indices $p_0, p_1, \ldots$ represent the
powers of the different components of the loop momentum $\bm{l}$ which should
not be mixed with Lorentz indices $\mu_1, \mu_2, \ldots$ of the corresponding
tensor {\tt OneLoopTens{\rm\it n}Pt}.

If the powers $t_n$ of the propagator terms in the denominator are omitted like
in (\ref{one-1}, \ref{one-2}, \ref{two-1}, \ref{two-2}, \ref{three-1},
\ref{three-2}) the program assumes the value 1 for each $t_n$.

\subsection{Output of {\md\tt OneLoop{\rm\it n}Pt}} \label{rex}

The results of the {\tt OneLoop{\rm\it n}Pt} functions are Laurent expansions in
terms of the ultraviolet regulator $\eps$. Therefore the output of the {\tt
OneLoop{\rm\it n}Pt} procedures consists of a list where the significant
coefficients -- ${\mcl O}(\eps^{-1})$, ${\mcl O}(\eps^0)$ -- of this expansion
are denoted. In two-loop calculations the term of ${\mcl O}(\eps^1)$ of the
one-loop functions is also of interest. For that purpose this coefficient is
also calculated if the qualifier {\tt more} is used in the function call:
\\ \nop
\bc
\btb{|l|l|}
\hline
Eingabe & Ausgabe \\
\hline
{\tt OneLoop1Pt}$(p,m,t)$ & 
$[\eps^{-1}\mbox{-{\em term}},\eps^0\mbox{-{\em term}}]$ \\
\hline
{\tt OneLoop1Pt}$(p,m,t,\mbox{\tt more})$ & 
$[\eps^{-1}\mbox{-{\em term}},\eps^0\mbox{-{\em term}},
\eps^1\mbox{-{\em term}}]$ \\
\hline
{\tt OneLoop1\_2Pt}$(p,m_1,m_2,t_1,t_2)$ & $[\eps^{-1}\mbox{-{\em
term}},\eps^0\mbox{-{\em term}}]$ \\
\hline
{\tt OneLoop1\_2Pt}$(p,m_1,m_2,t_1,t_2,\mbox{\tt more})$ & 
$[\eps^{-1}\mbox{-{\em term}},\eps^0\mbox{-{\em term}},
\eps^1\mbox{-{\em term}}]$ \\
\hline
{\tt OneLoop2Pt}$(p_0,p_\bot,q_{10},m_1,m_2,t_1,t_2)$ & $[\eps^{-1}\mbox{-{\em
term}},\eps^0\mbox{-{\em term}}]$ \\
\hline
{\tt OneLoop2Pt}$(p_0,p_\bot,q_{10},m_1,m_2,t_1,t_2,\mbox{\tt more})$ & 
$[\eps^{-1}\mbox{-{\em term}},\eps^0\mbox{-{\em term}},
\eps^1\mbox{-{\em term}}]$ \\
\hline
{\tt OneLoop1\_3Pt}$(p,m_1,m_2,m_3,t_1,t_2,t_3)$ &
$[\eps^{-1}\mbox{-{\em term}},\eps^0\mbox{-{\em term}}]$ \\
\hline
{\tt OneLoop1\_3Pt}$(p,m_1,m_2,m_3,t_1,t_2,t_3,\mbox{\tt
more})$ & $[\eps^{-1}\mbox{-{\em term}},\eps^0\mbox{-{\em term}},
\eps^1\mbox{-{\em term}}]$ \\
\hline
{\tt OneLoop2\_3Pt}$(p_0,p_\bot,q_{10},q_{20},m_1,m_2,m_3,t_1,t_2,t_3)$ &
$[\eps^{-1}\mbox{-{\em term}},\eps^0\mbox{-{\em term}}]$ \\
\hline
{\tt OneLoop2\_3Pt}$(p_0,p_\bot,q_{10},q_{20},m_1,m_2,m_3,t_1,t_2,t_3,\mbox{\tt
more})$ & $[\eps^{-1}\mbox{-{\em term}},\eps^0\mbox{-{\em term}},
\eps^1\mbox{-{\em term}}]$ \\
\hline
{\tt OneLoop3Pt}$(p_0,p_1,p_\bot,q_{10},q_{20},q_{21},m_1,m_2,m_3,
t_1,t_2,t_3)$ & 
$[\eps^{-1}\mbox{-{\em term}},\eps^0\mbox{-{\em term}}]$ \\
\hline
{\tt OneLoop3Pt}$(p_0,p_1,p_\bot,q_{10},q_{20},q_{21},$ & \\
$\hphantom{\mbox{\tt OneLoop3Pt}(p_0,p_1,p_\bot,}
m_1,m_2,m_3,t_1,t_2,t_3,\mbox{\tt more})$ & 
$[\eps^{-1}\mbox{-{\em term}},\eps^0\mbox{-{\em term}},
\eps^1\mbox{-{\em term}}]$ \\
\hline
\etb
\\[0.6cm]
\ec
In the output $\eps^{-1}$-{\em term} is meant to represent the divergent
part (the coefficient of $\frac{1}{\eps}$) whereas $\eps^{0}$-{\em term} is
describing the finite part and $\eps^{1}$-{\em term} the ${\mcl O}(\eps)$
contribution.

As long as the arguments of the {\tt OneLoop} functions are symbols the output
is given algebraically, whereas of course numbers inserted for the arguments
imply a numerical result. Like in the case of the {\tt EvalGraph} routines, in
numerical evaluations the program sets the value of $\varrho$, the imaginary
part of the propagators, five digits higher than the numerical accuracy of the
whole calculation, for example to $10^{-15}$ if one calculates with 20 digits.
In the algebraic case just {\tt rho} is returned. To get a compact result also
in algebraic calculations, the output is written by using several abbreviations:
\bea
\label{abbrev-1}
\mbox{\tt R2ex1}(x,y) & = & \sqrt{1-\frac{x}{y}} \,\, \left[\ln
\left(1-\sqrt{1-\frac{x}{y}}\,\right) - \ln \left(1+\sqrt{1-\frac{x}{y}}\,
\right) + i \pi \right] \\[0.4cm]
& & - \ln(-x) - i \pi \nonumber \\
\mbox{\tt R2ex2}(x,y) & = & \left(1+\sqrt{1-\frac{x}{y}}\,\right) \, \Li\left(1-
\frac{1-\sqrt{1-\frac{x}{y}}}{1+\sqrt{1-\frac{x}{y}}}\,\right) \nonumber \\
& & +\left(1+\sqrt{1-\frac{x}{y}}\,\right) \, \left[\ln\left(1-\sqrt{1-
\frac{x}{y}}\,\right)\right]^2 \nonumber \\
& & +\left(1-\sqrt{1-\frac{x}{y}}\,\right) \, \Li\left(1-
\frac{1+\sqrt{1-\frac{x}{y}}}{1-\sqrt{1-\frac{x}{y}}}\,\right) \nonumber \\
\label{abbrev-2}
& & +\left(1-\sqrt{1-\frac{x}{y}}\,\right) \, \left[\ln\left(1+\sqrt{1-
\frac{x}{y}}\,\right)\right]^2 \\[0.4cm]
& & +\,\frac{1}{2} \, \left(\ln y\right)^2 +2 \, \left(\ln x\right)^2 -2 \, \ln
x \, \ln y \nonumber \\[0.4cm]
& & +\left(\ln y - 2 \, \ln x \right) \, \left[\left(1+\sqrt{1-\frac{x}{y}}\,
\right) \, \ln\left(1-\sqrt{1-\frac{x}{y}}\,\right)\right. \nonumber \\
& & \left. + \left(1-\sqrt{1-\frac{x}{y}}\,
\right) \, \ln\left(1+\sqrt{1-\frac{x}{y}}\,\right)\right] \nonumber \\
& & -\,i\pi \, \frac{\sqrt{x-y}}{\sqrt{-y}} \, \left[2 \, \ln 2 + \ln(x-y)
\right]\nonumber \\[0.4cm]
\label{abbrev-3}
\mbox{\tt R3ex2}(x,y,z) & = & 2 \,\ln\left(1-\frac{x}{z}\right) \, \eta(x,z) + 2
\,\ln\left(1-\frac{y}{z}\right) \, \eta(y,z) \\
& & + 2 \,\Li\left(1-\frac{x}{z}\right) + 2 \,\Li\left(1-\frac{y}{z}\right) + 2
\,(\ln z)^2 \nonumber \\[0.4cm]
\label{abbrev-4}
\mbox{\tt R3ex3}(x,y,z) & = & 4\,\Spe\left(1-\frac{x}{z}\right) - 4\,\Tri
\left(1-\frac{x}{z}\right) + 4\,\Spe\left(1-\frac{y}{z}\right) - 4\,
\Tri\left(1-\frac{y}{z}\right) \nonumber \\
& & -\,4\,\eta\left(x,\frac{1}{z}\right)\left\{\Li\left(\frac{x}{z}\right) +
\ln\left(1-\frac{x}{z}\right)\ln\left(-\,\frac{x}{z}\right)+\frac{1}{2}
\left[\ln\left(1-\frac{x}{z}\right)\right]^2\right\} \nonumber \\
& & -\,4\,\eta\left(y,\frac{1}{z}\right)\left\{\Li\left(\frac{y}{z}\right) +
\ln\left(1-\frac{y}{z}\right)\ln\left(-\,\frac{y}{z}\right)+\frac{1}{2}
\left[\ln\left(1-\frac{y}{z}\right)\right]^2\right\} \nonumber \\
& & -\,4\,\ln z \left[\Li\left(1
-\frac{x}{z}\right)+\Li\left(1-\frac{y}{z}\right)\right] \nonumber \\
& & -\,4\,\ln z \left[\eta\left(x,\frac{1}{z}\right)\ln\left(1-\frac{x}{z}
\right) + \eta\left(y,\frac{1}{z}\right)\ln\left(1-\frac{y}{z}\right)\right]
\nonumber \\
& & + 2\,\Tri\left(\frac{z-y}{z-x}\right) - 2\,\Tri\left(\frac{(z-y)x}{(z-x)y}
\right) + 2\,\Tri\left(\frac{x}{y}\right) \\
& & + \left[\ln\left(\frac{z-y}{z-x}\right)\right]^2\,\eta\left(\frac{y-x}{z},
\frac{z}{z-x}\right) + 2\,\eta\left(y-x,\frac{1}{z}\right) \nonumber \\
& & \qquad\times\left\{\frac{1}{2}\left[\ln\left(
\frac{z-y}{z-x}\right)\right]^2-\ln\left(-\,\frac{z-y}{z-x}\right)
\eta\left(\frac{z-y}{z},\frac{z}{z-x}\right)\right\} \nonumber \\
& & + 2\,(\ln x - \ln y)\left[\Li\left(\frac{(z-y)x}{(z-x)y}\right)-\Li\left(
\frac{x}{y}\right)\right] \nonumber \\
& & - \ln\left(1-\frac{x}{z}\right)\,\left(\ln x - \ln y\right)^2 - 2\,\zeta(3)
- \frac{4}{3} \left(\ln z\right)^3 \nonumber \\
& & + 2 \left[\ln x - \ln y -\frac{1}{2}\ln\left(\frac{(z-y)x}{(z-x)y}\right)
\right]\ln\left(\frac{(z-y)x}{(z-x)y}\right)\eta\left(\frac{y-x}{y},
\frac{z}{z-x}\right)\,. \nonumber
\eea
These functions are related to the corresponding $\mcl R$ functions\footnote{The
function {\tt R3ex3}$(x,y,z)$ assumes that $x$ and $y$ have an imaginary part of
different sign which is always the case.} \cite{Car,Fr3}. They represent the
coefficients of the Taylor expansion in $\eps$ making use of (cf. \cite{Lew})
\beas
\zeta(n) & = & \sum_{k=1}^\infty\,\frac{1}{k^n} \\
\Li(z) & = & - \itg_0^z \frac{\ln(1-s)}{s}\,\,ds \\
\Tri(z) & = & \itg_0^z \frac{\Li(s)}{s}\,\,ds \\
\Spe(z) & = & \frac{1}{2}\,\itg_0^z\frac{\ln^2(1-s)}{s}\,\,ds \\
\eta(a,b) & = & 2\pi i \left[\theta(-\Im a) \theta(-\Im b) \theta(\Im (a b))
- \theta(\Im a) \theta(\Im b) \theta(-\Im (a b))\right]\,.
\eeas
The function \\[0.4cm]
\verb+evalRex(+$\langle${\it expression}$\rangle$\verb+);+ \\[0.4cm]
substitutes the {\tt R{\rm\it n}ex{\rm\it m}} functions in $\langle${\it
expression}$\rangle$ by the corresponding Polylogarithms.

\subsection{Output of {\md\tt OneLoopTens{\rm\it n}Pt}} \label{tens}

The functions {\tt OneLoopTens{\rm\it n}Pt} return a full tensor as output.

The output format looks similar to that of the {\tt EvalGraph} routines.
The following types of arguments -- here only demonstrated for the two-point
case -- are allowed for all {\tt OneLoopTens{\rm\it n}Pt} functions:
\bi
\item {\tt OneLoopTens2Pt}$(i)$: \\
      The rank $i$ tensor decomposition of the two-point
      function is given. The procedure returns a list consisting in the
      different coefficients, which are expressed in terms of the {\tt
      OneLoop{\rm\it n}Pt} tensor integrals, and the defining equation for the
      coefficients, for instance in the case $i=2$:
      \beas
      & & \left[\mbox{\tt C21} = - \,\frac{\mbox{\tt OneLoop2Pt}(0,2)}{3-2\eps}
      \,, \, \mbox{\tt C20} = - \,\frac{\mbox{\tt OneLoop2Pt}(0,2)}{q^2 \,
      (-3+2\eps)} \right. \\ 
      & & \hspace*{2cm} \left. + \frac{\mbox{\tt OneLoop2Pt}(2,0)}{q^2} \, , \,
      \mbox{\tt C20 q1(nu1) q1(nu2)} + \mbox{\tt C21 G(nu1,nu2)} \right]\,.
      \eeas
\item {\tt OneLoopTens2Pt$(i,\mbox{\tt full})$}: \\
      The function inserts the results of the {\tt OneLoop{\rm\it n}Pt} tensor
      integrals explicitly. Here again the case $i=2$:
      \beas
      & & \left[\mbox{\tt C21} = [\eps^{-1}\mbox{-{\em term}},
      \eps^0\mbox{-{\em term}}] \, , \, \mbox{\tt C20} = [\eps^{-1}\mbox{-{\em
      term}},\eps^0\mbox{-{\em term}}] \,, \right. \\
      & & \hspace*{2cm} \left. \mbox{\tt C20 q1(nu1) q1(nu2)} + \mbox{\tt C21
      G(nu1,nu2)}\vphantom{\eps^0}\right]\,.
      \eeas
\item {\tt OneLoopTens2Pt$(i,\bmm{q_1}{2},m_1,m_2,t_1,t_2)$}: \\
      The function returns the same as {\tt OneLoopTens2Pt$(i,\mbox{\tt
      full})$}, but expressed in the user-defined terms for $\bmm{q_1}{2},m_1$
      and $m_2$. Of course numerical values are also allowed. $t_1$ and $t_2$
      are optional. 
\item {\tt OneLoopTens2Pt$(i,\bmm{q_1}{2},m_1,m_2,t_1,t_2,\mbox{\tt more})$}: \\
      The function returns also the ${\mcl O}(\eps)$ contribution. Again, $t_1$
      and $t_2$ are optional:
      \beas
      & & \left[\mbox{\tt C21} = [\eps^{-1}\mbox{-{\em term}},
      \eps^0\mbox{-{\em term}},\eps^1\mbox{-{\em term}}] \, , \, \mbox{\tt C20}
      = [\eps^{-1}\mbox{-{\em term}},\eps^0\mbox{-{\em term}},\eps^1\mbox{-{\em
      term}}] \, , \right. \\
      & & \hspace*{2cm} \left.\mbox{\tt C20 q1(nu1) q1(nu2)} + \mbox{\tt C21
      G(nu1,nu2)}\vphantom{\eps^0}\right]\,.
      \eeas
\ei
The {\tt C20}, {\tt C21}, \ldots\ correspond to the coefficients of the
Passarino-Veltman procedure.

\subsection{Infrared divergences}

The program is regulating the infrared divergence of the three-point function as
well -- which may occur for instance if the momenta are on-shell and one mass is
set to 0. There is no distinction made between UV and IR divergences -- that
means that there is no infrared dimension parameter $\eps_{\mathrm{IR}}$. Both
kinds of divergences are described by $\eps = (4 - D)/2$.

\subsection{Special relations}

If special relations exist which fix the variables of the {\tt OneLoop}
functions or connect them with each other -- like for instance the on-shell
condition -- it is important first to declare these relations and then to call
the functions -- as already stressed in the case of the {\tt EvalGraph}
routines.

If the functions are evaluated first analytically and then numbers or relations
between variables are substituted in the result, it might happen in some special
cases that an error occurs -- something like ``{\tt division by 0}''. Usually
this sort of divergences is artifical due to the way the analytical result was
written before the insertion (cf. sect. \ref{rel}).

\section{The {\md\tt TwoLoop} procedures} \label{Twoloop}

\subsection{Input}

Unlike for the one-loop functions on the two-loop level for each $n$-point case
exist different types of integrals -- which correspond to different
topologies with the same number of external legs. To distinguish between the
different $n$-point functions with the same $n$ the {\tt TwoLoop} procedures
acquire an additional identification number:
\bi
\item {\tt TwoLoop2Pt1}$(p_0,p_\bot,r_0,r_\bot,s,q_{10},m_1,m_2,m_3,m_4,m_5,
      t_1,t_2,t_3,t_4,t_5)$ \\
      \vdots
\item {\tt TwoLoop2Pt8}$(p_0,p_\bot,r_0,r_\bot,s,q_{10},m_1,m_2,t_1,t_2)$
\item {\tt TwoLoopTens2Pt1}$(i_1,i_2,\bmm{q_1}{2},m_1,m_2,m_3,m_4,m_5,
      t_1,t_2,t_3,t_4,t_5)$ \\
      \vdots
\item {\tt TwoLoopTens2Pt8}$(i_1,i_2,\bmm{q_1}{2},m_1,m_2,t_1,t_2)$.
\ei
Like in the one-loop case each function needs the tensor degree, the external
momenta $\bm{q_n}$ and the masses $m_n$ as input. The powers $t_n$ are again
optional.

The functions {\tt TwoLoopTens{\rm\it n}Pt{\rm\it m}} require the squared
external momenta as input variables and return the complete tensor, whereas the
functions {\tt OneLoop{\rm\it n}Pt{\rm\it m}} correspond to the notation in
parallel and orthogonal space variables.

\newpage

\subsection{Notation}

The functions correspond to the following integrals:
{\psset{linewidth=0.048cm}\setlength{\unitlength}{0.8cm}\psset{unit=0.8cm}
\bi
\item Two-point functions: \\[0.4cm]
      Topology 1: \\[0.5cm]
      \nop\hfill
      \parbox{11.6cm}{\bc\SII
      \parbox{11cm}{\bp(11,5.5)
      \pscircle(5.5,3){2.5}
      \psarc(5.5,6.493){2.47}{225}{315}
      \psline{->}(5.4,5.47)(5.6,5.47)
      \psline{->}(5.4,0.53)(5.6,0.53)
      \psline{->}(5.6,4.03)(5.4,4.03)
      \psline{->}(7.821,3.857)(7.729,4.057)
      \psline{->}(3.271,4.057)(3.169,3.857)
      \pscircle[linewidth=0.1](3.03,3){0.1}
      \pscircle[linewidth=0.1](7.97,3){0.1}
      \pscircle[linewidth=0.1](7.243,4.75){0.1}
      \pscircle[linewidth=0.1](3.757,4.75){0.1}
      \psline{->}(0.5,3)(1.95,3)
      \psline(1.75,3)(3,3)
      \psline{->}(8,3)(9.45,3)
      \psline(9.25,3)(10.5,3)
      \put(1.55,3.27){$\bm{q_{1}}$}
      \put(9.05,3.27){$\bm{q_{1}}$}
      \put(5.5,5.7){$\bm{k}$}
      \put(5.1,0.03){$\bm{l+q_{1}}$}
      \put(8.1,3.9){$\bm{l}$}
      \put(2.8,3.9){$\bm{l}$}
      \put(5.1,3.54){$\bm{l+k}$}
      \put(5.35,0.8){$m_{1}$}
      \put(7.2,3.8){$m_{2}$}
      \put(3.4,3.8){$m_{3}$}
      \put(5.35,4.19){$m_{4}$}
      \put(5.35,5.1){$m_{5}$}
      \ep} 
      \ec} \vspace{-0.3cm} \nop
      \hfill\nop
      \bea
      \lefteqn{\mbox{\tt TwoLoop2Pt1}(p_0,p_\bot,r_0,r_\bot,s,q_{10},
      m_1,m_2,m_3,m_4,m_5,t_1,t_2,t_3,t_4,t_5)} & & \rule{12.5cm}{0cm} \nonumber
      \\[0.2cm]
      & = & \int \! d^{D}l \int \! d^{D}k \,\, \frac{(l_0)^{p_{0}} \,
      (l_{\bot})^{p_\bot} \, (k_{0})^{r_{0}} \, (k_{\bot})^{r_\bot} \, (z)^{s}}
      {[(\bm{l+q_{1}})^{2}-m_{1}^{2}+i\varrho]^{t_1} \, [\bm{l}^{2}-
      m_{2}^{2}+i\varrho]^{t_2} \, [\bm{l}^{2}-m_{3}^{2}+i\varrho]^{t_3}}
      \\[0.2cm]
      & & \hspace*{4cm} \times \, \frac{1}{[(\bm{l+k})^{2}-m_{4}^{2}+i\varrho]
      ^{t_4} \, [\bm{k}^{2}-m_{5}^{2}+i\varrho]^{t_5}} \nonumber \\[0.4cm]
      \lefteqn{\mbox{\tt TwoLoopTens2Pt1}(i_1,i_2,\bmm{q_1}{2},
      m_1,m_2,m_3,m_4,m_5,t_1,t_2,t_3,t_4,t_5)} \nonumber \\[0.2cm]
      & = & \int \! d^{D}l \int \! d^{D}k \,\, \frac{l_{\mu_1} \cdots
      l_{\mu_{i_1}}\, k_{\nu_1} \cdots k_{\nu_{i_2}}}
      {[(\bm{l+q_{1}})^{2}-m_{1}^{2}+i\varrho]^{t_1} \, [\bm{l}^{2}-
      m_{2}^{2}+i\varrho]^{t_2} \, [\bm{l}^{2}-m_{3}^{2}+i\varrho]^{t_3}}
      \\[0.2cm]
      & & \hspace*{4cm} \times \, \frac{1}{[(\bm{l+k})^{2}-m_{4}^{2}+i\varrho]
      ^{t_4} \, [\bm{k}^{2}-m_{5}^{2}+i\varrho]^{t_5}}\,. \nonumber
      \eea \\
      Topology 2: \\
      \nop\hfill
      \parbox{11.6cm}{\bc\SII
      \parbox{11cm}{\bp(11,5.5)
      \pscircle(5.5,3){2.5}
      \psline{->}(5.5,0.5)(5.5,3.2)
      \psline(5.5,5.5)(5.5,3)
      \psline{->}(3.823,4.817)(3.683,4.677)
      \psline{->}(3.683,1.323)(3.823,1.183)
      \psline{->}(7.177,4.817)(7.317,4.677)
      \psline{->}(7.317,1.323)(7.177,1.183)
      \pscircle[linewidth=0.1](5.5,5.47){0.1}
      \pscircle[linewidth=0.1](5.5,0.53){0.1}
      \pscircle[linewidth=0.1](3.03,3){0.1}
      \pscircle[linewidth=0.1](7.97,3){0.1}
      \psline{->}(0.5,3)(1.95,3)
      \psline(1.75,3)(3,3)
      \psline{->}(8,3)(9.45,3)
      \psline(9.25,3)(10.5,3)
      \put(1.55,3.27){$\bm{q_{1}}$}
      \put(9.05,3.27){$\bm{q_{1}}$}
      \put(7.4,0.95){$\bm{k-q_{1}}$}
      \put(7.4,4.75){$\bm{k}$}
      \put(3.5,4.75){$\bm{l}$}
      \put(2.7,0.95){$\bm{l+q_{1}}$}
      \put(4.4,2.85){$\bm{l+k}$}
      \put(3.8,1.45){$m_{1}$}
      \put(3.8,4.4){$m_{2}$}
      \put(5.715,2.85){$m_{3}$}
      \put(6.7,1.45){$m_{4}$}
      \put(6.7,4.4){$m_{5}$}
      \ep} 
      \ec} \vspace{-0.5cm} \nop
      \hfill\nop
      \bea
      \lefteqn{\mbox{\tt TwoLoop2Pt2}(p_{0},p_\bot,r_{0},r_\bot,s,q_{10},m_{1},
      m_{2},m_{3},m_{4},m_{5},t_1,t_2,t_3,t_4,t_5)} & & \rule{12.5cm}{0cm}
      \\[0.2cm]
      & = & \int \! d^{D}l \int \! d^{D}k \,\, \frac{(l_{0})^{p_{0}} \,
      (l_{\bot})^{p_{\bot}} \, (k_{0})^{r_{0}} \, (k_{\bot})^{r_{\bot}} \,
      (z)^{s}}{[(\bm{l+q_1})^{2}-m_{1}^{2}+i\varrho]^{t_1} \, [\bm{l}^{2}-m_{2}
      ^{2}+i\varrho]^{t_2} \, [(\bm{l+k})^{2}-m_{3}^{2}+i\varrho]^{t_3}}
      \nonumber \\[0.2cm]
      & & \hspace*{4cm} \times \, \frac{1}{[(\bm{k-q_{1}})^{2}-m_{4}^{2}
      +i\varrho]^{t_4} \, [\bm{k}^{2}-m_{5}^{2}+i\varrho]^{t_5}} \nonumber
      \\[0.4cm]
      \lefteqn{\mbox{\tt TwoLoopTens2Pt2}(i_1,i_2,\bmm{q_1}{2},m_{1},
      m_{2},m_{3},m_{4},m_{5},t_1,t_2,t_3,t_4,t_5)} \\[0.2cm]
      & = & \int \! d^{D}l \int \! d^{D}k \,\, \frac{l_{\mu_1} \cdots
      l_{\mu_{i_1}}\, k_{\nu_1} \cdots k_{\nu_{i_2}}}
      {[(\bm{l+q_1})^{2}-m_{1}^{2}+i\varrho]^{t_1} \, [\bm{l}^{2}-m_{2}^{2}+
      i\varrho]^{t_2} \, [(\bm{l+k})^{2}-m_{3}^{2}+i\varrho]^{t_3}} \nonumber
      \\[0.2cm]
      & & \hspace*{4cm} \times \, \frac{1}{[(\bm{k-q_{1}})^{2}-m_{4}^{2}
      +i\varrho]^{t_4} \, [\bm{k}^{2}-m_{5}^{2}+i\varrho]^{t_5}}\,. \nonumber
      \eea \\
      Topology 3: \\
      \nop\hfill
      \parbox{11.6cm}{\bc\SII
      \parbox{11cm}{\bp(11,5.5)
      \pscircle(5.5,3){2.5}
      \psarc(3,5.5){2.5}{270}{360}
      \psline{->}(5.4,0.53)(5.6,0.53)
      \psline{->}(3.823,4.817)(3.683,4.677)
      \psline{->}(7.317,4.677)(7.177,4.817)
      \psline{->}(4.697,3.663)(4.837,3.803)
      \pscircle[linewidth=0.1](3.03,3){0.1}
      \pscircle[linewidth=0.1](7.97,3){0.1}
      \pscircle[linewidth=0.1](5.5,5.47){0.1}
      \psline{->}(0.5,3)(1.95,3)
      \psline(1.75,3)(3,3)
      \psline{->}(8,3)(9.45,3)
      \psline(9.25,3)(10.5,3)
      \put(1.55,3.27){$\bm{q_{1}}$}
      \put(9.05,3.27){$\bm{q_{1}}$}
      \put(7.4,4.75){$\bm{l}$}
      \put(5.1,0.03){$\bm{l+q_{1}}$}
      \put(4.8,3.4){$\bm{k}$}
      \put(2.7,4.75){$\bm{l+k}$}
      \put(5.35,0.8){$m_{1}$}
      \put(6.7,4.4){$m_{2}$}
      \put(3.75,4.4){$m_{3}$}
      \put(4.1,3.823){$m_{4}$}
      \ep} 
      \ec} \vspace{-0.2cm} \nop
      \hfill\nop
      \bea
      \lefteqn{\mbox{\tt TwoLoop2Pt3}(p_{0},p_{\bot},r_{0},r_{\bot},s,q_{10},
      m_{1},m_{2},m_{3},m_{4},t_1,t_2,t_3,t_4)} & & \rule{12.5cm}{0cm} \nonumber
      \\[0.2cm]
      & = & \int \! d^{D}l \int \! d^{D}k \,\, \frac{(l_{0})^{p_{0}} \,
      (l_{\bot})^{p_{\bot}} \, (k_{0})^{r_{0}} \, (k_{\bot})^{r_{\bot}} \, (z)
      ^{s}}{[(\bm{l+q_{1}})^{2}-m_{1}^{2}+i\varrho]^{t_1} \, [\bm{l}^{2}
      -m_{2}^{2}+i\varrho]^{t_2}} \\[0.2cm]
      & & \hspace*{4cm} \times \, \frac{1}{[(\bm{l+k})^{2}-m_{3}^{2}+i\varrho]
      ^{t_3} \, [\bm{k}^{2}-m_{4}^{2}+i\varrho]^{t_4}} \nonumber \\[0.4cm]
      \lefteqn{\mbox{\tt TwoLoopTens2Pt3}(i_1,i_2,\bmm{q_1}{2},
      m_{1},m_{2},m_{3},m_{4},t_1,t_2,t_3,t_4)} \nonumber \\[0.2cm]
      & = & \int \! d^{D}l \int \! d^{D}k \,\, \frac{l_{\mu_1} \cdots
      l_{\mu_{i_1}}\,k_{\nu_1} \cdots k_{\nu_{i_2}}}
      {[(\bm{l+q_{1}})^{2}-m_{1}^{2}+i\varrho]^{t_1} \, [\bm{l}^{2}-m_{2}^{2}+
      i\varrho]^{t_2}} \\[0.2cm]
      & & \hspace*{4cm} \times \, \frac{1}{[(\bm{l+k})^{2}-m_{3}^{2}+i\varrho]
      ^{t_3} \, [\bm{k}^{2}-m_{4}^{2}+i\varrho]^{t_4}}\,. \nonumber
      \eea \\
      Topology 4: \\[0.5cm]
      \nop\hfill
      \parbox{11.6cm}{\bc\SII
      \parbox{11cm}{\bp(11,5.5)
      \pscircle(5.5,3){2.5}
      \psline{->}(5.6,5.47)(5.4,5.47)
      \psline{->}(5.4,0.53)(5.6,0.53)
      \pscircle[linewidth=0.1](3.03,3){0.1}
      \pscircle[linewidth=0.1](7.97,3){0.1}
      \psline{->}(1.75,3)(5.6,3)
      \psline{->}(5.3,3)(9.45,3)
      \psline{->}(0.5,3)(1.95,3)
      \psline(9.25,3)(10.5,3)
      \put(1.55,3.27){$\bm{q_{1}}$}
      \put(9.05,3.27){$\bm{q_{1}}$}
      \put(5.1,0.03){$\bm{l+q_{1}}$}
      \put(5.1,5.7){$\bm{l+k}$}
      \put(5.4,2.5){$\bm{k}$}
      \put(5.35,0.8){$m_{1}$}
      \put(5.35,5.1){$m_{2}$}
      \put(5.35,3.2){$m_{3}$}
      \ep} 
      \ec} \vspace{-0.2cm} \nop
      \hfill\nop

\newpage

      \bea
      \lefteqn{\mbox{\tt TwoLoop2Pt4}(p_{0},p_{\bot},r_{0},r_{\bot},s,q_{10},
      m_{1},m_{2},m_{3},t_1,t_2,t_3)} & & \rule{12.5cm}{0cm} \\[0.2cm]
      & = & \int \! d^{D}l \int \! d^{D}k \,\, \frac{(l_{0})^{p_{0}} \,
      (l_{\bot})^{p_{\bot}} \, (k_{0})^{r_{0}} \, (k_{\bot})^{r_{\bot}} \,
      (z)^{s}}{[(\bm{l+q_{1}})^{2}-m_{1}^{2}+i\varrho]^{t_1} \,
      [(\bm{l+k})^{2}-m_{2}^{2}+i\varrho]^{t_2} \, [\bm{k}^{2}-m_{3}^{2}
      +i\varrho]^{t_3}}\nonumber \\[0.4cm]
      \lefteqn{\mbox{\tt TwoLoopTens2Pt4}(i_1,i_2,\bmm{q_1}{2},
      m_{1},m_{2},m_{3},t_1,t_2,t_3)} \\[0.2cm]
      & = & \int \! d^{D}l \int \! d^{D}k \,\, \frac{l_{\mu_1} \cdots
      l_{\mu_{i_1}}\, k_{\nu_1} \cdots k_{\nu_{i_2}}}
      {[(\bm{l+q_{1}})^{2}-m_{1}^{2}+i\varrho]^{t_1} \, [(\bm{l+k})^{2}
      -m_{2}^{2}+i\varrho]^{t_2} \, [\bm{k}^{2}-m_{3}^{2}+i\varrho]^{t_3}}\,.
      \nonumber
      \eea \\
      Topology 5: \\[0.3cm]
      \nop\hfill
      \parbox{11.6cm}{\bc\SII
      \parbox{11cm}{\bp(11,6)
      \pscircle(5.5,4.47){1.5}
      \pscircle(5.5,1.53){1.5}
      \psline{->}(5.4,5.94)(5.6,5.94)
      \psline{->}(5.4,0.06)(5.6,0.06)
      \psline{->}(0.5,1.5)(2.7,1.5)
      \psline{->}(6.639,2.469)(6.439,2.669)
      \psline{->}(4.561,2.669)(4.361,2.469)
      \psline(2.5,1.5)(4.03,1.5)
      \psline{->}(6.97,1.5)(8.7,1.5)
      \psline(8.5,1.5)(10.5,1.5)
      \pscircle[linewidth=0.1](5.5,3){0.1}
      \pscircle[linewidth=0.1](4.03,1.5){0.1}
      \pscircle[linewidth=0.1](6.97,1.5){0.1}
      \put(2.3,1.8){$\bm{q_{1}}$}
      \put(8.3,1.8){$\bm{q_{1}}$}
      \put(5.5,6.17){$\bm{k}$}
      \put(5.1,-0.47){$\bm{l+q_{1}}$}
      \put(4.2,2.6){$\bm{l}$}
      \put(6.8,2.6){$\bm{l}$}
      \put(5.35,0.3){$m_{1}$}
      \put(6.1,2.2){$m_{2}$}
      \put(4.5,2.2){$m_{3}$}
      \put(5.4,5.57){$m_{4}$}
      \ep} 
      \ec}
      \hfill\nop
      \bea
      \lefteqn{\mbox{\tt TwoLoop2Pt5}(p_{0},p_{\bot},r_{0},r_{\bot},s,q_{10},
      m_{1},m_{2},m_{3},m_{4},t_1,t_2,t_3,t_4)} & & \rule{12.5cm}{0cm} \nonumber
      \\[0.2cm]
      & = & \int \! d^{D}l \int \! d^{D}k \,\, \frac{(l_{0})^{p_{0}} \,
      (l_{\bot})^{p_{\bot}} \, (k_{0})^{r_{0}} \, (k_{\bot})^{r_{\bot}} \,
      (z)^{s}}{[(\bm{l+q_{1}})^{2}-m_{1}^{2}+i\varrho]^{t_1} \, [\bm{l}^{2}-
      m_{2}^{2}+i\varrho]^{t_2}} \\[0.2cm]
      & & \hspace*{4cm} \times \, \frac{1}{[\bm{l}^{2}-m_{3}^{2}+i\varrho]^{t_3}
      \, [\bm{k}^{2}-m_{4}^{2}+i\varrho]^{t_4}} \nonumber \\[0.4cm]
      \lefteqn{\mbox{\tt TwoLoopTens2Pt5}(i_1,i_2,\bmm{q_1}{2},
      m_{1},m_{2},m_{3},m_{4},t_1,t_2,t_3,t_4)} \nonumber \\[0.2cm]
      & = & \int \! d^{D}l \int \! d^{D}k \,\, \frac{l_{\mu_1} \cdots
      l_{\mu_{i_1}}\,k_{\nu_1} \cdots k_{\nu_{i_2}}}
      {[(\bm{l+q_{1}})^{2}-m_{1}^{2}+i\varrho]^{t_1} \, [\bm{l}^{2}-
      m_{2}^{2}+i\varrho]^{t_2}} \\[0.2cm]
      & & \hspace*{4cm} \times \, \frac{1}{[\bm{l}^{2}-m_{3}^{2}+i\varrho]^{t_3}
      \, [\bm{k}^{2}-m_{4}^{2}+i\varrho]^{t_4}}\,. \nonumber 
      \eea \\
      Topology 6: \\[0.2cm]
      \nop\hfill
      \parbox{11.6cm}{\bc\SII
      \parbox{11cm}{\bp(11,4.5)
      \pscircle(4.03,3){1.5}
      \pscircle(6.97,3){1.5}
      \pscircle[linewidth=0.1](2.56,3){0.1}
      \pscircle[linewidth=0.1](8.44,3){0.1}
      \pscircle[linewidth=0.1](5.5,3){0.1}
      \psline{->}(4.13,4.47)(3.93,4.47)
      \psline{->}(3.93,1.53)(4.13,1.53)
      \psline{->}(6.87,4.47)(7.07,4.47)
      \psline{->}(7.07,1.53)(6.87,1.53)
      \psline{->}(0.5,3)(1.6,3)
      \psline(1.4,3)(2.47,3)
      \psline{->}(8.53,3)(9.8,3)
      \psline(9.6,3)(10.5,3)
      \put(1.25,3.27){$\bm{q_{1}}$}
      \put(9.35,3.27){$\bm{q_{1}}$}
      \put(6.87,4.7){$\bm{k}$}
      \put(6.47,1.03){$\bm{k-q_{1}}$}
      \put(3.93,4.7){$\bm{l}$}
      \put(3.53,1.03){$\bm{l+q_{1}}$}
      \put(3.83,1.8){$m_{1}$}
      \put(3.83,4.1){$m_{2}$}
      \put(6.77,1.8){$m_{3}$}
      \put(6.77,4.1){$m_{4}$}
      \ep} 
      \ec} \vspace{-0.8cm} \nop
      \hfill\nop
      \bea
      \lefteqn{\mbox{\tt TwoLoop2Pt6}(p_{0},p_{\bot},r_{0},r_{\bot},s,q_{10},
      m_{1},m_{2},m_{3},m_{4},t_1,t_2,t_3,t_4)} & & \rule{12.5cm}{0cm} \nonumber
      \\[0.2cm]
      & = & \int \! d^{D}l \int \! d^{D}k \,\, \frac{(l_{0})^{p_{0}} \,
      (l_{\bot})^{p_{\bot}} \, (k_{0})^{r_{0}} \, (k_{\bot})^{r_{\bot}} \,
      (z)^{s}}{[(\bm{l+q_{1}})^{2}-m_{1}^{2}+i\varrho]^{t_1} \, [\bm{l}^{2}-
      m_{2}^{2}+i\varrho]^{t_2}} \\[0.2cm]
      & & \hspace*{4cm} \times \, \frac{1}{[(\bm{k-q_{1}})^{2}-m_{3}^{2}
      +i\varrho]^{t_3} \, [\bm{k}^{2}-m_{4}^{2}+i\varrho]^{t_4}} \nonumber
      \\[0.4cm]
      \lefteqn{\mbox{\tt TwoLoopTens2Pt6}(i_1,i_2,\bmm{q_1}{2},
      m_{1},m_{2},m_{3},m_{4},t_1,t_2,t_3,t_4)} \nonumber \\[0.2cm]
      & = & \int \! d^{D}l \int \! d^{D}k \,\, \frac{l_{\mu_1} \cdots
      l_{\mu_{i_1}}\,k_{\nu_1} \cdots k_{\nu_{i_2}}}
      {[(\bm{l+q_{1}})^{2}-m_{1}^{2}+i\varrho]^{t_1} \, [\bm{l}^{2}-
      m_{2}^{2}+i\varrho]^{t_2}} \\[0.2cm]
      & & \hspace*{4cm} \times \, \frac{1}{[(\bm{k-q_{1}})^{2}-m_{3}^{2}
      +i\varrho]^{t_3} \, [\bm{k}^{2}-m_{4}^{2}+i\varrho]^{t_4}}\,. \nonumber
      \eea \\
      Topology 7: \\[0.3cm]
      \nop\hfill
      \parbox{11.6cm}{\bc\SII
      \parbox{11cm}{\bp(11,6)
      \pscircle(5.5,4.47){1.5}
      \pscircle(5.5,1.53){1.5}
      \psline{->}(5.4,5.94)(5.6,5.94)
      \psline{->}(5.4,0.06)(5.6,0.06)
      \psline{->}(0.5,1.5)(2.7,1.5)
      \psline{->}(4.561,2.669)(4.361,2.469)
      \psline(2.5,1.5)(4.03,1.5)
      \psline{->}(5.5,3)(8.7,3)
      \psline(8.5,3)(10.5,3)
      \pscircle[linewidth=0.1](5.5,3){0.1}
      \pscircle[linewidth=0.1](4.03,1.5){0.1}
      \put(2.3,1.8){$\bm{q_{1}}$}
      \put(8.3,3.3){$\bm{q_{1}}$}
      \put(5.5,6.17){$\bm{k}$}
      \put(5.1,-0.47){$\bm{l+q_{1}}$}
      \put(4.2,2.6){$\bm{l}$}
      \put(5.35,0.3){$m_{1}$}
      \put(4.5,2.2){$m_{2}$}
      \put(5.4,5.57){$m_{3}$}
      \ep} 
      \ec} \vspace{0.2cm} \nop
      \hfill\nop
      \bea
      \lefteqn{\mbox{\tt TwoLoop2Pt7}(p_{0},p_{\bot},r_{0},r_{\bot},s,q_{10},
      m_{1},m_{2},m_{3},t_1,t_2,t_3)} & & \rule{12.5cm}{0cm} \\[0.2cm]
      & = & \int \! d^{D}l \int \! d^{D}k \,\, \frac{(l_{0})^{p_{0}} \,
      (l_{\bot})^{p_{\bot}} \, (k_{0})^{r_{0}} \, (k_{\bot})^{r_{\bot}} \,
      (z)^{s}}{[(\bm{l+q_{1}})^{2}-m_{1}^{2}+i\varrho]^{t_1} \, [\bm{l}^{2}
      -m_{2}^{2}+i\varrho]^{t_2} \, [\bm{k}^{2}-m_{3}^{2}+i\varrho]^{t_3}}
      \nonumber \\[0.4cm]
      \lefteqn{\mbox{\tt TwoLoopTens2Pt7}(i_1,i_2,\bmm{q_1}{2},
      m_{1},m_{2},m_{3},t_1,t_2,t_3)} \\[0.2cm]
      & = & \int \! d^{D}l \int \! d^{D}k \,\, \frac{l_{\mu_1} \cdots
      l_{\mu_{i_1}}\,k_{\nu_1} \cdots k_{\nu_{i_2}}}
      {[(\bm{l+q_{1}})^{2}-m_{1}^{2}+i\varrho]^{t_1} \, [\bm{l}^{2}-m_{2}^{2}+
      i\varrho]^{t_2} \, [\bm{k}^{2}-m_{3}^{2}+i\varrho]^{t_3}}\,. \nonumber
      \eea

\newpage

      Topology 8: \\[0.4cm]
      \nop\hfill
      \parbox{11.6cm}{\bc\SII
      \parbox{11cm}{\bp(11,6)
      \pscircle(5.5,4.47){1.5}
      \pscircle(5.5,1.53){1.5}
      \psline{->}(5.4,5.94)(5.6,5.94)
      \psline{->}(5.4,0.06)(5.6,0.06)
      \psline{->}(0.5,3)(3.2,3)
      \psline{->}(3,3)(8.2,3)
      \psline(8,3)(10.5,3)
      \pscircle[linewidth=0.1](5.5,3){0.1}
      \put(2.8,3.3){$\bm{q_{1}}$}
      \put(7.8,3.3){$\bm{q_{1}}$}
      \put(5.5,6.17){$\bm{k}$}
      \put(5.1,-0.47){$\bm{l+q_{1}}$}
      \put(5.35,0.3){$m_{1}$}
      \put(5.4,5.57){$m_{2}$}
      \ep} 
      \ec} \vspace{0.2cm} \nop
      \hfill\nop
      \bea
      \lefteqn{\mbox{\tt TwoLoop2Pt8}(p_{0},p_{\bot},r_{0},r_{\bot},s,q_{10},
      m_{1},m_{2},t_1,t_2)} & & \rule{12.5cm}{0cm} \\[0.2cm]
      & = & \int \! d^{D}l \int \! d^{D}k \,\, \frac{(l_{0})^{p_{0}} \,
      (l_{\bot})^{p_{\bot}} \, (k_{0})^{r_{0}} \, (k_{\bot})^{r_{\bot}} \,
      (z)^{s}}{[(\bm{l+q_{1}})^{2}-m_{1}^{2}+i\varrho]^{t_1} \, [\bm{k}^{2}
      -m_{2}^{2}+i\varrho]^{t_2}} \nonumber \\[0.4cm]
      \lefteqn{\mbox{\tt TwoLoopTens2Pt8}(i_1,i_2,\bmm{q_1}{2},
      m_{1},m_{2},t_1,t_2)} \\[0.2cm]
      & = & \int \! d^{D}l \int \! d^{D}k \,\, \frac{l_{\mu_1} \cdots
      l_{\mu_{i_1}}\,k_{\nu_1} \cdots k_{\nu_{i_2}}}
      {[(\bm{l+q_{1}})^{2}-m_{1}^{2}+i\varrho]^{t_1} \, [\bm{k}^{2}-m_{2}^{2}
      +i\varrho]^{t_2}}\,. \nonumber
      \eea
\ei}
Like in the one-loop case for each function holds
\bea
\lefteqn{\mbox{\tt TwoLoop2Pt}n(p_0,p_\bot,r_0,r_\bot,s,q_{10},
m_1,\ldots,m_k)} \nonumber \\
& = & \mbox{\tt TwoLoop2Pt}n(p_0,p_\bot,r_0,r_\bot,s,q_{10},m_1,
\ldots,m_k,1,\ldots,1) \\[0.4cm]   
\lefteqn{\mbox{\tt TwoLoopTens2Pt}n(i_1,i_2,\bmm{q_1}{2},
m_1,\ldots,m_k)} \nonumber \\
& = & \mbox{\tt TwoLoopTens2Pt}n(i_1,i_2,\bmm{q_1}{2},m_1,
\ldots,m_k,1,\ldots,1)\,.
\eea
This means that if the powers $t_n$ of the propagator terms in the denominator
are omitted the program assumes the value 1 for each $t_n$.

The abbreviations (\ref{split1}) we made in the one-loop case for parallel and
orthogonal space notation are still valid.

There are several relations between different two-point functions. Some of them
we would like to mention here:
\bea
\lefteqn{\mbox{\tt TwoLoop2Pt1}(p_0,p_\bot,r_0,r_\bot,s,q_{10},m_1,m_2,m_3,m_4,
m_5)} \nonumber \\
& = & \frac{1}{m_{3}^{2}-m_{2}^{2}} \, \left[\mbox{\tt TwoLoop2Pt3}(p_{0},
p_{\bot},r_{0},r_{\bot},s,q_{10},m_{1},m_{3},m_{4},m_{5})
\right. \\
& & \hspace*{2cm} \left. - \mbox{\tt TwoLoop2Pt3}(p_{0},p_{\bot},r_{0},
r_{\bot},s,q_{10},m_{1},m_{2},m_{4},m_{5})\right] \nonumber
\\[0.4cm]
\lefteqn{\mbox{\tt TwoLoop2Pt5}(p_0,p_\bot,r_0,r_\bot,s,q_{10},m_1,m_2,m_3,
m_4)} \nonumber \\
& = & \frac{1}{m_{3}^{2}-m_{2}^{2}} \, \left[\mbox{\tt TwoLoop2Pt7}(p_{0},
p_{\bot},r_{0},r_{\bot},s,q_{10},m_{1},m_{3},m_{4}) \right.
\\
& & \hspace*{2cm} \left. - \mbox{\tt TwoLoop2Pt7}(p_{0},p_{\bot},r_{0},
r_{\bot},s,q_{10},m_{1},m_{2},m_{4}) \right]\,. \nonumber
\eea
Both relations are only valid for $m_2\not=m_3$, otherwise one has
\bea
\lefteqn{\mbox{\tt TwoLoop2Pt1}(p_0,p_\bot,r_0,r_\bot,s,q_{10},m_1,m_2,m_2,m_4,
m_5,t_1,t_2,t_3,t_4,t_5)} \nonumber \\
& = & \mbox{\tt TwoLoop2Pt3}(p_{0},p_{\bot},r_{0},r_{\bot},s,q_{10},
m_{1},m_{2},m_{4},m_{5},t_1,t_2+t_3,t_4,t_5) \\
\lefteqn{\mbox{\tt TwoLoop2Pt5}(p_0,p_\bot,r_0,r_\bot,s,q_{10},m_1,m_2,m_2,
m_4,t_1,t_2,t_3,t_4)} \nonumber \\
& = & \mbox{\tt TwoLoop2Pt7}(p_{0},p_{\bot},r_{0},r_{\bot},s,q_{10},m_{1},
m_{2},m_{4},t_1,t_2+t_3,t_4)\,.
\eea
The two-point functions numbered with 1-4 are nonfactorizable whereas the
functions 5-8 are products of one-loop integrals, for instance
\bea
\lefteqn{\mbox{\tt TwoLoop2Pt6}(p_{0},p_{\bot},r_{0},r_{\bot},s,q_{10},
      m_{1},m_{2},m_{3},m_{4},t_1,t_2,t_3,t_4)} \\
& = & \,\left\{\ba{ll}
\Ds{\frac{(\frac{1}{2},\frac{s}{2})}{(\frac{D-1}{2},\frac{s}{2})} \,\,
\mbox{\tt OneLoop2Pt}(p_{0},p_{\bot},q_{10},m_{1},m_{2},t_1,t_2)} \\
\hphantom{\Ds{\frac{(\frac{1}{2},\frac{s}{2})}{(\frac{D-1}{2},
\frac{s}{2})}}} \times \, \Ds{\mbox{\tt OneLoop2Pt}(r_{0},r_{\bot},-q_{10}
,m_{3},m_{4},t_3,t_4)} & \qquad \mbox{if $s$ even} \\[0.4cm]
\hspace*{2.5cm} 0 & \qquad \mbox{if $s$ odd}
\ea\right. \nonumber \\[1cm]
\lefteqn{\mbox{\tt TwoLoop2Pt7}(p_{0},p_{\bot},r_{0},r_{\bot},s,q_{10},
m_{1},m_{2},m_{3},t_1,t_2,t_3)} \\
& = & \,\left\{\ba{ll}
\Ds{\frac{(\frac{1}{2},\frac{s}{2})}{(\frac{D-1}{2},\frac{s}{2})} \,
(-1)^{\frac{r_{\bot}}{2}} \, \frac{(D-1) \cdots (D-3+r_{\bot})}{D \cdots
(D-2+r_{0}+r_{\bot})} \, \frac{r_{0}!}{(\frac{r_{0}}{2})!2^{\frac{r_{0}}
{2}}}} \\[0.4cm]
\hphantom{\Ds{\frac{(\frac{1}{2},\frac{s}{2})}{(\frac{D-1}{2},\frac{s}{2})}}}
\times \, \Ds{\mbox{\tt OneLoop2Pt}(p_{0},p_{\bot},q_{10},
m_{1},m_{2},t_1,t_2)} \\[0.4cm]
\hphantom{\Ds{\frac{(\frac{1}{2},\frac{s}{2})}{(\frac{D-1}{2},\frac{s}{2})}}}
\times \, \Ds{\mbox{\tt OneLoop1Pt}(r_{0}+r_{\bot},m_{3},t_3)} &
\qquad \mbox{if $s, r_{0}, r_{\bot}$ even} \\[0.4cm]
\hspace*{1.5cm} 0 & \qquad \mbox{else}\,.
\ea\right. \nonumber
\eea

\subsection{Output of {\md\tt TwoLoop{\rm\it n}Pt{\rm\it m}}} \label{twoout}

The output -- like in the one-loop case -- is a list which contains the relevant
coefficients of the Laurent expansion in $\eps$. Ultraviolet divergences at the
two-loop level occur as terms of ${\mcl O}(\eps^{-2})$ and ${\mcl
O}(\eps^{-1})$. Therefore the notation of the output starts with two divergent
coefficients.

For the two-point functions 1-4 the finite part cannot be given completely
analytically for all mass cases. Therefore these functions have two entries in
the output list which describe the finite part. The first one is a list itself
denoting an integral representation, the second one gives the analytically
calculable terms.

The integral representation is evaluated directly with the help of \vegas if two
arguments are added. Both arguments must be lists which contain each two
numbers. These numbers carry the input information for \vegas. Within an
ordinary \maple session one has in addition to assign the variable
\verb+NumPath+ as described in sect.~\ref{num}.
\\ \nop
\bc
\btb{|l|l|}
\hline
input & output \\
\hline
{\tt TwoLoop2Pt{\rm\it m}}$(p_0,p_\bot,r_0,r_\bot,s,q_{10},$ &
$[\eps^{-2}\mbox{-{\em term}},\eps^{-1}\mbox{-{\em term}},$ \\
$\hphantom{\mbox{\tt TwoLoop2Ptm}}
m_1,\ldots,m_k,t_1,\ldots,t_k)$ & 
\hspace*{1cm}$\mbox{{\em numeric}},\eps^0\mbox{-{\em term}}]$ \\
\hline
{\tt TwoLoop2Pt{\rm\it m}}$(p_0,p_\bot,r_0,r_\bot,s,q_{10},m_1,\ldots,m_k,$ &
$[\eps^{-2}\mbox{-{\em term}},\eps^{-1}\mbox{-{\em term}},$ \\
$\hphantom{\mbox{\tt TwoLoop2Ptm}}
t_1,\ldots,t_k,[l_1,l_2],[n_1,n_2])$ & 
\hspace*{1cm}$[\mbox{{\em factor}},[\mbox{\em result},\mbox{\em error}]],
\eps^0\mbox{-{\em term}}]$ \\
\hline
\etb \\[0.6cm]
\ec
{\em numeric} describes the integral representation. This list consists of
the following entries
\bdm
[\mbox{\em factor},\mbox{\em integrand},\mbox{\tt x=-infinity..infinity,
y=-infinity..infinity}]
\edm
where {\em integrand} is the integrand of the integral representation which has
to be integrated over $x$ and $y$ in the interval $(-\infty,\infty)$. {\em
factor} is an analytical factor which multiplies the integral representation.

The variables $l_1,l_2,n_1,n_2$ represent the grids where the integrand
is evaluated: first $l_1$ iterations with $l_2$ points -- \verb+itmx+$=l_1$ and
\verb+ncall+$=l_2$ are passed to \vegas -- and then $n_1$ iterations with $n_2$
points -- this corresponds to \verb+itmx+$=n_1$ and \verb+ncall+$=n_2$. Real and
imaginary part are integrated one after the other.

The other entries of the output list have the same behaviour as in the one-loop
case: As long as the arguments of the {\tt TwoLoop} functions are symbols the
output is given algebraically, whereas numbers inserted for the arguments imply
a numerical result. In the numerical case, the program sets the value of
$\varrho$, the imaginary part of the propagators, five digits higher than the
numerical accuracy of the whole calculation, for example to $10^{-15}$ if one
calculates with 20 digits. In the algebraic case just ``{\tt rho}'' is returned.

In the case where all internal particles are massless the result is of course
known analytically. In this case {\em numeric} is just \verb+[0]+.

The two-point functions 5-8 are products of one-loop integrals and therefore
evaluated completely analytically. The output list has no entry for numerical
evaluation and looks simply like:
\\ \nop
\bc
\btb{|l|l|}
\hline
input & output \\
\hline
{\tt TwoLoop2Pt{\rm\it m}}$(p_0,p_\bot,r_0,r_\bot,s,q_{10},$ & \\
$\hphantom{\mbox{\tt TwoLoop2Ptm}}
m_1,\ldots,m_k,t_1,\ldots,t_k)$ &
$[\eps^{-2}\mbox{-{\em term}},\eps^{-1}\mbox{-{\em term}},
\eps^0\mbox{-{\em term}}]$ \\
\hline
{\tt TwoLoop2Pt{\rm\it m}}$(p_0,p_\bot,r_0,r_\bot,s,q_{10},m_1,\ldots,m_k,$ & \\
$\hphantom{\mbox{\tt TwoLoop2Ptm}}
t_1,\ldots,t_k,[l_1,l_2],[n_1,n_2])$ &
$[\eps^{-2}\mbox{-{\em term}},\eps^{-1}\mbox{-{\em term}},
\eps^0\mbox{-{\em term}}]$ \\
\hline
\etb \\[0.6cm]
\ec
The optional input of two lists for the \vegas input data doesn't affect the
output since no numerical integration is performed here. The ${\mcl O}(\eps^1)$
is not relevant in two-loop calculations. Therefore the qualifier {\tt more}
does not exist for two-loop functions.

The divergent part of two-loop integrals is closely related to one-loop
integrals. Therefore the same abbreviations (\ref{abbrev-1}, \ref{abbrev-2}) as
for one-loop  integrals are appearing. They are substituted if the {\tt evalRex}
command is used (cf. sect. \ref{rex}).

\subsection{Output of {\md\tt TwoLoopTens{\rm\it n}Pt{\rm\it m}}}

The functions {\tt TwoLoopTens{\rm\it n}Pt{\rm\it m}} return a full tensor as
output. The same types of arguments are allowed as for the corresponding {\tt
OneLoopTens{\rm\it n}Pt} functions described in sect.~\ref{tens} -- except the
qualifier {\tt more}.

\section{The {\md\tt OneLoopLib} procedures} \label{lib}

For different calculations of the same {\tt OneLoop} function it is of course
rather inconvenient to start the whole program again. Therefore our package
includes the functions
\bi
\item {\tt OneLoopLib2Pt}($i,j$)
\item {\tt OneLoopLib3Pt}($i,j$)
\ei
which generate a library of all {\tt OneLoop2Pt} and {\tt OneLoop3Pt} functions
respectively up to the tensor rank $i$. All powers of the denominators from 1 to
$j$ are considered. The second parameter $j$ is optional. If it is omitted the
procedures assume the value 1 for $j$, so that no powers of denominators higher
than one are calculated. These functions are then read by the proper procedures
described before, so that they in any case speed up.

If you use the {\tt OneLoopLib} functions within an ordinary \maple session and
want to write this library in any other than the actual directory you
have to assign the variable {\tt LibPath}: \\[0.4cm]
\verb+LibPath:=+{$\langle$\it path$\rangle$}\verb+;+ \\[0.4cm]
Now the library will be written to $\langle${\it path}$\rangle$. The procedures
will only look for the library if the entry {\sf Oneloop Library} of the
{\sf Options} menu is selected or if {\tt LibPath} is assigned. If the
procedures search for an integral which is not contained in the library it will 
be calculated and stored in the library on the fly. Each integral corresponds to
one file. In appendix \ref{alltable} we list the convention for the file names.

\section{Numerical integration with \md\vegas} \label{num}

For numerical integration a parallelized implementation of the \vegas algorithm
for \Cxx invented by R. Kreckel is provided with \xloops. To use the routines
within an ordinary \maple session one has to declare the variable
\verb+NumPath+: \\[0.4cm]
\verb+NumPath:=+{$\langle$\it path$\rangle$}\verb+;+ \\[0.4cm]
{$\langle$\it path$\rangle$} describes the directory where the integration
routines are located. If \xloops is installed properly this is the subdirectory
{\tt cxx} of the \xloops directory.

The routines are customized for those {\tt TwoLoop} procedures which cannot be
calculated analytically. The most elegant way to perform those numerical
integrations is described in sect.~\ref{output} and~\ref{twoout}. In addition
there is also the possibility to perform an integration separately. For that
purpose there exists the command \\[0.4cm]
\verb+NumIntC(+{$\langle$\it integrand$\rangle$}\verb+,[+$m_1,
m_2$\verb+],[+$n_1,n_2$\verb+]);+ \\[0.4cm]
{$\langle$\it integrand$\rangle$} denotes the expression which has to be
integrated numerically. \xloops expects the integration variables to be named
{\tt x} and {\tt y} -- as they are called by the {\tt EvalGraph2} and {\tt
TwoLoop} procedures. The variables $m_1,m_2,n_1,n_2$ represent the grids where
the integrand is evaluated: first $m_1$ iterations with $m_2$ points --
\verb+itmx+$=m_1$ and \verb+ncall+$=m_2$ are passed to \vegas -- and then $n_1$
iterations with $n_2$ points -- this corresponds to \verb+itmx+$=n_1$ and
\verb+ncall+$=n_2$. Real and imaginary part are integrated one after the other.
If only the real part is of interest one has to specify \\[0.4cm]
\verb+NumIntC(+{$\langle$\it integrand$\rangle$}\verb+,[+$m_1,
m_2$\verb+],[+$n_1,n_2$\verb+],Re);+ \\[0.4cm]
if only the imaginary part shall be evaluated one writes \\[0.4cm]
\verb+NumIntC(+{$\langle$\it integrand$\rangle$}\verb+,[+$m_1,
m_2$\verb+],[+$n_1,n_2$\verb+],Im);+ \\[0.4cm]
If \verb+NumIntC+ is called only with the first argument \xloops takes the
values $m_1=20$, $m_2=1000$, $n_1=5$ and $n_2=10000$.

The output has the following format:
\bdm
[\langle\mbox{\it result}\rangle,\langle\mbox{\it error}\rangle]\,.
\edm
{$\langle$\it result$\rangle$} describes the -- possibly complex -- numerical
result, {$\langle$\it error$\rangle$} is the uncertainty ($1\sigma$) of the
numerical integration which is returned by \vegas.

\verb+NumIntC+ performs the necessary internal steps ("`compile, link"')
and starts the \Cxx program as subprocess. \maple waits until the subprocess
terminates and finally re-reads the result.

The numerical integration%
\index{numerical integration@numerical integration\\ \nop} with \vegas should
be used only once by each user. Otherwise there will occur conflicts because
several numerical integrations would use the same \Cxx files.

\bfg[h]\bc
\epsfig{file=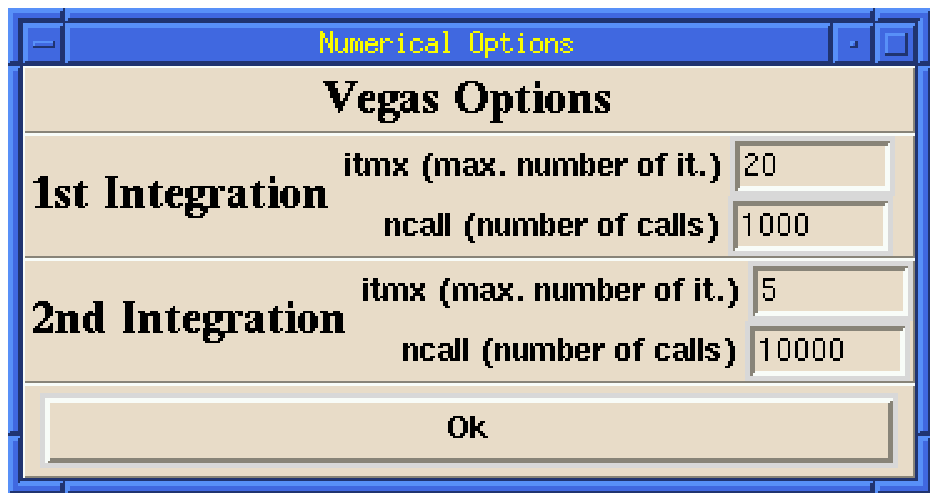,width=8cm}    
\caption{Options for \vegas} \label{fig:vegas}
\ec\efg
The entry {\sf Numeric} of the {\sf Options} menu enables the user to change
the parameters for \vegas. These parameters are set in a window displayed in
fig. \ref{fig:vegas}.

\chapter{Conclusion and outlook}

Our aim was to introduce a program package that gives any computer user the
possibility to calculate particle processes without much effort in
manpower and hardware. At this stage of development the program is still
understood to be under construction. In fact we are planning to
incorporate several additional features:
\bi
\item One-loop four-, five- and six-point functions.
\item Two-loop three- and four-point functions.
\item Calculation of complete processes, generation of all diagrams which
  contribute to a process in a given order.
\item Incorporation of other models, for instance supersymmetric theories.
\item An interface for adding other models.
\item A plotting device for automatical plotting of results.
\item Automatical renormalization, determination of $Z$ factors and renormalized
  functions.
\ei

\section*{Acknowledgements}

We would like to thank R. Stemler for his help in developing several
procedures and A. Frink for his contribution and advice in devoloping the
numerics and for valuable tests.

\begin{appendix}

\printappendix{chapter}

\chapter{Examples} \label{test}

\section{Installation} \label{instapp}

A typical run of the {\tt configure} script looks like the following:

{\SII
\begin{verbatim}
./configure 

 This is the XLOOPS installation script ! 

Which Release of Maple V are you using [1/2/3/4] ? 3

Enter the name of the WWW Browser executable,
 which XLOOPS should use ! netscape

Do you have Isolatin-1 character encoding [y/n] ? y

If you feel unsure, answer the following question with 'n'.
XLOOPS can work in single processor mode on multi processor machines as well.
Do you have multiprocessing (and POSIX thread library installed) [y/n] ? y
How many processors has your computer ? 2

On which platform are you installing xloops ? 
 Linux [l] Digital OSF [d] SUN Solaris [s] IBM AIX [a] Other Unix [o] l

Instead of creating the library you can also get a prebuilt library 
from the XLOOPS homepage and unpack it. 
Shall I create the library of one-loop functions
 (takes approx. 1 hour) [y/n] ? n

Now testing for Maple output bug while reading the library !   
----------------------------------------------------------------------------
    |\^/|     Maple V Release 3 (Universitaet Mainz)
._|\|   |/|_. Copyright (c) 1981-1994 by Waterloo Maple Software and the
 \  MAPLE  /  University of Waterloo. All rights reserved. Maple and Maple V
 <____ ____>  are registered trademarks of Waterloo Maple Software.
      |       Type ? for help.
> read(`loops.ma`);
Warning: new definition for   &*


       »»»»»»»»»»»»»»»»»» This is XLOOPS Version 1.0 «««««««««««««««««««

              by Lars Brücher, Johannes Franzkowski, Dirk Kreimer

             ©1993-1997 Johannes Gutenberg Universität Mainz, Germany

bytes used=1001220, alloc=851812, time=0.03
bytes used=2011736, alloc=1310480, time=0.03
> LibPath := `../lib/` ;
                               LibPath := ../lib/

> EvalGraph1(3,[higgs,higgs,higgs,higgs,higgs,higgs],full);
bytes used=3011944, alloc=1834672, time=0.03
> quit;
bytes used=3662072, alloc=1834672, time=0.03
----------------------------------------------------------------------------
Did you see any result of the Maple command 'EvalGraph' used above [y/n] ? n

 System configuration completed, doing now some checks ! 

                 This may take a while !

              Creating the index file for tcl !   

Now testing the Makefile for numerical integrations generated for your System ! 

rm -f main funct.o numint.o div0.o vegas.o gauss.o utils.o 
g++     -DGNU_COMPLEX -O3 -fomit-frame-pointer -finline-functions -m486 -malign-
functions=4 -malign-jumps=4 -malign-loops=4 -fexpensive-optimizations -funroll-l
oops -ffast-math -c funct.cxx
g++     -DGNU_COMPLEX -O3 -fomit-frame-pointer -finline-functions -m486 -malign-
functions=4 -malign-jumps=4 -malign-loops=4 -fexpensive-optimizations -funroll-l
oops -ffast-math -c numint.cxx
g++     -DGNU_COMPLEX -O3 -fomit-frame-pointer -finline-functions -m486 -malign-
functions=4 -malign-jumps=4 -malign-loops=4 -fexpensive-optimizations -funroll-l
oops -ffast-math -c div0.cxx
g++  -DGNU_COMPLEX -O3 -fomit-frame-pointer -finline-functions -m486 -malign-fun
ctions=4 -malign-jumps=4 -malign-loops=4 -fexpensive-optimizations -funroll-loop
s -ffast-math -D_REENTRANT -DPVEGAS_POSIX -c vegas.c
g++     -DGNU_COMPLEX -O3 -fomit-frame-pointer -finline-functions -m486 -malign-
functions=4 -malign-jumps=4 -malign-loops=4 -fexpensive-optimizations -funroll-l
oops -ffast-math -c gauss.cxx
g++     -DGNU_COMPLEX -O3 -fomit-frame-pointer -finline-functions -m486 -malign-
functions=4 -malign-jumps=4 -malign-loops=4 -fexpensive-optimizations -funroll-l
oops -ffast-math -c utils.cxx
g++  -o main funct.o numint.o div0.o vegas.o gauss.o utils.o  -lm -lpthread  mai
n.cxx
Division by 0 Errors will be ignored
Test: 1/0
Initializing SR-sequences with seed 876920052
 Input parameters for vegas:  ndim=   2  ncall=      968       2 thread(s)
                              ittot=    1  itmx=    5       22^1 hypercubes
                              nprn=  0  ALPH= 1.50
                              mds=  1  nd= 200                npg=2
                               xl[ 1]=           0 xu[ 1]=           1
                               xl[ 2]=           0 xu[ 2]=           1
 iteration no.   1 : integral =  -1.077957e-06 +/-    9.2e-07
 all iterations:   integral = -1.077957e-06+/-  9.2e-07 chi**2/IT n =         0
 iteration no.   2 : integral =   1.703926e-07 +/-    7.1e-07
 all iterations:   integral = -2.975584e-07+/-  5.6e-07 chi**2/IT n =       1.2
 iteration no.   3 : integral =   1.407306e-09 +/-    4.3e-07
 all iterations:   integral = -1.079265e-07+/-  3.4e-07 chi**2/IT n =      0.67
 iteration no.   4 : integral =  -6.310369e-08 +/-      4e-07
 all iterations:   integral = -8.938296e-08+/-  2.6e-07 chi**2/IT n =      0.45
 iteration no.   5 : integral =  -2.702396e-07 +/-    3.2e-07
 all iterations:   integral = -1.618004e-07+/-    2e-07 chi**2/IT n =      0.38
 Input parameters for vegas:  ndim=   2  ncall=     9800       2 thread(s)
                              ittot=    1  itmx=    3       70^1 hypercubes
                              nprn=  0  ALPH= 1.50
                              mds=  1  nd= 200                npg=2
\end{verbatim}}

\newpage

{\SII
\begin{verbatim}
                               xl[ 1]=           0 xu[ 1]=           1
                               xl[ 2]=           0 xu[ 2]=           1
 iteration no.   1 : integral =  -4.470697e-08 +/-    6.2e-08
 all iterations:   integral = -4.470697e-08+/-  6.2e-08 chi**2/IT n =         0
 iteration no.   2 : integral =   2.028027e-08 +/-    5.3e-08
 all iterations:   integral = -6.763662e-09+/-    4e-08 chi**2/IT n =      0.63
 iteration no.   3 : integral =  -2.663549e-08 +/-    5.1e-08
 all iterations:   integral = -1.435286e-08+/-  3.2e-08 chi**2/IT n =      0.36
Total Number of ignored Divisions by 0: 0
 Input parameters for vegas:  ndim=   2  ncall=      968       2 thread(s)
                              ittot=    1  itmx=    5       22^1 hypercubes
                              nprn=  0  ALPH= 1.50
                              mds=  1  nd= 200                npg=2
                               xl[ 1]=           0 xu[ 1]=           1
                               xl[ 2]=           0 xu[ 2]=           1
 iteration no.   1 : integral =   7.704579e-06 +/-    1.3e-06
 all iterations:   integral =  7.704579e-06+/-  1.3e-06 chi**2/IT n =         0
 iteration no.   2 : integral =   8.286222e-06 +/-    6.8e-07
 all iterations:   integral =  8.160574e-06+/-  6.1e-07 chi**2/IT n =      0.16
 iteration no.   3 : integral =   9.681104e-06 +/-    6.2e-07
 all iterations:   integral =  8.902334e-06+/-  4.3e-07 chi**2/IT n =       1.6
 iteration no.   4 : integral =   8.987195e-06 +/-      5e-07
 all iterations:   integral =  8.938647e-06+/-  3.3e-07 chi**2/IT n =       1.1
 iteration no.   5 : integral =   8.732733e-06 +/-    3.7e-07
 all iterations:   integral =   8.84922e-06+/-  2.5e-07 chi**2/IT n =      0.85
 Input parameters for vegas:  ndim=   2  ncall=     9800       2 thread(s)
                              ittot=    1  itmx=    3       70^1 hypercubes
                              nprn=  0  ALPH= 1.50
                              mds=  1  nd= 200                npg=2
                               xl[ 1]=           0 xu[ 1]=           1
                               xl[ 2]=           0 xu[ 2]=           1
 iteration no.   1 : integral =   8.890832e-06 +/-      1e-07
 all iterations:   integral =  8.890832e-06+/-    1e-07 chi**2/IT n =         0
 iteration no.   2 : integral =    8.87138e-06 +/-    5.3e-08
 all iterations:   integral =  8.875575e-06+/-  4.7e-08 chi**2/IT n =     0.029
 iteration no.   3 : integral =   8.798005e-06 +/-    4.7e-08
 all iterations:   integral =  8.836898e-06+/-  3.4e-08 chi**2/IT n =      0.68
Total Number of ignored Divisions by 0: 0

 Make seems to work, so Makefile is correct ! 

If you had problems please edit the Makefile in the subdirectory ./cxx 
until it compiles correctly. Please report your system configuration and 
Makefile changes to xloops@thep.physik.uni-mainz.de ! 

Do you agree sending the XLOOPS team a mail [Y/n] ? Y

    XLOOPS installed successfully, invoke with ./xloops !
\end{verbatim}}

\newpage

\section{Examples with the {\md\sf Xwindows} frontend}

To demonstrate the input and output of \xloops some typical
examples are given. The Feynman diagrams in this chapter are Postscript 
pictures produced with \xloops. 

\subsection{Output of one-loop integrals}

\subsubsection{Analytic examples}

\parbox{9cm}{
At first we want to show the different ways of evaluating the
diagrams. As an example we have chosen a simple self-energy diagram. The
advantage of this diagram is the relatively short result. The diagram
on the right shows the process, a nondiagonal $Z$ self-energy
with a $W$ bubble.}
\hfill
\parbox{5cm}{\epsfig{file=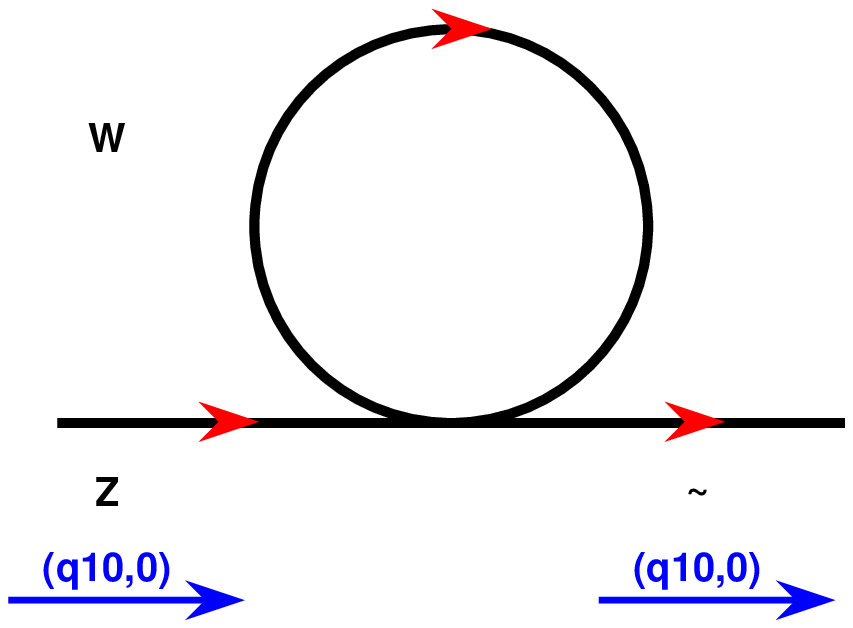,width=5cm}}\\
With \eval the following result is obtained:

{\SII
\begin{verbatim}
G1 := [
                  2                         2   2           2        2   2 2
     C1 = 1/16 I e  cos(tw) (1 + eps Ln(4 Pi  MU ) + 1/2 eps  Ln(4 Pi  MU ) )

                                             /            4
         (6 I - 4 I eps) Oneloop1Pt(0, Mw)  /  (sin(tw) Pi ),
                                           /

     C1 G(nu1, nu2)]
\end{verbatim}}

\noindent
As expected, the structure of form factors, indicated by the last
element of the list above, is just proportional to $g^{\mu\nu}$.
The pre-factor {\tt C1} is determined by an equation, which shows the
corresponding {\tt Oneloop} functions. In this example just {\tt
Oneloop1Pt(0, Mw)} appears. As pre-factor the coupling {\tt e}, the
Weinberg angle {\tt tw} and the renormalization parameter {\tt MU}
occur. 

With \evalf the {\tt OneLoop} function can be evaluated further:

{\SII
\begin{verbatim}
GF2 := [
                  2           2
               I e  cos(tw) Mw
   C1 = [- 3/8 ----------------,
                            2
                  sin(tw) Pi

                          2   2    2   2       2   2
       1/16 I (- 6 Ln(4 Pi  MU ) Pi  Mw  + 4 Pi  Mw

                        2                                  2   2   2
            - 6 (- Ln(Mw  - I rho) - Ln(Pi) - gamma + 1) Pi  Mw ) e  cos(tw)

              /            4
             /  (sin(tw) Pi )                                               ],
            /

   C1 G(nu1, nu2)]
\end{verbatim}}

\noindent 
Now the constant {\tt C1} is evaluated to a list, which elements 
represent the coefficients of the Laurent series in the dimensional 
regulator $\eps = (4 - D)/2$. The first element of the list shows 
the divergence of the integral, the second element the convergent part.

Additionally the next order in dimensional regularization can be evaluated.
Clicking on \evalm shows the result:

{\SII
\begin{verbatim}
GM3 := [
               2           2                     2   2    2   2       2   2          2
            I e  cos(tw) Mw        I (- 6 Ln(4 Pi  MU ) Pi  Mw  + 4 Pi  Mw  - 6 %1) e  cos(tw)
C1 = [- 3/8 ----------------, 1/16 -----------------------------------------------------------,
                         2                                       4
               sin(tw) Pi                              sin(tw) Pi

        2                     2   2 2   2   2
1/16 I e  cos(tw) (- 3 Ln(4 Pi  MU )  Pi  Mw

              2   2       2   2                       2         2
     + Ln(4 Pi  MU ) (4 Pi  Mw  - 6 %1) - 6 (1/2 Ln(Mw  - I rho)

                                   2                      2
     - (- Ln(Pi) - gamma + 1) Ln(Mw  - I rho) + 1/2 Ln(Pi)

                                     2            2                2   2
     - (- gamma + 1) Ln(Pi) + 1/12 Pi  + 1/2 gamma  - gamma + 1) Pi  Mw  + 4 %1

        /            4
    )  /  (sin(tw) Pi )
      /

],

C1 G(nu1, nu2)]
                      2                                  2   2
%1 :=         (- Ln(Mw  - I rho) - Ln(Pi) - gamma + 1) Pi  Mw
\end{verbatim}}

\noindent
In the above result the list for {\tt C1} was just enhanced by an additinal
element, the ${\cal O}(\eps^1)$. 

\subsubsection{Numerical examples}

In the following subsection we want to show some numerical examples,
where different kinds of form factors occur.
The values for couplings and particle masses where inserted with
{\sf Insert Particle Properties} from the {\sf Options} menu. The
momentum was set to {\tt q10:=Mz0;} and the renormalization parameter 
was chosen to be {\tt MU:=1;}. All results are given in natural units
$\hbar=c=1$ and the energy scale GeV.   

\bc\btb{ccc}
   \parbox{4.5cm}{\epsfig{file=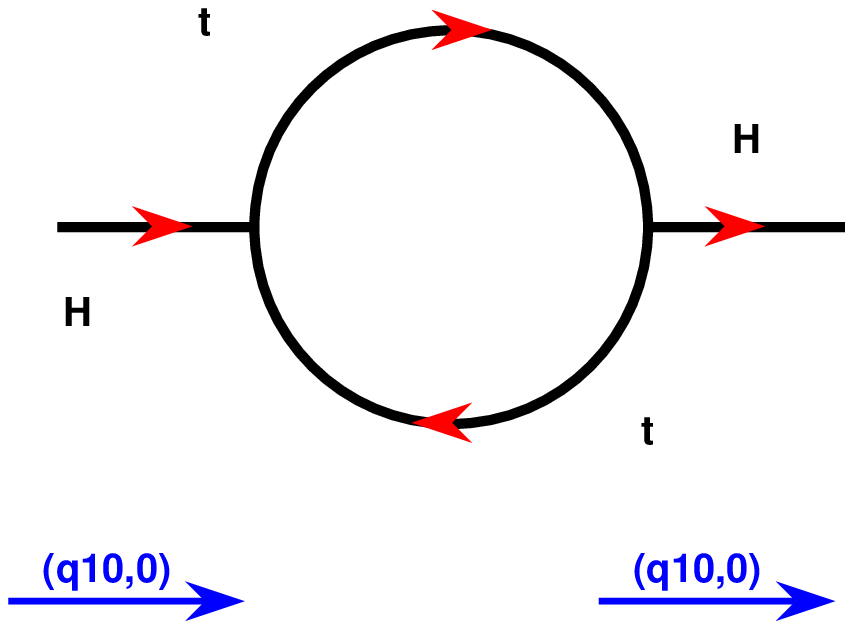,width=4.5cm}\\
      \vspace*{-0.8cm} \bc\SII (a) \ec}
  &
   \parbox{4.5cm}{\epsfig{file=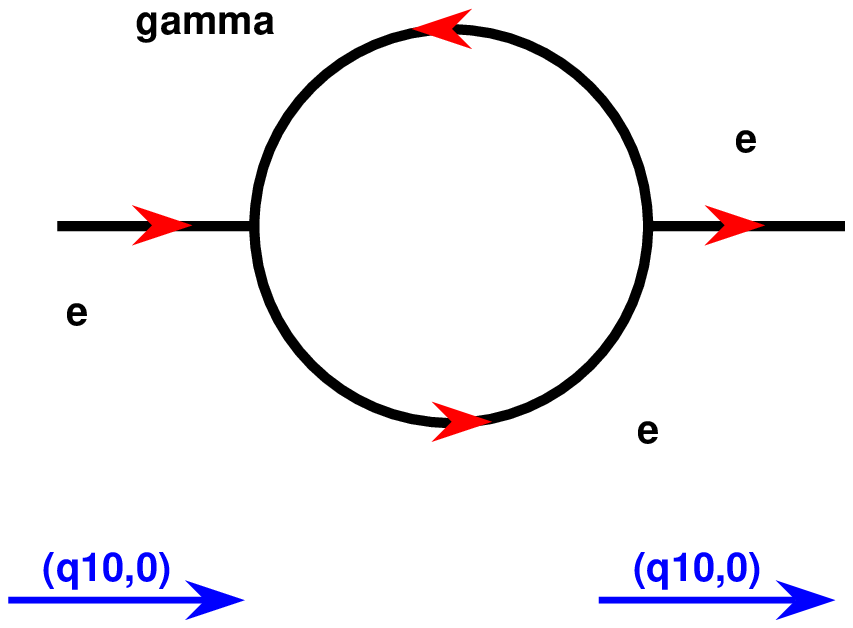,width=4.5cm}\\
      \vspace*{-0.8cm} \bc\SII (b) \ec}
  &
   \parbox{4.5cm}{\epsfig{file=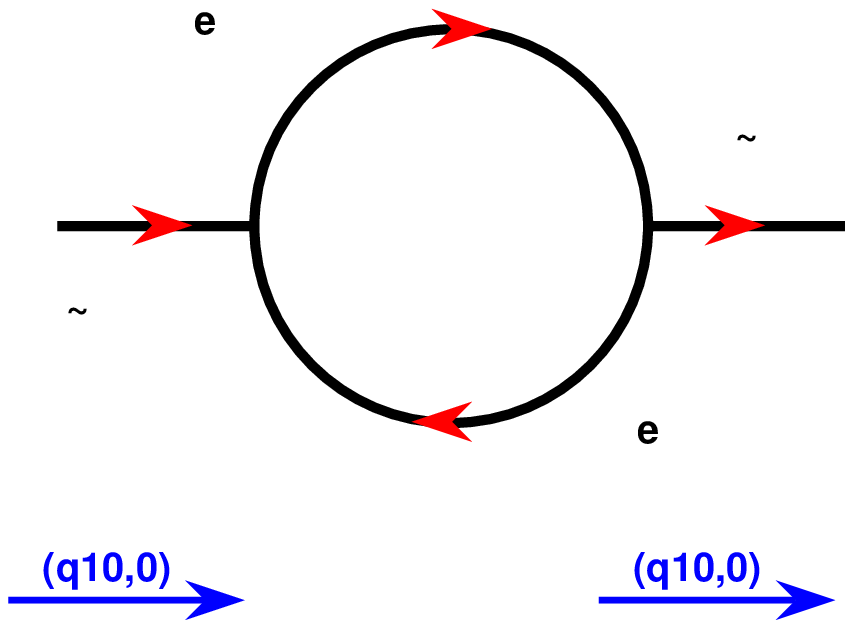,width=4.5cm}\\
      \vspace*{-0.8cm} \bc\SII (c) \ec}
  \\
 \etb\ec

\newpage

\noindent
The results, when evaluated with \evalf, read: \\
(a)

{\SII
\begin{verbatim}
GF4 := [
   
   C1 = [ - 2143.7993127296539430 I,

                                      -15
            - .33537943559210349312*10    + 16785.676157258815796 I], C1]
\end{verbatim}}

\noindent
(b)

{\SII
\begin{verbatim}
GF5 := [

     C1 = [ - .11869586429575443972*10   I,

                                  -5                           -5
          .37289405527131873878*10   + .66133100616139736842*10   I],

     C2 =

      [.052942866215076463492 I,  - .16632491956127147989 - .32145019412334161258 I],

     C1 (1 &* ONE) + C2 (1 &* Dg0)]
\end{verbatim}}

\noindent
(c)

{\SII
\begin{verbatim}
GF6 := [

     C1 =

       [ - 6.4357348171046949024 I, 20.218457221868161079 + 34.784995718164094540 I],

     C2 = [.00077427320705022066450 I,

          - .0024324510191403821401 - .0041849285213475828278 I],

     C1 G(nu1, nu2) + C2 q1(nu1) q1(nu2)]
\end{verbatim}}

Figure (a) shows a scalar self-energy. So the result just consists of
the scalar constant {\tt C1}. Like in the previous example {\tt C1} 
is expanded as a series in $\eps$. Figure (b) shows a fermionic self-energy, 
which has Dirac structure. So the form factors are the unity matrix {\tt ONE}
and {\tt Dg0}. The pre-factors {\tt C1} and
{\tt C2} are again represented as lists of coefficients of the Laurent series 
in $\eps$. Finally figure (c) shows a vector boson self-energy. The result
is proportional to the metric tensor {\tt G(nu1, nu2)} and to the product
of the external momenta {\tt q1(nu1) q1(nu2)}. Again two constants,
{\tt C1} and {\tt C2}, are needed.

\subsection{Two-loop diagrams}

Now some two-loop diagrams shall be evaluated numerically.
The values are chosen as in the previous section, if no comment is made.
First we want to show some two-loop diagrams which factorize, starting with
a photon self-energy:
\bc
\epsfig{file=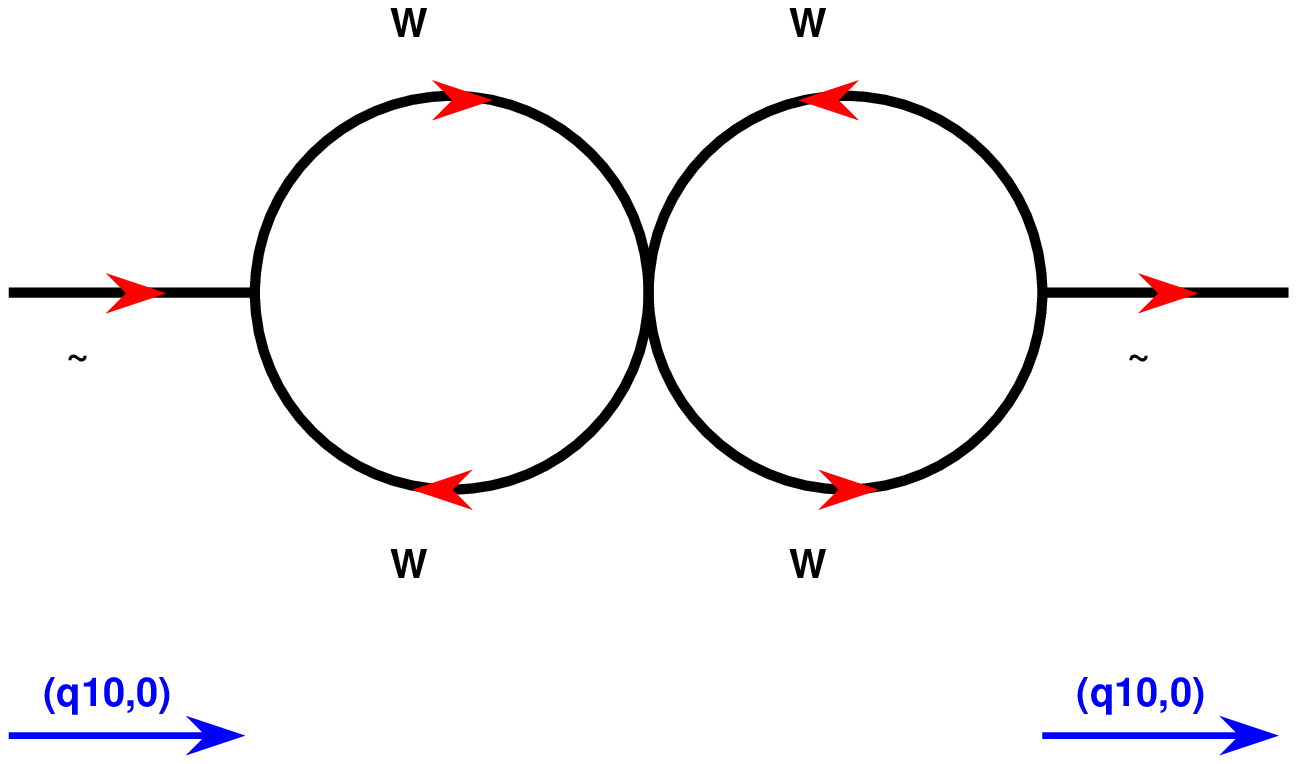,width=7cm}
\ec

{\SII
\begin{verbatim}
GF7 := [C1 = [ - .16744033026458920315 I,

                                  -19
          .70833637543010397531*10    + 2.1988161494279602884 I,

                                     -17
           - .42232923827218965144*10    - 14.715169445826507291 I],

         C2 = [.000020136974316008130172 I,

                                        -23
              - .85187071577025180931*10    - .00026443751189864126887 I,

                                     -21
             .50790828055269934758*10    + .0017696990248291598135 I],

         C1 G(nu1, nu2) + C2 q1(nu1) q1(nu2)]
\end{verbatim}}

The typical structure of a photon self-energy,
\bdm
\mbox{\tt  C1 G(nu1, nu2) + C2 q1(nu1) q1(nu2)}\,,
\edm
can be seen. The difference to the one-loop self-energy (c) is just the fact,
that this graph is more divergent. So the constants 
{\tt C1} and {\tt C2} have an additional list entry, the order
$\eps^{-2}$.

Next we demonstrate a three-point function. All momenta are on-shell, e.g.
{\tt q10:=Mz0; q20:=Mz0/2; q21:=sqrt(Mz0\^{}2-Melec\^{}2);} is passed
to \maple with {\sf Insert Maple Command}. 
\bc
\epsfig{file=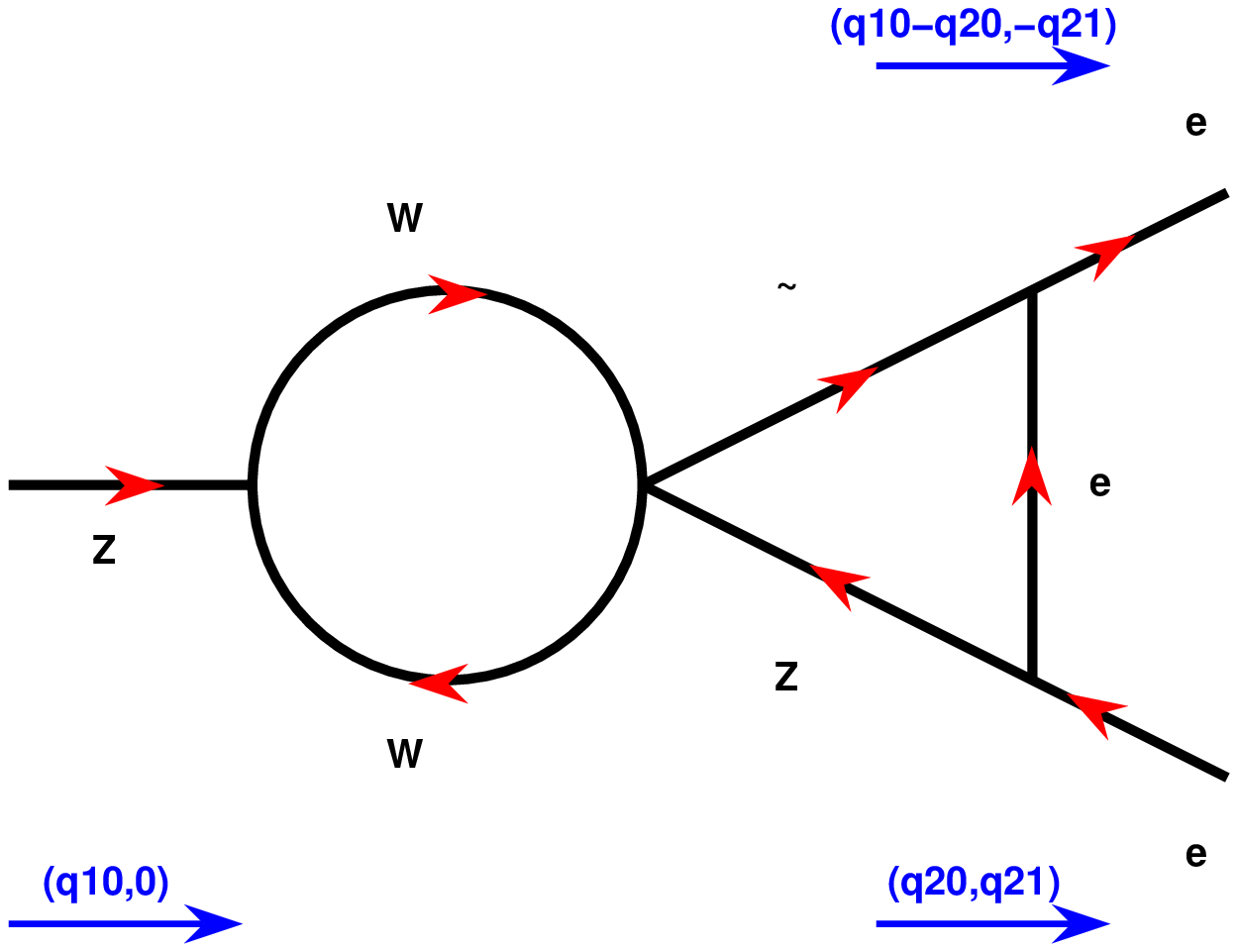,width=7cm}
\ec

{\SII
\begin{verbatim}
GF8 := [
                                      -35                           -25
 [C1 = [0,  - .82353974706799645789*10    - .64017338869752026115*10    I,

                                 -8                           -9
       - .30410069590311173109*10   - .86001332568491139242*10   I],

                                      -34                           -25
     C2 = [0, .16428685998560205727*10    - .72770813279248241968*10    I,

                                    -7                           -8
          - .31775075936778308993*10   - .89861644838148378070*10   I],

                                      -27                           -27
     C3 = [0, .25907616430369099002*10    + .17084563249491568053*10    I,

                                 -10                           -10
         .50023692425121327568*10    - .38336659269131090697*10    I],

                                      -27                           -26
     C4 = [0, .56375581291706017052*10    + .83584350815468097846*10    I,

                                 -9                           -9
         .52269088721608778944*10   - .40057463723369682265*10   I],

                                      -27                           -27
     C5 = [0, .11444962108462501382*10    - .21574341953884228136*10    I,

                                    -26                           -26
          - .13104643398124069502*10    + .25219418322926334904*10    I],

                                      -27                           -27
     C6 = [0, .19581694212974079513*10    - .32129543956635657630*10    I,

                                    -26                           -26
          - .22477792468575032187*10    + .37675015987648266663*10    I],

                                      -31                           -31
     C7 = [0, .13476989938404571673*10    + .21449295663578860996*10    I,

                        -23                           -14
         .10998568100*10    - .26575450025692144978*10    I],

                                         -31                           -31
     C8 = [0,  - .13476989941451368157*10    - .21449295669672453982*10    I,

                           -22                           -13
          - .11492263298*10    + .27768333121127352587*10    I],

                                      -26                           -26
     C9 = [0, .31582291657035101225*10    - .30101578572250827460*10    I,

                                 -9                           -10
         .10838466684439346062*10   + .56893589844240048516*10    I],

                                          -36                           -26
     C10 = [0,  - .98315817920409517794*10    - .13705392377889147738*10    I,

                                 -8                           -9
         .11324969215000662291*10   + .59447352866059701005*10   I],

                                       -27                           -26
     C11 = [0, .76454675644007305666*10    - .48786422812941166168*10    I,
\end{verbatim}}

\newpage

{\SII
\begin{verbatim}
                                 -9                           -9
         .25011846212167898923*10   - .19168329634264528256*10   I],

                                          -26                           -25
     C12 = [0,  - .20628807340956139895*10    - .21423923511009696554*10    I,

                                 -8                           -8
         .26134544360394034972*10   - .20028731861370353818*10   I],

     C1 (1 &* Dg(nu1)) + C2 &*(1, Dg5, Dg(nu1)) + C3 (1 &* Dg0) q2(nu1)

          + C4 &*(1, Dg5, Dg0) q2(nu1) + C5 (1 &* Dg1) q2(nu1)

          + C6 &*(1, Dg5, Dg1) q2(nu1) + C7 (1 &* ONE) q1(nu1)

          + C8 (1 &* Dg5) q1(nu1) + C9 (1 &* Dg0) q1(nu1)

          + C10 &*(1, Dg5, Dg0) q1(nu1) + C11 (1 &* Dg1) q1(nu1)

          + C12 &*(1, Dg5, Dg1) q1(nu1)                                ]
\end{verbatim}}

Of course the result has more form factors as the diagrams evaluated
before. The reason for this is, that as external particles vector bosons
as well as fermions occur.

\chapter{Technical details}

\section{File structure}

\bc
\epsfig{file=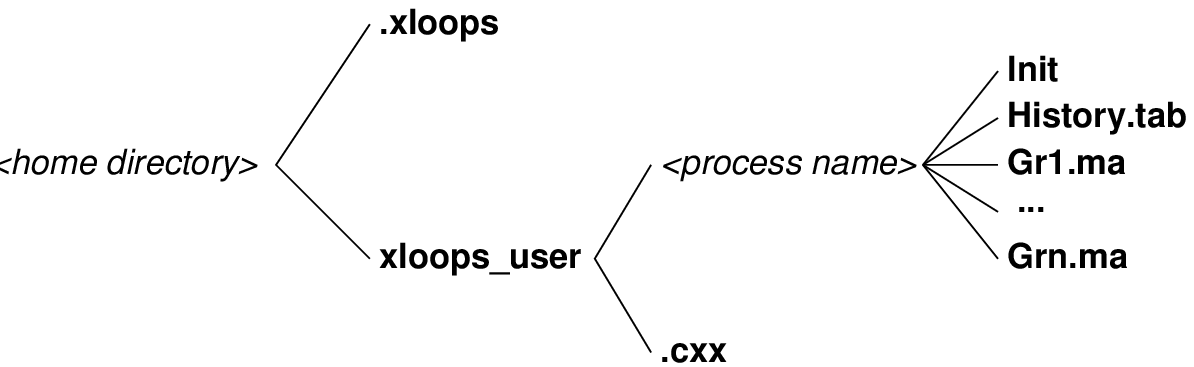}    
\ec

To understand which files \xloops uses for loading and saving of processes,
it is necessary to have a short glance on the file structure, which \xloops
writes into every user's home directory. First of all \xloops creates 
the subdirectory {\tt xloops\_user}. In this directory \xloops writes all 
user-specific data. If the user now starts to input a new process, \xloops
creates in
{\tt xloops\_user} a new subdirectory, which has the same name as the process.
In this subdirectory \xloops writes to basic files: the file {\tt Init} with
all options entered for this process (e.g. the model), and the file
{\tt History.tab}, where for every diagram a corresponding number
and a particle list is saved. If one graph from this list is calculated,
\xloops creates a new file to save the result. The name of this file includes
the number of the diagram in the particle list. If the user now
tries to evaluate this graph again, \xloops looks for the corresponding file
and reads the result.

\section{The one-loop library} \label{alltable}

To avoid multiple evaluation of one-loop integrals \xloops can read these
integrals from a pre-built library. This library is produced with by the 
{\tt OneLoopLib} procedures and written to the directory, which is indicated
by the variable {\tt LibPath}. If the {\sf Xwindows} interface is used, this
variable points to a common directory, where all integrals are saved. 
Otherwise the user can set this variable to the directory, where he has
his own library stored. If some integrals in the library are missing, 
\xloops creates them automatically.\footnote{Currently only supported on 
{\sf Unix} platforms.}

In practice it is necessary to store
not only the general mass case of each integral. Several special cases are also
needed. The correspondence is
\bc
\tt
B$p_0p_1t_1t_2n_1$.1LOOP $\Longleftrightarrow$
OneLoop2Pt$(p_0,p_1,q,m_1,m_2,t_1,t_2)$ \\[0.4cm]
C$p_0p_1p_2t_1t_2t_3n_1n_2i_1i_2i_3i_4i_5i_6$.1LOOP $\Longleftrightarrow$
OneLoop3Pt$(p_0,p_1,p_2,q_{1},q_{20},q_{21},m_1,m_2,m_3,t_1,t_2,t_3)\,.$
\ec
The additional numbers $n_1,n_2$ and the optional
numbers $i_1,i_2,i_3,i_4,i_5,i_6$ specify the different kinematical arrangements
which have to be taken into account. In the following tables we list the
conditions which belong to each case:

\bc
\addtolength{\tabcolsep}{-1.7pt}
Two-point functions: \\[0.4cm]
\btb{|l|l|c|c|c|}
\hline
\multicolumn{2}{|c|}{conditions} &       & method of & effective \\
1 & 2                            & $n_1$ & solution  & {\it n}-point \\
\hline
\hline
$q=0$       & $m_1=m_2$          & triv. & $D$       & 1 \\
            & ---                &       & $P$       & 1 \\
\hline
$m_1=0$     & ---                & 1     & $R$       & 2 \\
$m_2=0$     &                    & 2     & $R$       & 2 \\
\hline
$q^2+m_1^2-m_2^2=0$ & ---        & 3     & $R$       & 2 \\
$q^2-m_1^2+m_2^2=0$ &            & 4     & $R$       & 2 \\
\hline
$q+m_1+m_2=0$ & ---        & 5     & $R$       & 2 \\
$q-m_1+m_2=0$ &            & 6     & $R$       & 2 \\
$q+m_1-m_2=0$ & ---        & 7     & $R$       & 2 \\
$q-m_1-m_2=0$ &            & 8     & $R$       & 2 \\
\hline
\etb \\[0.8cm]  
Three-point functions: \\[0.4cm]
\btb{|l|l|l|l|c|c|c|}
\hline
\multicolumn{4}{|c|}{conditions}                        &          & method of & effective \\
1          & 2            & 3         & 4               & $n_1n_2$ & solution & {\it n}-point \\
\hline
\hline
$q_{21}=0$ & $q_{20}=0$   & $q_1=0$ & $m_1=m_2=m_3$ & triv. & $2\times D$ & 1 \\
\cline{4-7}
           &              &           & $m_1=m_2$       & triv. & $P$ $D$ & 1 \\
           &              &           & $m_1=m_3$       &       &         & 1 \\
           &              &           & $m_2=m_3$       &       &         & 1 \\
\cline{4-7}
           &              &           & ---             & triv. & $P$     & 1 \\
\cline{3-7}
           &              & $m_2=m_3$ & $q_1+m_1+m_3=0$ & 11    & $D$ $R$ & 2 \\
           &              &           & $q_1+m_1-m_3=0$ & 12    &         &   \\
           &              &           & $q_1-m_1+m_3=0$ & 13    &         &   \\
           &              &           & $q_1-m_1-m_3=0$ & 14    &         &   \\
\cline{4-7}
           &              &           & ---             & 10    & $D$     & 2 \\
\cline{2-7}
           & $q_1=q_{20}$ & $m_1=m_2$ & $q_1+m_1+m_3=0$ & 21    & $D$ $R$ & 2 \\
           &              &           & $q_1+m_1-m_3=0$ & 22    &         &   \\
           &              &           & $q_1-m_1+m_3=0$ & 23    &         &   \\
           &              &           & $q_1-m_1-m_3=0$ & 24    &         &   \\
\cline{4-7}
           &              &           & ---             & 20    & $D$     & 2 \\
\cline{2-7}
           & $q_{20}=0$   & ---       & ---             & 40    & $P$     & 2 \\
\cline{2-7}
           & $q_1=0$      & ---       & ---             & 50    & $P$ $L$ & 2 \\
\cline{2-7}
           & $q_1=q_{20}$ & ---       & ---             & 60    & $P$     & 2 \\
\cline{2-7}
           & \multicolumn{3}{l|}
{$(q_{20}^2-m_2^2)q_1-(q_1^2-m_1^2)q_{20}-m_3^2(q_{20}-q_1)=0$} & 70 & $P$ & 2 \\
\cline{2-7}
           & ---          & ---       & ---             & 80       & $P$  & 2 \\
\hline
$q_1=0$ & $m_1=m_3$ & \multicolumn{2}{l|}{$\sqrt{q_{20}^2-q_{21}^2}+m_1+m_2=0$} & 31 & $L$ $D$ $R$ & 2 \\
        &           & \multicolumn{2}{l|}{$\sqrt{q_{20}^2-q_{21}^2}+m_1-m_2=0$} & 32 & & \\
        &           & \multicolumn{2}{l|}{$\sqrt{q_{20}^2-q_{21}^2}-m_1+m_2=0$} & 33 & & \\
        &           & \multicolumn{2}{l|}{$\sqrt{q_{20}^2-q_{21}^2}-m_1-m_2=0$} & 34 & & \\
\cline{3-7}
           &              & ---       & ---             & 30     & $L$ $D$ & 2 \\
\cline{2-7}
           & ---          & ---       & ---             & 90     & $P$ $L$ & 2 \\
\hline
\etb

\btb{|l|l|l|l|c|c|c|}
\hline
\multicolumn{4}{|c|}{conditions}                        &          & method of & effective \\
1          & 2            & 3         & 4               & $n_1n_2$ & solution  & {\it n}-point \\
\hline
\hline
$q_{20}-q_1=q_{21}$ & $q_{20}=-q_{21}$ & --- & --- & --- & & 3 \\
\cline{2-7}
 & $m_3\not=0$                         & --- & --- & C0  & $S$ & 3 \\
\hline
$q_{20}-q_1=-q_{21}$ & $q_{20}=q_{21}$ & --- & --- & --- & & 3 \\
\cline{2-7}
 & $m_3\not=0$                         & --- & --- & D0  & $S$ & 3 \\
\hline
$q_{20}=q_{21}$ & $m_2\not=0$          & --- & --- & E0  & $S$ & 3 \\
\hline
$q_{20}=-q_{21}$ & $m_2\not=0$         & --- & --- & F0  & $S$ & 3 \\
\hline
$q_{20}^2-q_{21}^2=m_2^2$ & $q_1^2=m_1^2$ & $m_3=0$ & --- & G0 & $S$ & 3 \\
\hline
$(q_{20}-q_1)^2-q_{21}^2=m_1^2$ & $q_{20}^2-q_{21}^2=m_3^2$ & $m_2=0$ & --- & H0 & $S$ & 3 \\
\hline
$(q_{20}-q_1)^2-q_{21}^2=m_2^2$ & $q_1^2=m_3^2$ & $m_1=0$ & --- & I0 & $S$ & 3 \\
\hline
\etb \\[0.8cm]
\ec
The table shows in the first columns the conditions to the masses $m_i$
and the external momenta for a special case. Then 
the number occuring in the name of the file, in which the special case is saved
is given. In the following two columns the method of solving this integrals
and the effective {\it n}-point function to solve is given. The abbreviations
for the method of solution have the following meaning: \\
\btb{rl}
 $D$ & Differentiation to the squares of masses.\\
 $P$ & Solving by partial fraction decomposition.\\
 $R$ & Special case of the \Rfc, either vanishing or equal arguments.\\
 $L$ & Solution by Lorentz transformation and  \\
     & reduction of parallel space. \\
 $S$ & Summands vanish in the sum of residua,\\
     & e.g. some residua are 0.\\
\etb

For the tensor integrals of the three point function the following
additional cases occur:
\bc
\btb{|l|l|l|c|c|c|}
\hline
\multicolumn{3}{|c|}{conditions} &                    & method of & effective \\
1          & 2            & 3    &                    & solution & {\it n}-point \\
\hline
\hline
$m_1=0$ & ---     & ---     & $i_1=4$ & $R$ & 3 \\
\cline{2-6}
        & $m_2=0$ & ---     & $i_1=6$ & $R$ & 3 \\
\cline{3-6}
        &         & $m_3=0$ & $i_1=7$ & $R$ & 3 \\
\cline{2-6}
        & $m_3=0$ & ---     & $i_1=5$ & $R$ & 3 \\
\hline
$m_2=0$ & ---     & ---     & $i_1=2$ & $R$ & 3 \\
\cline{2-6}
        & $m_3=0$ & ---     & $i_1=3$ & $R$ & 3 \\
\hline
$m_3=0$ & ---     & ---     & $i_1=1$ & $R$ & 3 \\
\hline
$q_1^2-m_1^2+m_3^2=0$    & $m_1\not=0$   & --- & $i_2=1$ & $R$ & 3 \\
$q_1^2+m_1^2-m_3^2=0$    &               &     & $i_2=2$ & $R$ & 3 \\
\hline
$q_1^2-m_2^2+m_3^2=0$    & $m_2\not=0$   & --- & $i_3=1$ & $R$ & 3 \\
$q_1^2+m_2^2-m_3^2=0$    &               &     & $i_3=2$ & $R$ & 3 \\
\hline
$q_{20}^2-m_2^2+m_3^2=0$ & $m_2\not=0$   & --- & $i_4=1$ & $R$ & 3 \\
$q_{20}^2+m_2^2-m_3^2=0$ &               &     & $i_4=2$ & $R$ & 3 \\
$q_{20}^2-q_{21}^2-m_2^2+m_3^2=0$ &      &     & $i_4=3$ & $R$ & 3 \\
$q_{20}^2-q_{21}^2+m_2^2-m_3^2=0$ &      &     & $i_4=4$ & $R$ & 3 \\
\hline
$q_1^2-m_1^2+m_2^2=0$    & $m_1\not=0$   & --- & $i_5=1$ & $R$ & 3 \\
$q_1^2+m_1^2-m_2^2=0$    &               &     & $i_5=2$ & $R$ & 3 \\
$(q_{20}-q_1)^2-m_1^2+m_2^2=0$    &      &     & $i_5=3$ & $R$ & 3 \\
$(q_{20}-q_1)^2+m_1^2-m_2^2=0$    &      &     & $i_5=4$ & $R$ & 3 \\
\hline
$q_{20}^2-q_{21}^2-m_1^2+m_2^2=0$ & $m_2\not=0$ & --- & $i_6=1$ & $R$ & 3 \\
$q_{20}^2-q_{21}^2+m_1^2-m_2^2=0$ &             &     & $i_6=2$ & $R$ & 3 \\
\hline
\etb
\ec
To make the notation clearer, we give the following examples:
\\ \nop
\bc
{\tt\btb{ll} 
B$p_0p_1t_1t_2$.1LOOP       & OneLoop2Pt$(p_0,p_1,q,m_1,m_2,t_1,t_2)$ \\
B$p_0p_1t_1t_2$1.1LOOP      & OneLoop2Pt$(p_0,p_1,q,0,m_2,t_1,t_2)$ \\
B$p_0p_1t_1t_2$2.1LOOP      & OneLoop2Pt$(p_0,p_1,q,m_1,0,t_1,t_2)$ \\
B$p_0p_1t_1t_2$3.1LOOP
     & OneLoop2Pt$(p_0,p_1,\sqrt{m_2^2-m_1^2},m_1,m_2,t_1,t_2)$ \\
B$p_0p_1t_1t_2$4.1LOOP
     & OneLoop2Pt$(p_0,p_1,\sqrt{m_1^2-m_2^2},m_1,m_2,t_1,t_2)$ \\[0.5cm]
C$p_0p_1p_2t_1t_2t_3$.1LOOP \qquad  
     & OneLoop3Pt$(p_0,p_1,p_2,q_{1},q_{20},q_{21},m_1,m_2,m_3,t_1,t_2,t_3)$ \\
C$p_0p_1p_2t_1t_2t_3$1.1LOOP \qquad  
     & OneLoop3Pt$(p_0,p_1,p_2,q_{1},q_{20},0,m_1,m_2,m_3,t_1,t_2,t_3)$ \\
C$p_0p_1p_2t_1t_2t_3$2.1LOOP \qquad  
     & OneLoop3Pt$(p_0,p_1,p_2,q_{1},q_{20},q_{21},q_1,\sqrt{q_{20}^2-q_{21}^2},
       0,t_1,t_2,t_3)$ \\
C$p_0p_1p_2t_1t_2t_3$3.1LOOP \qquad  
     & OneLoop3Pt$(p_0,p_1,p_2,q_{1},q_{20},q_{1}-q_{20},m_1,m_2,m_3,
       t_1,t_2,t_3)$ \\
C$p_0p_1p_2t_1t_2t_3$4.1LOOP \qquad  
     & OneLoop3Pt$(p_0,p_1,p_2,q_{1},q_{20},-q_{20},m_1,m_2,m_3,t_1,t_2,t_3)$ \\
C$p_0p_1p_2t_1t_2t_3$5.1LOOP \qquad  
     & OneLoop3Pt$(p_0,p_1,p_2,q_{1},q_{20},q_{20}-q_{1},m_1,m_2,m_3,
       t_1,t_2,t_3)$ \\
C$p_0p_1p_2t_1t_2t_3$6.1LOOP \qquad  
     & OneLoop3Pt$(p_0,p_1,p_2,q_{1},q_{20},q_{20},m_1,m_2,m_3,t_1,t_2,t_3)$ \\
C$p_0p_1p_2t_1t_2t_3$7.1LOOP \qquad  
     & OneLoop3Pt$(p_0,p_1,p_2,q_{1},0,0,m_1,m_2,m_3,t_1,t_2,t_3)$ \\
C$p_0p_1p_2t_1t_2t_3$8.1LOOP \qquad  
     & OneLoop3Pt$(p_0,p_1,p_2,q_{1},q_{1},0,m_1,m_2,m_3,t_1,t_2,t_3)$ \\
C$p_0p_1p_2t_1t_2t_3$9.1LOOP \qquad  
     & OneLoop3Pt$(p_0,p_1,p_2,0,q_{20},q_{21},m_1,m_2,m_3,t_1,t_2,t_3)$ \\
\etb}
\ec

\section{Run time and numerical stability} \label{run}

Typical run times for the different integrals are displayed in the following
table. We used two different systems: (i) a DEC Alpha 8400 workstation (ii) a
PC with i586 chip, 32~MB RAM and 133 MHz frequency running Linux. 
All systems are working with \maplev. Without using the library we
found the following run times:

\bc
\begin{tabular}{|c|c|c|c|c|}
\hline
 & \multicolumn{2}{c|}{DEC Alpha 8400} & \multicolumn{2}{c|}{i586 133MHz} \\
\hline
tensor rank & numerical & algebraical & numerical & algebraical \\
\hline
\hline
\multicolumn{5}{|c|}{One-loop two-point functions} \\
\hline
 0 &  0.02 s & 0.04 s & 0.04 s & 0.06 s \\
 1 &  0.04 s & 0.21 s & 0.06 s & 0.26 s \\
 2 &  0.07 s & 0.28 s & 0.07 s & 0.48 s \\
 3 &  0.08 s & 0.41 s & 0.09 s & 0.76 s \\
\hline
\hline
\multicolumn{5}{|c|}{One-loop three-point functions} \\
\hline
 0 & 2.54 s & 1.83 s &  4.83 s &  3.31 s \\
 1 & 2.57 s & 2.23 s &  5.09 s &  3.69 s \\
 2 & 3.01 s & 3.62 s &  5.67 s &  6.14 s \\
 3 & 3.78 s & 4.16 s &  7.61 s &  34.58s \\
\hline
\end{tabular} \\[0.6cm]
\ec

It should be emphasized that these are the times which are necessary to 
generate the functions once and forever using the {\tt OneLoopLib{\rm\it n}Pt}
routines. If they once are stored in the {\tt LibPath} directory they may be
read in quickly. With assigned {\tt LibPath} we get:

\newpage
\bc
\begin{tabular}{|c|c|c|c|c|}
\hline
 & \multicolumn{2}{c|}{DEC Alpha 8400}  & \multicolumn{2}{c|}{i586 133MHz} \\
\hline
tensor rank & numerical & algebraical & numerical & algebraical \\
\hline
\hline
\multicolumn{5}{|c|}{One-loop two-point functions} \\
\hline
 0 & 0.34 s & 0.03 s & 0.08 s & 0.02 s \\
 1 & 0.39 s & 0.05 s & 0.09 s & 0.03 s \\
 2 & 0.43 s & 0.06 s & 0.19 s & 0.03 s \\
 3 & 0.47 s & 0.06 s & 0.20 s & 0.04 s \\
\hline
\hline
\multicolumn{5}{|c|}{One-loop three-point functions} \\
\hline
 0 & 1.30 s & 0.09 s & 4.77 s & 0.47 s \\
 1 & 1.32 s & 0.10 s & 4.81 s & 0.57 s \\
 2 & 1.36 s & 0.12 s & 5.20 s & 0.74 s \\
 3 & 1.88 s & 0.45 s & 6.72 s & 1.71 s \\
\hline
\end{tabular}
\ec

The accuracy and stability of numerical results may suffer from cancellations 
of large, approximately equal dilogarithms. Since \maple supports calculations 
of arbitrary precision, it is always possible to increase the number of digits 
to improve accuracy -- instead of Fortran programs which are limited to 
Fortran's restricted accuracy.

Usually we calculated with 20 or 40 digits. For all practical purposes 
our numerical results were not influenced significantly -- only in the last 
four or five digits -- if the number of digits increased.

\section{The \xloops distribution} \label{dist}

The distribution (packed in {\tt xloops.tgz}) consists of the following files:

{\SII
\begin{verbatim}
README           graphen0.tcl     graphenp.tcl     particles.txt
XLwidget.tcl     graphen1.tcl     help.tcl         xlinit.tcl
configure*       graphen2.tcl     history.tcl      xloops.prototyp
convert.tcl      graphen2a.tcl    maple.tcl
graphen.tcl      graphen2b.tcl    model.tcl

MapleVR1:
cfcn.ma             make_lib.map        numint.ma_unix      simple.ma
evalproc.ma         mess.isolatin       numint.ma_vms       test_maple_bug.map
fmrules.ma          mess.ma             oneloop.ma          twoloop.ma
fmuser.ma           mess.no_iso         pv.ma               unvalue.ma
loops.ma            numint.ma           r.ma                values.ma

MapleVR3:
cfcn.ma             make_lib.map        numint.ma_unix      simple.ma
evalproc.ma         mess.isolatin       numint.ma_vms       test_maple_bug.map
fmrules.ma          mess.ma             oneloop.ma          twoloop.ma
fmuser.ma           mess.no_iso         pv.ma               unvalue.ma
loops.ma            numint.ma           r.ma                values.ma

cxx:
div0.cxx          gauss.h           makehead.solaris  numint2.vms*
div0.h            incl.cxx          makethread.0      nvegas.c
dpc1.in           main.cxx          makethread.1      pvegas.c
dpc2.in           makefile.rest     makethread.2      utils.cxx
dpc3.in           makefile.vms      numint.cxx        utils.h
dpc4.in           makehead.AIX      numint.prototyp   vegas.c
funct.cxx         makehead.Linux    numint1.unix*
funct.h           makehead.generic  numint1.vms
gauss.cxx         makehead.osf      numint2.unix*

lib:
B00110.1loop

manual:
manual.html

xbms:
Hp1l1t0.xbm     Hp3l2t9.xbm     Hp4l2t9.xbm     p3l0t2.xbm      p4l2t13.xbm
Hp1l2t0.xbm     Hp4l0t0.xbm     attention.xbm   p3l1t0.xbm      p4l2t14.xbm
Hp1l2t1.xbm     Hp4l1t0.xbm     file.xbm        p3l1t1.xbm      p4l2t15.xbm
Hp1l2t2.xbm     Hp4l1t1.xbm     help.xbm        p3l1t4.xbm      p4l2t16.xbm
Hp2l0t0.xbm     Hp4l1t2.xbm     insmath.xbm     p3l1t5.xbm      p4l2t17.xbm
Hp2l1t0.xbm     Hp4l2t0.xbm     leftarrow.xbm   p3l2t0.xbm      p4l2t18.xbm
Hp2l1t1.xbm     Hp4l2t1.xbm     p1l1t0.xbm      p3l2t1.xbm      p4l2t19.xbm
Hp2l2t0.xbm     Hp4l2t10.xbm    p1l1t1.xbm      p3l2t10.xbm     p4l2t2.xbm
Hp2l2t1.xbm     Hp4l2t11.xbm    p1l2t0.xbm      p3l2t11.xbm     p4l2t20.xbm
Hp2l2t10.xbm    Hp4l2t12.xbm    p1l2t1.xbm      p3l2t12.xbm     p4l2t21.xbm
Hp2l2t11.xbm    Hp4l2t13.xbm    p1l2t2.xbm      p3l2t13.xbm     p4l2t22.xbm
Hp2l2t2.xbm     Hp4l2t14.xbm    p1l2t3.xbm      p3l2t2.xbm      p4l2t23.xbm
Hp2l2t3.xbm     Hp4l2t15.xbm    p1proc.xbm      p3l2t3.xbm      p4l2t24.xbm
Hp2l2t4.xbm     Hp4l2t16.xbm    p2l0t0.xbm      p3l2t4.xbm      p4l2t25.xbm
Hp2l2t5.xbm     Hp4l2t17.xbm    p2l0t1.xbm      p3l2t5.xbm      p4l2t26.xbm
Hp2l2t6.xbm     Hp4l2t18.xbm    p2l1t0.xbm      p3l2t6.xbm      p4l2t27.xbm
Hp2l2t7.xbm     Hp4l2t19.xbm    p2l1t1.xbm      p3l2t7.xbm      p4l2t28.xbm
Hp3l0t0.xbm     Hp4l2t2.xbm     p2l1t2.xbm      p3l2t8.xbm      p4l2t3.xbm
Hp3l1t0.xbm     Hp4l2t20.xbm    p2l1t3.xbm      p3l2t9.xbm      p4l2t4.xbm
Hp3l1t1.xbm     Hp4l2t21.xbm    p2l2t0.xbm      p3proc.xbm      p4l2t5.xbm
Hp3l2t0.xbm     Hp4l2t22.xbm    p2l2t1.xbm      p4l0t0.xbm      p4l2t6.xbm
Hp3l2t1.xbm     Hp4l2t23.xbm    p2l2t10.xbm     p4l0t3.xbm      p4l2t7.xbm
Hp3l2t10.xbm    Hp4l2t24.xbm    p2l2t11.xbm     p4l1t0.xbm      p4l2t8.xbm
Hp3l2t11.xbm    Hp4l2t25.xbm    p2l2t2.xbm      p4l1t1.xbm      p4l2t9.xbm
Hp3l2t12.xbm    Hp4l2t26.xbm    p2l2t3.xbm      p4l1t2.xbm      p4proc.xbm
Hp3l2t13.xbm    Hp4l2t27.xbm    p2l2t4.xbm      p4l1t6.xbm      rightarrow.xbm
Hp3l2t2.xbm     Hp4l2t28.xbm    p2l2t5.xbm      p4l1t7.xbm      schraff.xbm
Hp3l2t3.xbm     Hp4l2t3.xbm     p2l2t6.xbm      p4l1t8.xbm      xloops.gif
Hp3l2t4.xbm     Hp4l2t4.xbm     p2l2t7.xbm      p4l2t0.xbm      xloops.xbm
Hp3l2t5.xbm     Hp4l2t5.xbm     p2l2t8.xbm      p4l2t1.xbm
Hp3l2t6.xbm     Hp4l2t6.xbm     p2l2t9.xbm      p4l2t10.xbm
Hp3l2t7.xbm     Hp4l2t7.xbm     p2proc.xbm      p4l2t11.xbm
Hp3l2t8.xbm     Hp4l2t8.xbm     p3l0t0.xbm      p4l2t12.xbm
\end{verbatim}}






\end{appendix}

\bibliography{hep}

\printindex

\end{document}